\documentclass[lettersize,journal]{IEEEtran}
\usepackage{amsmath,amsfonts}
\usepackage{algorithmic}
\usepackage{array}
\usepackage{textcomp}
\usepackage{stfloats}
\usepackage{url}
\usepackage{verbatim}
\usepackage{graphicx}
\usepackage{cite}
\usepackage{multirow}
\usepackage[table,xcdraw]{xcolor}
\usepackage{caption}
\usepackage{subfigure}
\usepackage{tablefootnote}
\usepackage{makecell}

\usepackage{bbding}

\begin{document}
\title{MaskCRT: Masked Conditional Residual Transformer for Learned Video Compression}

\author{\textcolor{black}{Yi-Hsin Chen, Hong-Sheng Xie, Cheng-Wei Chen, Zong-Lin Gao, Martin Benjak, \\Wen-Hsiao Peng,~\IEEEmembership{Senior Member,~IEEE}, and Jörn Ostermann,~\IEEEmembership{Fellow,~IEEE}} \thanks{\textcolor{black}{Yi-Hsin Chen, Hong-Sheng Xie, Cheng-Wei Chen, Zong-Lin Gao, and Wen-Hsiao Peng are affiliated with the Department of Computer Science, National Yang Ming Chiao Tung University, Hsinchu, Taiwan (e-mail: wpeng@cs.nctu.edu.tw).}}
\thanks{\textcolor{black}{Martin Benjak and Jörn Ostermann are affiliated with the Institut für Informationsverarbeitung, Leibniz Universität Hannover, Hannover, Germany.}}
\thanks{This work is supported by National Science and Technology Council, Taiwan (112-2634-F-A49-007-, 110-2221-E-A49-065-MY3, 111-2923-E-A49-007-MY3), MediaTek, and National Center for High-performance Computing.}
}

\IEEEpubid{\begin{minipage}{\textwidth}\ \\[12pt] \centering
  Copyright © 2024 IEEE. Personal use of this material is permitted. \\ However, permission to use this material for any other purposes must be obtained from the IEEE by sending an email to pubs-permissions@ieee.org.
\end{minipage}} 


\maketitle

\begin{abstract}
Conditional coding has lately emerged as the mainstream approach to learned video compression. However, a recent study shows that it may perform worse than residual coding when the information bottleneck arises. Conditional residual coding was thus proposed, creating a new school of thought to improve on conditional coding. Notably, conditional residual coding relies heavily on the assumption that the residual frame has a lower entropy rate than that of the intra frame. Recognizing that this assumption is not always true due to dis-occlusion phenomena or unreliable motion estimates, we propose a masked conditional residual coding scheme. It learns a soft mask to form a hybrid of conditional coding and conditional residual coding in a pixel adaptive manner. We introduce a Transformer-based conditional autoencoder. Several strategies are investigated with regard to how to condition a Transformer-based autoencoder for inter-frame coding, a topic that is largely under-explored. Additionally, we propose a channel transform module (CTM) to decorrelate the image latents along the channel dimension, with the aim of using the simple hyperprior to approach similar compression performance to the channel-wise autoregressive model. Experimental results confirm the superiority of our masked conditional residual transformer (termed MaskCRT) to both conditional coding and conditional residual coding. On commonly used datasets, MaskCRT shows comparable BD-rate results to VTM-17.0 under the low delay P configuration in terms of PSNR-RGB \textcolor{black}{and outperforms VTM-17.0 in terms of MS-SSIM-RGB}. It also opens up a new research direction for advancing learned video compression.

\end{abstract}


\section{Introduction}
\label{sec:intro}

\begin{figure}[t]
    \centering
    \includegraphics[width=\linewidth]{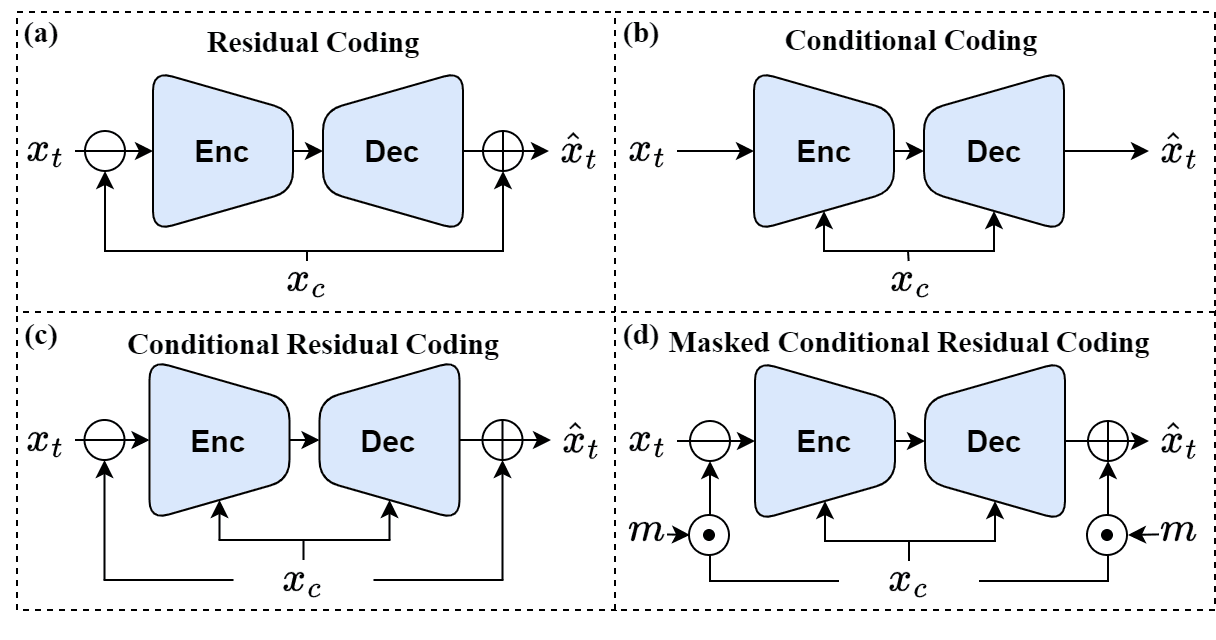}
    \caption{\textcolor{black}{Categorization of different types of inter-frame codecs based on how the temporal predictor $x_c$ is utilized.}}
    \label{fig:codec_type}
\end{figure}

Most learned video codecs~\cite{dvc, liu2022end, chen2019learning, ssf, nvc, rafc, elfvc, mlvc, mmsp, gd, dcvc, tcm, hem, canf, dcvc_dc, bcanf, cond_res_coding, tcsvt23, fvc, coarsetofine, alphavc} follow the traditional hybrid-based coding architecture to reduce temporal redundancy between video frames by temporal predictive coding. That is, to encode a video frame $x_t$ (see Fig.~\ref{fig:codec_type}(a)), the previously coded frame is leveraged to generate a temporal predictor $x_c$ in the form of a temporally warped frame. Generally, the greater the similarity between $x_t$ and $x_c$, the smaller the compressed bitstream. Notably, temporal predictive coding can be conducted in the pixel~\cite{dvc, liu2022end, chen2019learning, ssf, nvc, rafc, elfvc, mlvc, mmsp, gd, dcvc, tcm, hem, canf, dcvc_dc, bcanf, cond_res_coding, tcsvt23} or feature domain~\cite{fvc, coarsetofine, alphavc}. 

How to make most use of the temporal predictor $x_c$ has been the central theme in both learned and traditional video codecs. The utilization of $x_c$ for inter-frame coding leads to different types of inter-frame codecs. The first generation of learned video codecs~\cite{dvc, liu2022end, chen2019learning, ssf, nvc, rafc, elfvc, mlvc, fvc, coarsetofine, alphavc, tcsvt23} adopt residual coding (Fig.~\ref{fig:codec_type}(a)), which encodes the frame differences $x_t-x_c$ between $x_t$ and $x_c$ in the pixel or feature domain. More recently, the second generation of learned video codecs~\cite{mmsp, dcvc, tcm, hem, dcvc_dc, gd, canf, bcanf} (Fig.~\ref{fig:codec_type}(b)), known as conditional coding schemes, emerged, using $x_c$ to condition the inter-frame codec and allowing it to be utilized in a non-linear fashion to encode the input frame $x_t$. Under an ideal setting, the entropy rate $H(x_t|x_c)$ of conditional coding is shown to be smaller than or equal to that $H(x_t-x_c)$ of residual coding~\cite{mmsp}; that is, conditional coding is never worse than residual coding and has great potential to outperform residual coding. As such, it has gone mainstream and is widely adopted by state-of-the-art learned video codecs~\cite{canf, bcanf, tcm, hem, dcvc_dc}.

However, one issue faced by conditional coding in a practical setting is information bottlenecks~\cite{pcs22}. The root cause comes from the information loss during the feature extraction process in formulating the conditioning signal on the decoder side. Most conditional coding involves a neural network in extracting features from $x_c$ or the previously coded frame as the conditioning signal. In the extreme case, where $x_t$ is identical to $x_c$, no additional information should ideally be signaled in the bitstream. However, due to the potential information loss during the feature extraction process, \IEEEpubidadjcol there is no guarantee that the input frame can be reconstructed solely from the conditioning signal (e.g. the extracted features of $x_c$). Additional corrections must then be signaled, causing conditional coding to perform worse than residual coding. The effect of information bottlenecks becomes most apparent when $x_t,x_c$ are highly correlated such as in slow-motion sequences and when much information is lost during the feature extraction process. As a result, some conditional codecs~\cite{dcvc, tcm, hem, dcvc_dc} have to keep a large number of feature maps to remedy the information loss. This induces not only a huge buffer but also high memory bandwidth when encoding high-definition, high frame-rate videos. The memory bandwidth issue has been largely overlooked.          

Recently, Brand~\emph{et~al.}~\cite{cond_res_coding} propose a new school of thought, known as conditional residual coding (Fig.~\ref{fig:codec_type}(c)), aiming at enjoying the merits of both residual and conditional coding without having to store a large number of feature maps. The idea is to encode the prediction residue $x_t-x_c$ with a conditional codec. They show that the conditional entropy rate $H(x_t-x_c|\tilde{x}_c)$ of the prediction residue $x_t-x_c$ is always smaller than or equal to the conditional entropy rate $H(x_t|\tilde{x}_c)$ even in the presence of information bottlenecks, where $\tilde{x}_c$ is a lossy version of $x_c$. \textcolor{black}{The caveat is that the prediction residue $x_t–x_c$ must have a smaller entropy rate than that of the intra frame $x_t$. This assumption, although generally true, can be violated in the regions of a video frame where dis-occlusion occurs or the co-located regions of $x_c$ are poorly motion compensated due to unreliable motion estimates.}


\textcolor{black}{Recognizing that the assumption $H(x_t – x_c) \leq H(x_t)$ is not always true, we introduce a masked conditional residual Transformer, termed MaskCRT, for learned video compression. MaskCRT incorporates a pixel-level soft mask $m$, the values of which range between 0 and 1, to enable a hybrid of conditional coding and conditional residual coding in a pixel-adaptive manner. As depicted in Fig.~\ref{fig:codec_type}(d), the input to our conditional inter-frame codec is $x_t - m \odot x_c$ with $\odot$ denoting pixel-wise multiplication. It amounts to a mixture $(1-m) \odot x_t + m \odot (x_t - x_c)$ of the intra frame $x_t$ and the prediction residue $x_t - x_c$. We learn a neural network to predict $m$, in order to save the overhead needed to signal it.} 

Furthermore, MaskCRT has a Transformer-based conditional autoencoder. 
Transformers have recently proven to be promising alternatives to convolutional neural networks for learned image and video compression~\cite{qualcomm_transformer, tinylic, transformer_cvpr22, transformer_cvpr23, aict, vct, mimt}. However, Transformer-based conditional video codecs are still largely under-explored. We investigate several strategies to condition a Transformer-based autoencoder.

\textcolor{black}{Another striking feature of our MaskCRT is the use of a channel transform module (CTM) to improve the entropy coding efficiency without resorting to any channel-wise autoregressive model. This is a departure from the mainstream approaches, which trade increasingly more complex entropy models for simpler main autoencoders. The channel-wise autoregressive models~\cite{minnen2020channel}, although more parallel-friendly than the conventional spatial-wise autoregressive model~\cite{minnen2018joint}, increases the model complexity and model size considerably. CTM offers a mechanism to decorrelate the image latents along the channel dimension to the extent that the simple hyperprior model is able to achieve comparable compression performance to the more complex channel-wise autoregressive model. This line of thinking opens up a new research direction that is also advocated in a recent publication~\cite{nips23}, which attempts to decorrelate adjacent feature samples linearly along the spatial dimension.}



To sum up, this work has the following contributions:

\begin{itemize}[]
\item MaskCRT presents a novel masked conditional residual coding scheme that results in a spatially adaptive hybrid of conditional coding and conditional residual coding. MaskCRT shows better compression performance than the single use of conditional coding or conditional residual coding. \textcolor{black}{The idea can readily be applied to other state-of-the-art conditional codecs.} 

\item MaskCRT is one of the few earliest attempts to explore how to condition a Transformer-based video codec. To justify our design choice, we present several strategies for implementing a Transformer-based conditional autoencoder. 

\item MaskCRT introduces a channel transform module (CTM) to decorrelate image latents along the channel dimension. With CTM, the simple hyperprior is able to achieve comparable compression performance to a more complex channel-wise autoregressive model.

\item MaskCRT achieves comparable BD-rate results to VTM-17.0 under the low delay P configuration in terms of PSNR-RGB \textcolor{black}{and outperforms VTM-17.0 in terms of MS-SSIM-RGB}.

\end{itemize}

The remainder of this paper is organized as follows: Section~\ref{sec:related} reviews the recent progress in learned video compression. Section~\ref{sec:method} presents our MaskCRT. Section~\ref{sec:experiment} addresses the ablation experiments, rate-distortion comparison, and complexity analyses. Finally, we provide concluding remarks in Section~\ref{sec:conclusion}.
\section{Related Work}
\label{sec:related}
\subsection{Learned Video Compression}
Since the advent of DVC~\cite{dvc}, the first end-to-end learned video codec, the coding frameworks of learned codecs have evolved into two major types: residual coding and conditional coding. Residual coding shares a similar coding architecture to traditional video codecs, replacing key modules, such as motion estimation, motion compensation, residual/motion coding, and in-loop filtering, with neural network-based components. It uses a variational autoencoder (VAE) to compress the temporal prediction residue in either the pixel or feature domain. Recent research in this school of thought focuses on improving the temporal predictor quality~\cite{ssf}, performing deformable or multi-scale motion compensation in the feature domain~\cite{fvc, nvc}, improving motion coding efficiency~\cite{rafc, coarsetofine}, and introducing spatio-temporal context models~\cite{hem, dcvc_dc} for better entropy coding. 

In comparison with residual coding, conditional coding encodes an input frame using a conditional VAE without first evaluating a residual signal. The conditioning signal is usually the motion-compensated frame $x_c$ (Fig.~\ref{fig:codec_type}) or the motion-compensated features extracted from the previously decoded frame~\cite{dcvc}. Unlike residual coding~\cite{dvc, liu2022end, chen2019learning, ssf, nvc, rafc, elfvc, mlvc, fvc, coarsetofine, alphavc, tcsvt23}, which forms a linear prediction of the input frame $x_t$ based on $x_c$ in the pixel (or feature) domain, conditional coding allows $x_c$ to be utilized in a non-linear fashion. How to construct a conditional VAE is the central theme of many conditional coding schemes. Ladune~\emph{et al.}~\cite{mmsp} concatenate $x_t$ and $x_c$ in the input space for encoding, and their latent representations for decoding. Brand~\emph{et al.}~\cite{gd} introduce the operations of generalized difference and generalized sum for conditional coding. Li~\emph{et al.}~\cite{dcvc} adopt the motion-compensated features as the conditioning signal for both the encoder and decoder. The idea is extended in their DCVC family of works~\cite{dcvc_dc, hem, tcm} to leverage multi-scale features as a better condition. There are also the augmented normalizing flow (ANF)-based conditional codecs~\cite{canf, bcanf}, replacing conditional VAE with conditional ANF.

\subsection{Combination of Residual and Conditional Coding}

Brand~\emph{et al.}~\cite{pcs22} note that conditional coding may suffer from information bottlenecks due to the generally lossy nature of the feature extraction process in obtaining the conditioning signal for the decoder. When the information bottleneck occurs, conditional coding is shown to perform worse than residual coding. With this in mind, Brand~\emph{et al.}~\cite{fau_hybrid} propose to switch between residual coding and conditional coding. They generate and encode jointly the latent representations for both schemes based on a single shared VAE. Two distinct outputs are then decoded and chosen adaptively at the frame or region level for reconstruction. Obviously, encoding two distinct types of image latents into a single bitstream introduces redundancy. 

To overcome this problem, they recently propose a conditional residual coding framework (Fig.~\ref{fig:codec_type}(c)) in ~\cite{cond_res_coding}, encoding the prediction residue $x_t – x_c$ with the conditional codec in~\cite{modenet}. It is obvious that the entropy rate of conditional residual coding is always smaller than or equal to that of unconditional residual coding. Furthermore, under the assumption $H(x_t – x_c) \leq H(x_t)$, it is shown that conditional residual coding outperforms conditional coding, whether the information bottleneck exists or not. Notably, when the prediction signal $x_c$ is less correlated with the input frame $x_t$, the effect of the information bottleneck is less significant. In such a case, conditional residual coding and conditional coding perform comparably to each other. However, when $x_c,x_t$ are highly correlated, conditional residual coding becomes advantageous. The former often occurs in fast-motion sequences, while the latter is common in static sequences. \textcolor{black}{Although theoretically promising, conditional residual coding may fail to achieve the desired compression performance when the assumption $H(x_t – x_c) \leq H(x_t)$ is violated, which motivates our MaskCRT.}




\subsection{Context Models}
Advanced context models~\cite{minnen2018joint, checkerboard, minnen2020channel, elic, mlic} that leverage the spatial and/or channel-wise context for better entropy coding are being researched extensively for learned image and video coding. The recent trend has been towards using more sophisticated context models in exchange for simpler main autoencoders. Minnen~\emph{et al.}~\cite{minnen2018joint} present the first autoregressive model to employ the spatial correlation of the image latents for entropy coding. However, the need to query the context model sequentially for decoding every feature sample makes it impractical. For parallel-friendly decoding, He~\emph{et al.}~\cite{checkerboard} introduce a checkerboard pattern in dividing image latents along the spatial dimension into multiple slices. In each slice, the coding probabilities of all the feature samples are obtained in a parallel manner. In a similar vein, Minnen~\emph{et al.}~\cite{minnen2020channel} split the latents along the channel dimension, forming a channel-wise autoregressive model. Both ideas are further combined in some recent spatial-channel context models~\cite{elic, dcvc_dc, mlic}. 

Recently, Ali~\emph{et al.}~\cite{nips23} introduce a correlation loss during training, with the aim of decorrelating linearly the image latents along the spatial dimension. This allows the simple hyperprior~\cite{hyperprior} model, which assumes image latents are independently distributed, to perform comparably with the 2-slice checkerboard context model~\cite{checkerboard}. \textcolor{black}{While their focus is on the spatial linear correlation, our CTM is aimed at de-correlating along the channel dimension, where the inherent correlation is usually non-linear.} In common, both open up a new dimension of thinking that attempts to simplify the entropy coding model, which is often the implementation bottleneck.





\begin{figure}[t]
    \centering
    \includegraphics[width=\linewidth]{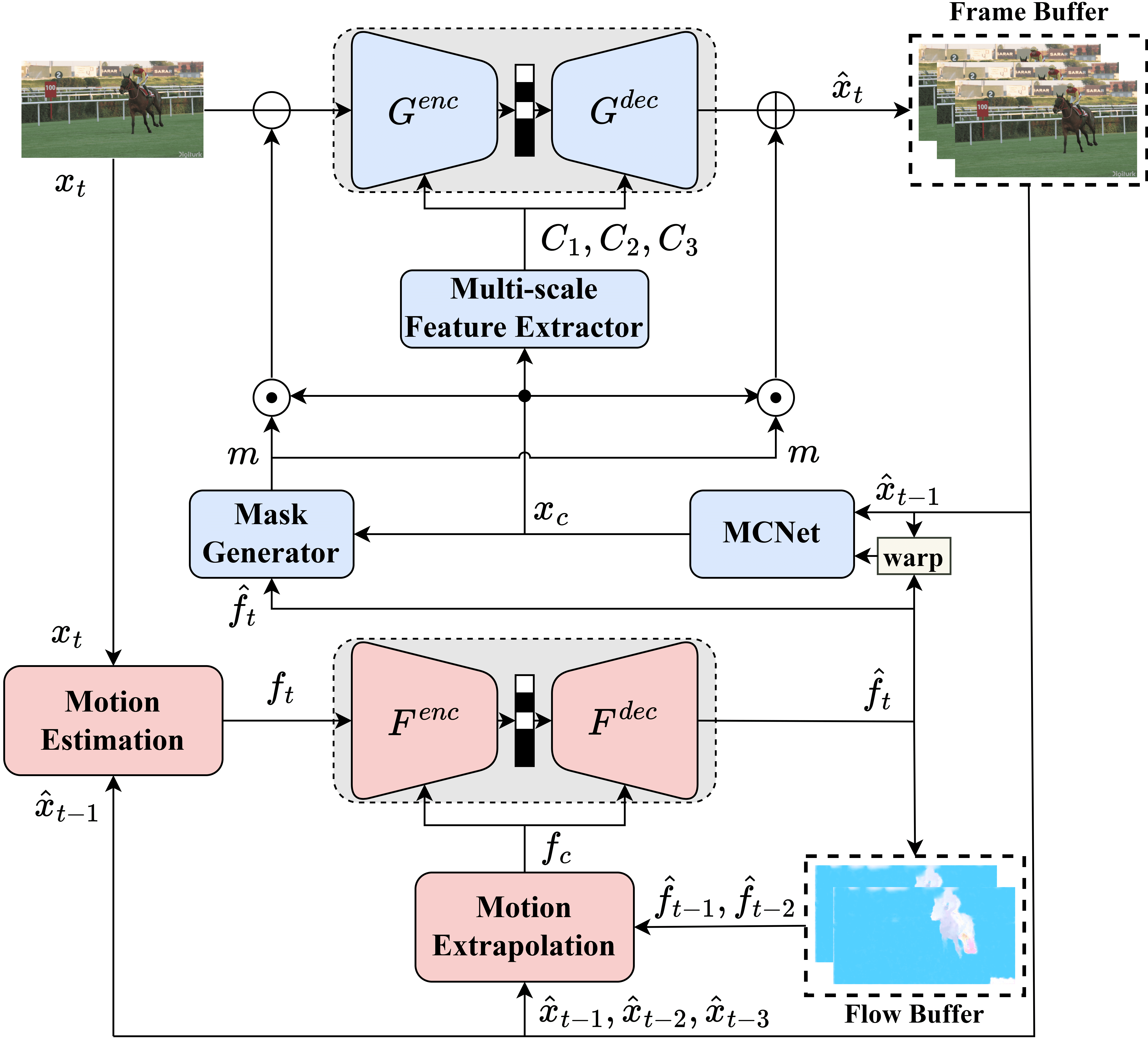}
    \caption{Overview of the proposed MaskCRT framework.}
    \label{fig:overview}
\end{figure}

\section{Proposed Method}
\label{sec:method}

\subsection{System Overview}
Fig.~\ref{fig:overview} illustrates our MaskCRT framework. It includes a conditional motion codec $\{F^{enc}, F^{dec}\}$ and a masked conditional inter-frame codec $\{G^{enc}, G^{dec}\}$. Encoding a video frame $x_t \in \mathbb{R}^{3 \times W \times H}$ of width $W$ and height $H$ begins by estimating an optical flow map $f_t \in \mathbb{R}^{2 \times W \times H}$ with respect to its decoded reference frame $\hat{x}_{t-1}$. $f_t$ is then coded based on a conditioning signal $f_c \in \mathbb{R}^{2 \times W \times H}$, which is a flow map predictor synthesized from the three decoded frames $\hat{x}_{t-1}, \hat{x}_{t-2}, \hat{x}_{t-3}$ and two decoded flow maps $\hat{f}_{t-1}, \hat{f}_{t-2}$ \textcolor{black}{by using the same flow extrapolation network in \cite{canf}}. Subsequently, the coded flow map $\hat{f}_t$ is utilized for warping $\hat{x}_{t-1}$. \textcolor{black}{With this warped frame and $\hat{x}_{t-1}$ as inputs, the motion compensation network (MCNet) outputs the temporal predictor $x_c \in \mathbb{R}^{3 \times W \times H}$.} 
In particular, $x_c$ and $\hat{f}_t$ are employed to generate a \textit{soft} mask $m \in \mathbb{R}^{3 \times W \times H}$, where its elements are real values between 0 and 1 that guide the trade-off between conditional coding and conditional residual coding at the pixel level. The three channels of $m$ share the same mask values. The input to our masked conditional inter-frame codec is evaluated as a weighted sum $(1-m) \odot x_t + m \odot (x_t - x_c)$ of $x_t$ and the prediction residue $x_t - x_c$, where $\odot$ denotes element-wise multiplication. It further reduces to $x_t - m \odot x_c$. We note that our conditional inter-frame codec behaves in \textit{conditional residual coding} mode if $m$ is set uniformly to one, and in \textit{conditional coding} mode if $m$ is zero element-wise. The network-generated soft mask $m$ allows \textcolor{black}{for} a hybrid mode. On the decoder side, the input frame is reconstructed by adding $m \odot x_c$ to the output of the decoder $G^{dec}$. As depicted in Fig.~\ref{fig:overview}, we adopt the multi-scale features $C_1 \in \mathbb{R}^{96 \times W/2 \times H/2}, C_2 \in \mathbb{R}^{128 \times W/8 \times H/8}, C_3 \in \mathbb{R}^{192 \times W/16 \times H/16}$ extracted from the temporal predictor $x_c$ as the conditioning signal for both the inter-frame encoder $G^{enc}$ and decoder $G^{dec}$.

\begin{figure}[t!]
    \centering
    \includegraphics[width=\linewidth]{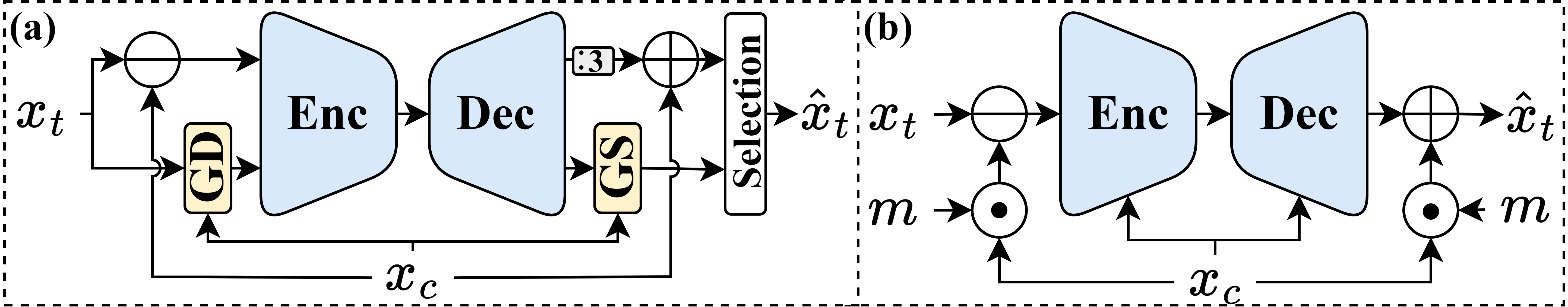}
    \caption{Illustration of (a) the hybrid of residual and conditional coding proposed in \cite{fau_hybrid}, and (b) our MaskCRT. The notation ":3" represents taking the first three channels out of the decoded signal. GD and GS refer to generalized difference and generalized sum, respectively.}

    \label{fig:hybrid_based}
\end{figure}
\begin{figure*}[t!]
    \centering
    \includegraphics[width=\linewidth]{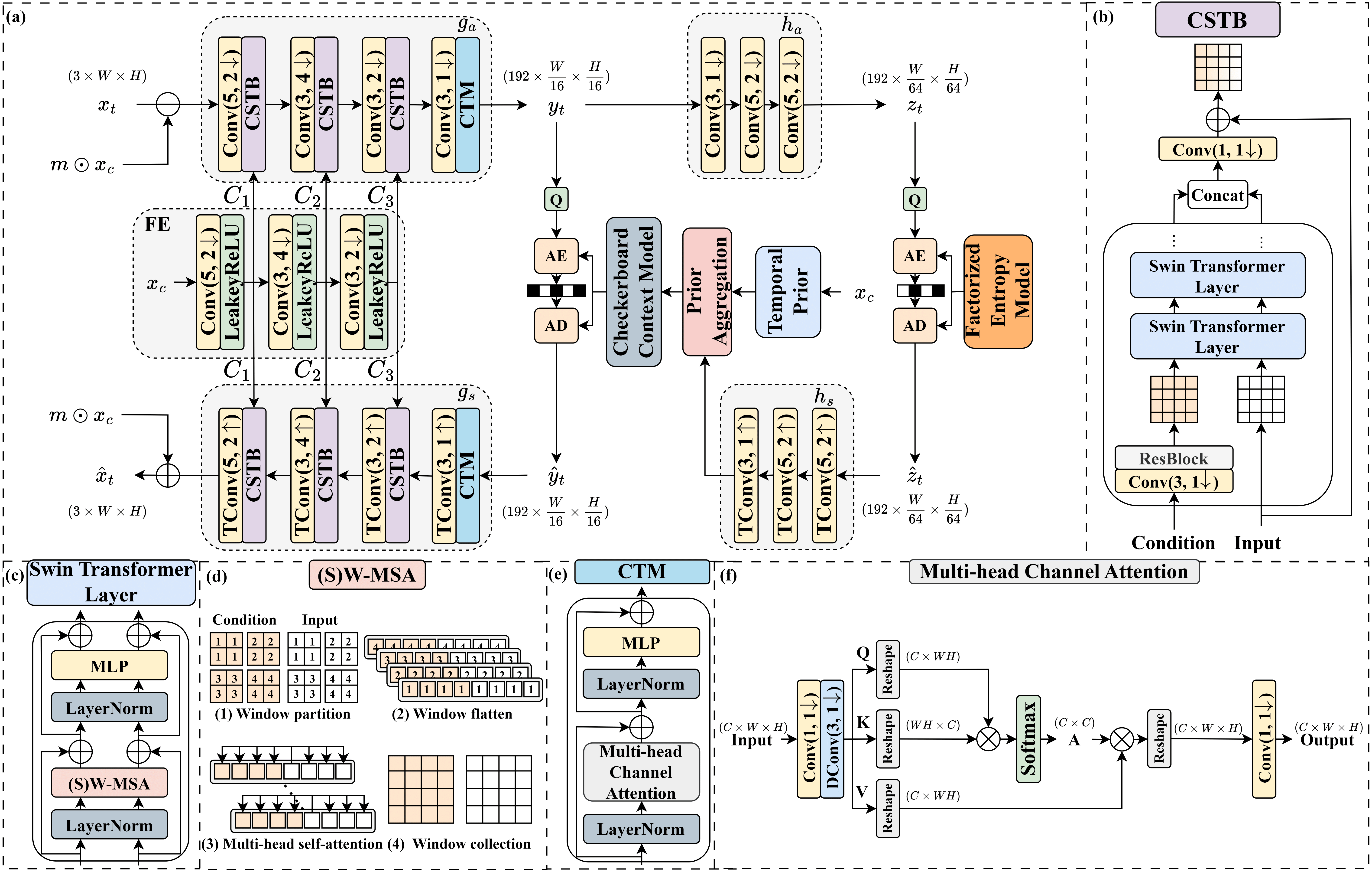}
    \caption{Illustration of our proposed MaskCRT: (a) the network architecture of MaskCRT, where FE is the multi-scale feature extractor used to generate the conditioning signal, (b) the design specifics of our conditional Swin-Transformer blocks (CSTB), (c) our Swin-Transformer layer, (d) the operations of window-based multi-head self-attention in our CSTB, (e) the network architecture of our channel transform module (CTM), and (f) the multi-head channel attention in CTM.}
    \label{fig:inter_coder}

\end{figure*}
\subsection{Masked Conditional Residual Coding}

\textcolor{black}{We motivate our masked conditional residual coding by the potential violation of the assumption that the entropy $H(x_t–x_c)$ of the prediction residue $x_t–x_c$ is always smaller than that $H(x_t)$ of the intra frame $x_t$. Being generally true, this assumption $H(x_t–x_c) \leq H(x_t)$ ensures that conditional residual coding is better than or equal to conditional coding, whether the bottleneck issue is in effect or not~\cite{cond_res_coding}. That is, $H(x_t–x_c|\tilde{x}_c) \leq H(x_t|\tilde{x}_c)$ where $\tilde{x}_c$ is a lossy version of $x_c$. However, when $H(x_t–x_c) \geq H(x_t)$, the same line of derivation in~\cite{cond_res_coding} yields the complete opposite, i.e.~$H(x_t–x_c|\tilde{x}_c) \geq H(x_t|\tilde{x}_c)$.}

\textcolor{black}{Indeed, temporal prediction can be ineffective to the extent that the prediction residue $x_t - x_c$ has a higher entropy rate than that of the intra frame $x_t$, i.e. without temporal prediction. An obvious case is the occurrence of dis-occlusion in $x_t$, where the motion in the dis-occluded regions is undefined. As such, part of $x_t$ cannot find a good match in $x_c$. Likewise, this holds true in the regions of $x_t$ where the optical flow map $f_t$ is incorrectly estimated or coarsely quantized, leading to a poor-quality temporal predictor $x_c$.}

\textcolor{black}{Recognizing that $H(x_t–x_c) \leq H(x_t)$ may not hold consistently across the entire video frame, we introduce a pixel-level mask $m$ to switch between conditional coding and conditional residual coding. Because $m$ is designed to be a soft mask, the input to our conditional inter-frame codec is a mixture $(1-m) \odot x_t + m \odot (x_t - x_c)$ of the intra frame $x_t$ and the prediction residue $(x_t - x_c)$. Apparently, conditional coding and conditional residual coding are just two extreme special cases of our scheme, the masks of which are uniformly 0 and 1, respectively. Notably, a pixel-adaptive hybrid scheme arises when a non-uniform $m$ is applied.}

A direct application of the same argument in~\cite{cond_res_coding} shows that the masked conditional residual coding, although conceptually simple, is superior to conditional residual coding in the presence of the bottleneck issue when $H(x_t–m \odot x_c) \leq H(x_t - x_c)$. That is, $H(x_t–m \odot x_c|\tilde{x}_c) \leq H(x_t–x_c|\tilde{x}_c)$ when $H(x_t–m \odot x_c) \leq H(x_t - x_c)$. By the same token, we have $H(x_t–m \odot x_c|\tilde{x}_c) \leq H(x_t|\tilde{x}_c)$ when $H(x_t–m \odot x_c) \leq H(x_t)$. These results stress the importance of choosing $m$ properly, to ensure $H(x_t–m \odot x_c) \leq H(x_t)$ and $H(x_t–m \odot x_c) \leq H(x_t - x_c)$. 

To save the overhead needed to signal the pixel-level mask $m$, we adopt a neural network to predict a spatially-varying soft mask based on the decoded flow map $\hat{f}_t$ and the temporal predictor $x_c$, both are causally available to the encoder and decoder at inference time. $\hat{f}_t$ and $x_c$ provide helpful clues to whether or not a coding pixel is at object boundaries, where dis-occlusion or unreliable motion often occurs. 

In passing, we note that Brand~\emph{et al.}~\cite{fau_hybrid} present a hybrid scheme that uses a quadtree-based binary mask to switch between conditional coding and residual coding (see Fig.~\ref{fig:hybrid_based}(a)). In comparison, our mask is a pixel-level, soft mask, allowing the intra frame $x_t$ and the prediction residue $x_t-x_c$ to be mixed using spatially-varying weightings (see Fig.~\ref{fig:hybrid_based}(b)). Conceptually, our scheme offers an infinite number of coding modes at each pixel location because the mask value can take any value between 0 and 1. In addition, unlike \cite{fau_hybrid}, which signals the quadtree-based binary mask in the bistream, our scheme incorporates a neural network in the encoder and decoder to predict the mask in a synchronized way. Last but not least, \cite{fau_hybrid} encodes two inputs $x_t$ and $x_t-x_c$ jointly, with dedicated outputs for conditional coding and residual coding, respectively. Inevitably, this approach introduces redundancy. To overcome this issue, our scheme computes the mixture $(1-m) \odot x_t + m \odot (x_t - x_c)$ as the only input to the inter-frame codec and outputs only its reconstruction.

\subsection{Masked Conditional Residual Transformer (MaskCRT)}
\label{sec:masked_cond_res_transformer}
We now explain in detail our masked conditional residual transformer (MaskCRT), a conditional residual codec built on the Swin-Transformer~\cite{swin}. Recently, Transformer-based autoencoders emerge as appealing alternatives to convolutional neural network (CNN)-based autoencoders due to their high content adaptivity. Zhu~\emph{et al.}~\cite{qualcomm_transformer} demonstrate their superiority to CNN-based autoencoders in terms of the compression performance, model size and decoding complexity. \textcolor{black}{Unlike their residual-based coding framework, MaskCRT is one of the first few Transformer-based conditional video coding schemes. How to condition a Transformer-based inter-frame codec remains largely under-explored.}

Fig.~\ref{fig:inter_coder}(a) depicts the network architecture of MaskCRT, where the conditional Swin-Transformer blocks (CSTB) are the key components used to condition the encoding and decoding of the masked prediction residue $x_t - m \odot x_c$ on the conditioning signal $x_c$. The CSTB stacks several Swin-Transformer layers to facilitate information exchange between the input and conditioning signals (Fig.~\ref{fig:inter_coder}(b)). From left to right in the main encoder and decoder, the number of Swin-Transformer layers is 4, 4, and 2, respectively. Notably, we have both the input and conditioning signals play symmetric roles. Information exchange between them is achieved by window-based multi-head self-attention rather than cross-attention. In Fig.~\ref{fig:inter_coder}(c), our Swin-Transformer layer updates both inputs simultaneously. The two outputs are then fused into one set of tokens using $1 \times 1$ convolution, as depicted in Fig.~\ref{fig:inter_coder}(b). It is worth noting that $x_c$ has been motion compensated (i.e.~temporally aligned with the input signal). The window-based multi-head self-attention is thus applied to tokens extracted respectively from the input and conditioning signals in the co-located windows (Fig.~\ref{fig:inter_coder}(d)). 





\begin{figure}[t!]
    \centering
    \includegraphics[width=\linewidth]{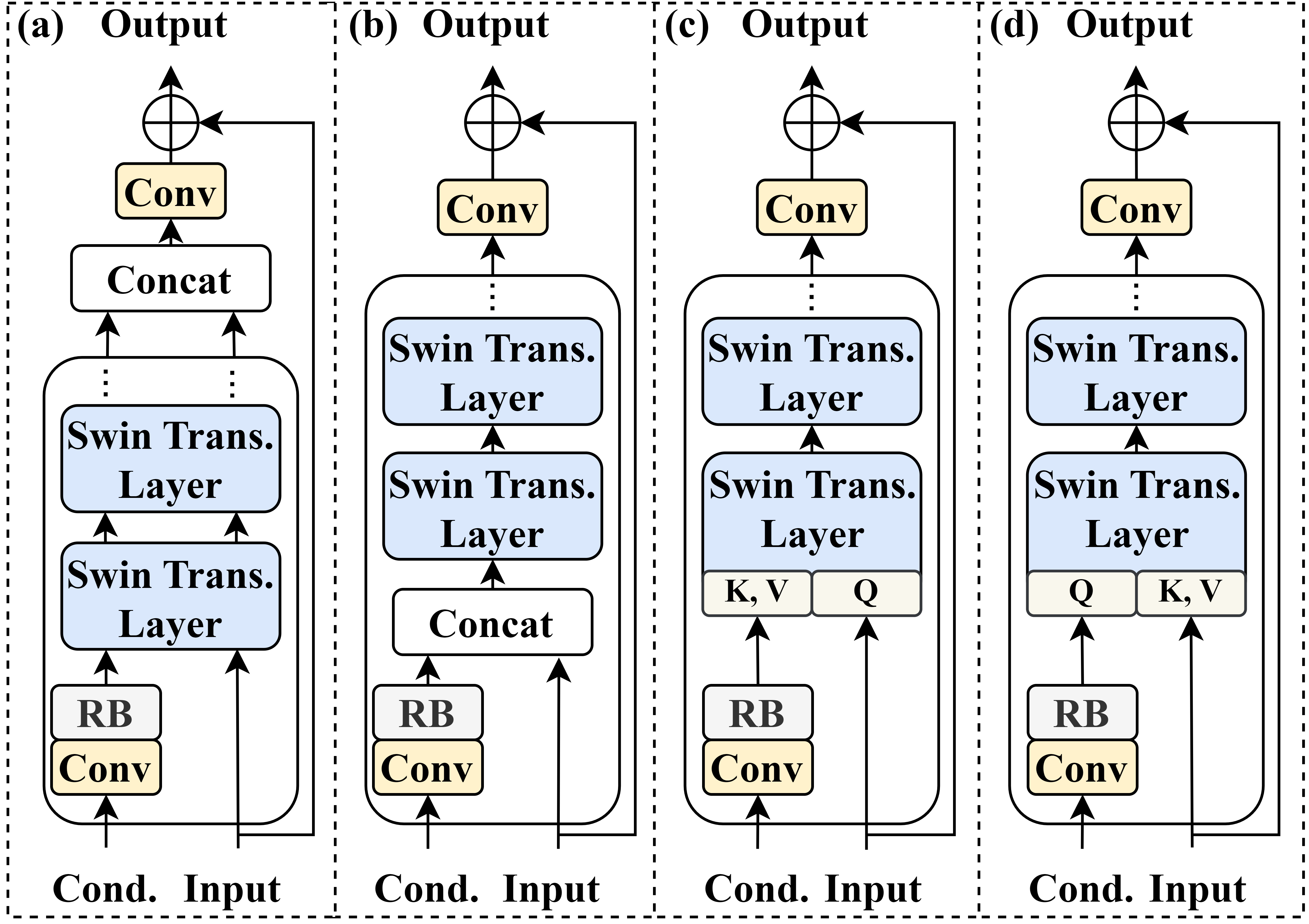}
    \caption{\textcolor{black}{Various implementations of CSTB: (a) the proposed method, (b) concatenating the input and conditioning signals, (c) formulating queries with the input signal, and (d) formulating queries with the conditioning signal. RB is a residual block.}}
    \label{fig:Variants_of_CSTB}
\end{figure}

\textcolor{black}{To justify our design choice, we investigate several other ways, as depicted in Fig.~\ref{fig:Variants_of_CSTB}, to condition a Transformer-based autoencoder. Some of them share parallels with Transformer-based conditional image generators~\cite{tmm23, maskgit, mital2023}. Similar to CNN-based conditional coding~\cite{mmsp, canf, bcanf, dcvc, hem, tcm, dcvc_dc}, Fig.~\ref{fig:Variants_of_CSTB}(b) straightforwardly concatenates the input and conditioning signals along the channel dimension, with the resulting signal at different spatial locations turned into tokens for subsequent self-attention. Compared with our proposed method in Fig.~\ref{fig:Variants_of_CSTB}(a), which allows tokens constructed separately from the input and conditioning signals to exchange information freely via window-based self-attention, the design in Fig.~\ref{fig:Variants_of_CSTB}(b) has fairly restricted interactions between the co-located input and conditioning signals. Its performance is expected to be suboptimal when the conditioning signal is not temporally well aligned with the input signal due to less accurate motion. 
Fig.~\ref{fig:Variants_of_CSTB}(c) follows from \cite{mital2023} to implement cross-attention for information exchange, formulating queries from the input signal and keys/values from the conditioning signal. Fig.~\ref{fig:Variants_of_CSTB}(d) formulates queries and keys/values in the other way around. Unlike our symmetric design, where the input and conditioning signals are processed without distinction in the Swin-Transformer layer, the cross-attention designs in Figs.~\ref{fig:Variants_of_CSTB}(c) and~\ref{fig:Variants_of_CSTB}(d) are asymmetric. Our ablation study in Section~\ref{sec:ablation} confirms the superiority of the proposed method, where its symmetric design allows for the inherent correlations existing within and across the two inputs.}

\begin{table*}[t!]
\centering
\caption{\textcolor{black}{Training procedure. MENet, MWNet, MCNet represent the motion estimation network, the motion extrapolation network, and the motion compensation network, respectively. EPA is the error propagation aware training in \cite{EPA}.}}
\label{tab:training}
\begin{tabular}{ccllcc}

\hline
\textcolor{black}{Phase}                                                      & \textcolor{black}{Number of Frames} & \textcolor{black}{Training Modules}                      & \textcolor{black}{Loss}                                                             & \textcolor{black}{lr}   & \textcolor{black}{Epoch} \\
\hline
\textcolor{black}{Motion Coding}                                              & \textcolor{black}{3}                & \textcolor{black}{MWNet and motion codec}                & \textcolor{black}{$R^{motion}_t + \lambda \times D(x_t, warp(x_{t-1}, \hat{f}_t))$}   & \textcolor{black}{1e-4} & \textcolor{black}{10}     \\\hline
\textcolor{black}{Motion Compensation}                                        & \textcolor{black}{3}                & \textcolor{black}{MCNet}                                 & \textcolor{black}{$R^{motion}_t + \lambda \times D(x_t, \hat{x}_c)$}                  & \textcolor{black}{1e-4} & \textcolor{black}{5}      \\\hline
\multicolumn{1}{c}{\multirow{2}{*}{\textcolor{black}{Inter-frame Coding}}}    & \textcolor{black}{3}                & \textcolor{black}{Inter-frame codec and mask generator}  & \textcolor{black}{$R_t + \lambda \times D(x_t, \hat{x}_t)$}                           & \textcolor{black}{1e-4} & \textcolor{black}{10}     \\
\multicolumn{1}{c}{}                                                         & \textcolor{black}{5}                & \textcolor{black}{Inter-frame codec and mask generator}  & \textcolor{black}{$R_t + \lambda \times D(x_t, \hat{x}_t)$}                           & \textcolor{black}{1e-4} & \textcolor{black}{5}     \\\hline
\textcolor{black}{Finetuning}                                                 & \textcolor{black}{5}                & \textcolor{black}{All modules except MENet}             & \textcolor{black}{$R_t + \lambda \times D(x_t, \hat{x}_t)$}                           & \textcolor{black}{1e-4} & \textcolor{black}{3}     \\\hline
\multicolumn{1}{c}{\multirow{2}{*}{\textcolor{black}{Finetuning with EPA}}}   & \textcolor{black}{5}                & \textcolor{black}{All modules except MENet}             & \textcolor{black}{$R_t + \lambda \times D(x_t, \hat{x}_t)$}                           & \textcolor{black}{1e-4} & \textcolor{black}{3}     \\
\multicolumn{1}{c}{}                                                         & \textcolor{black}{5}                & \textcolor{black}{All modules}                          & \textcolor{black}{$R_t + \lambda \times D(x_t, \hat{x}_t)$}                           & \textcolor{black}{1e-5} & \textcolor{black}{3}     \\\hline
\end{tabular}
\end{table*}

\subsection{Channel Transform Module (CTM)}
\textcolor{black}{The channel transform module (CTM) is aimed at de-correlating the image latents $y_t$ along the channel dimension, with the hopes of achieving comparable compression performance to the channel-wise autoregressive model while using the simple hyperprior. We focus on the channel dimension because in the case of P-frame coding, the image latents exhibit a relatively low spatial correlation. Moreover, channel-wise context models usually incur a larger model size and a higher number of multiply–accumulate operations than spatial context models.}


Fig.~\ref{fig:inter_coder}(e) illustrates the architecture of CTM. Inspired by~\cite{restormer} for image restoration tasks, our CTM implements a Transformer block to perform multi-head channel attention (Fig.~\ref{fig:inter_coder}(f)). However, in our task, it is meant to be a learnable mechanism for reducing the channel correlation of the image latents $y_t$. As such, it acts in the latent space as the final encoding step. For the inverse operation, another CTM is placed on the decoder side as the first decoding step (Fig.~\ref{fig:inter_coder}(a)). \textcolor{black}{Unlike~\cite{nips23}, which regularizes the model training with an additional correlation loss to reduce the linear correlation between feature samples along the spatial dimension, our CTM offers a mechanism to reduce the potentially non-linear correlation between feature maps in the channel dimension.} We train CTM end-to-end together with the other components, including the hyperprior, by minimizing the conventional rate-distortion cost. In a sense, CTM relies on the factorial assumption of the hyperprior--namely, $p(y_t|\hat{z_t})$ is fully factorial along the channel and spatial dimensions--to achieve channel-wise decorrelation without any additional regularization loss. That is, the rate term with the hyperprior model achieves its minimum only when the image latents are fully factorial. 



\section{Experiments}
\label{sec:experiment}
\begin{figure*}[t!]
    \centering
    \subfigure{
        \centering
        \includegraphics[height=0.245\linewidth, trim= 0 0 60 50, clip]{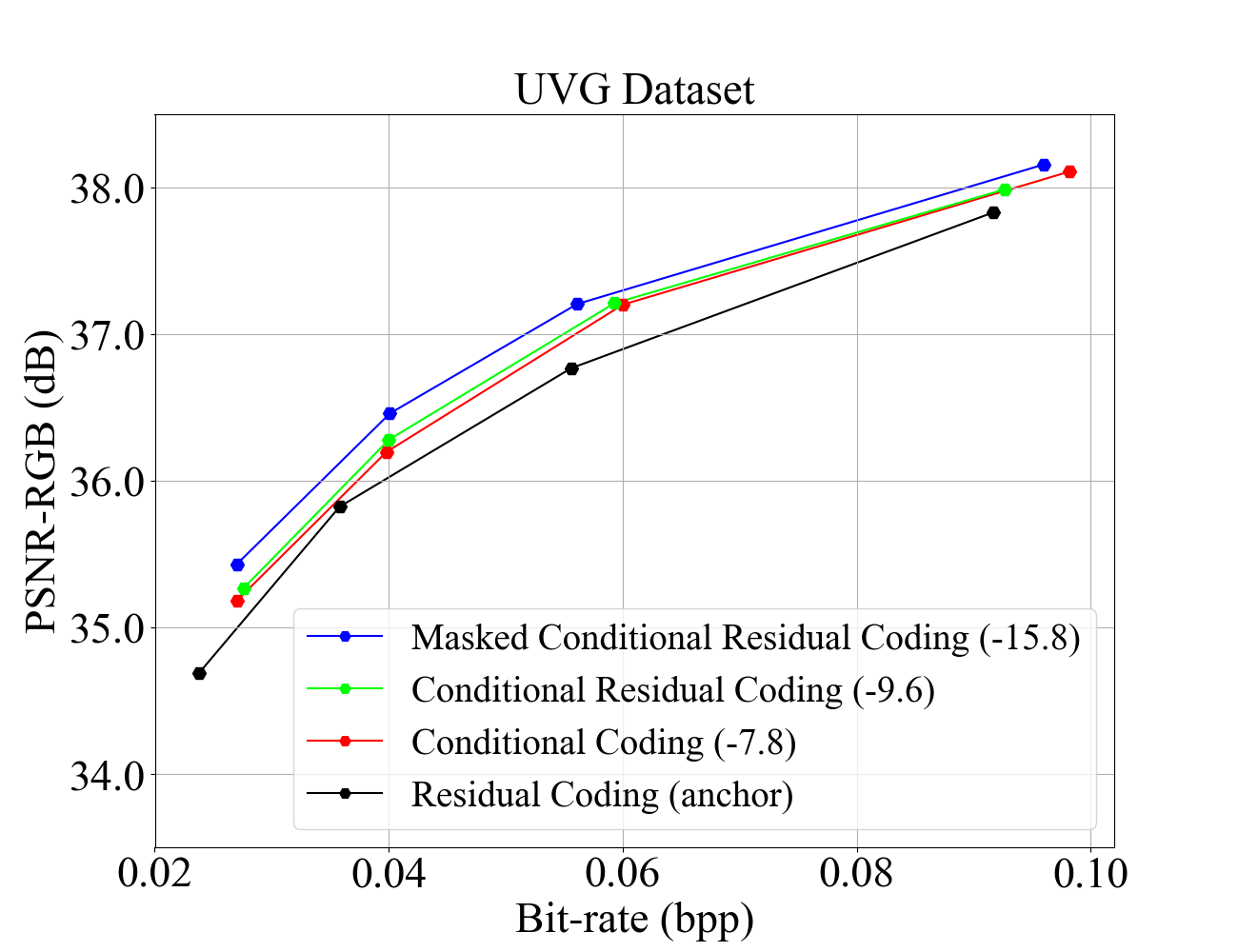}
        } \hspace{-0.4cm}
    \subfigure{
        \centering
        \includegraphics[height=0.245\linewidth, trim= 0 0 60 50, clip]{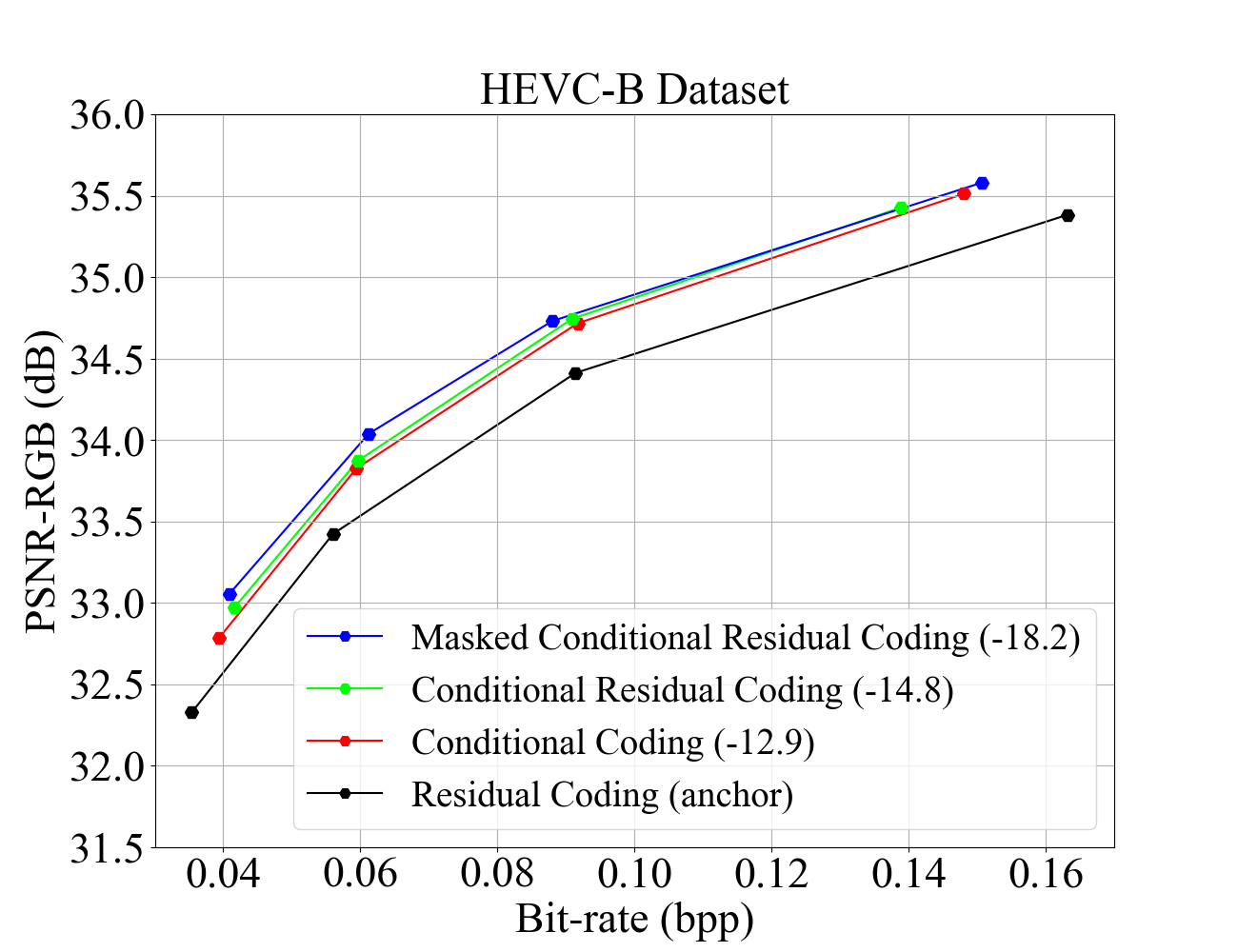}
        } \hspace{-0.4cm}
    \subfigure{
        \centering
        \includegraphics[height=0.245\linewidth, trim= 0 0 60 50, clip]{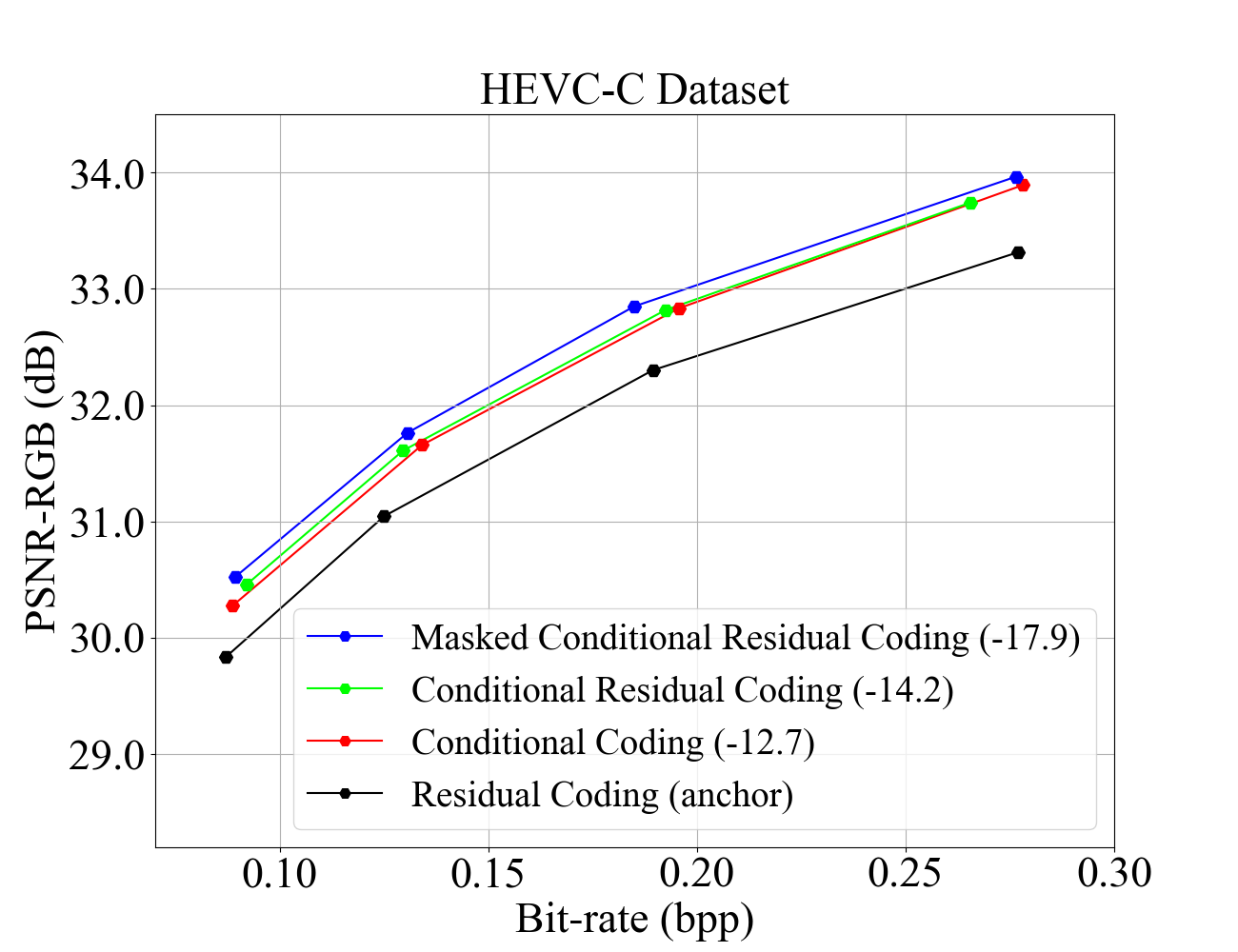}
        } 
    \subfigure{
        \centering
        \includegraphics[height=0.245\linewidth, trim= 0 0 60 50, clip]{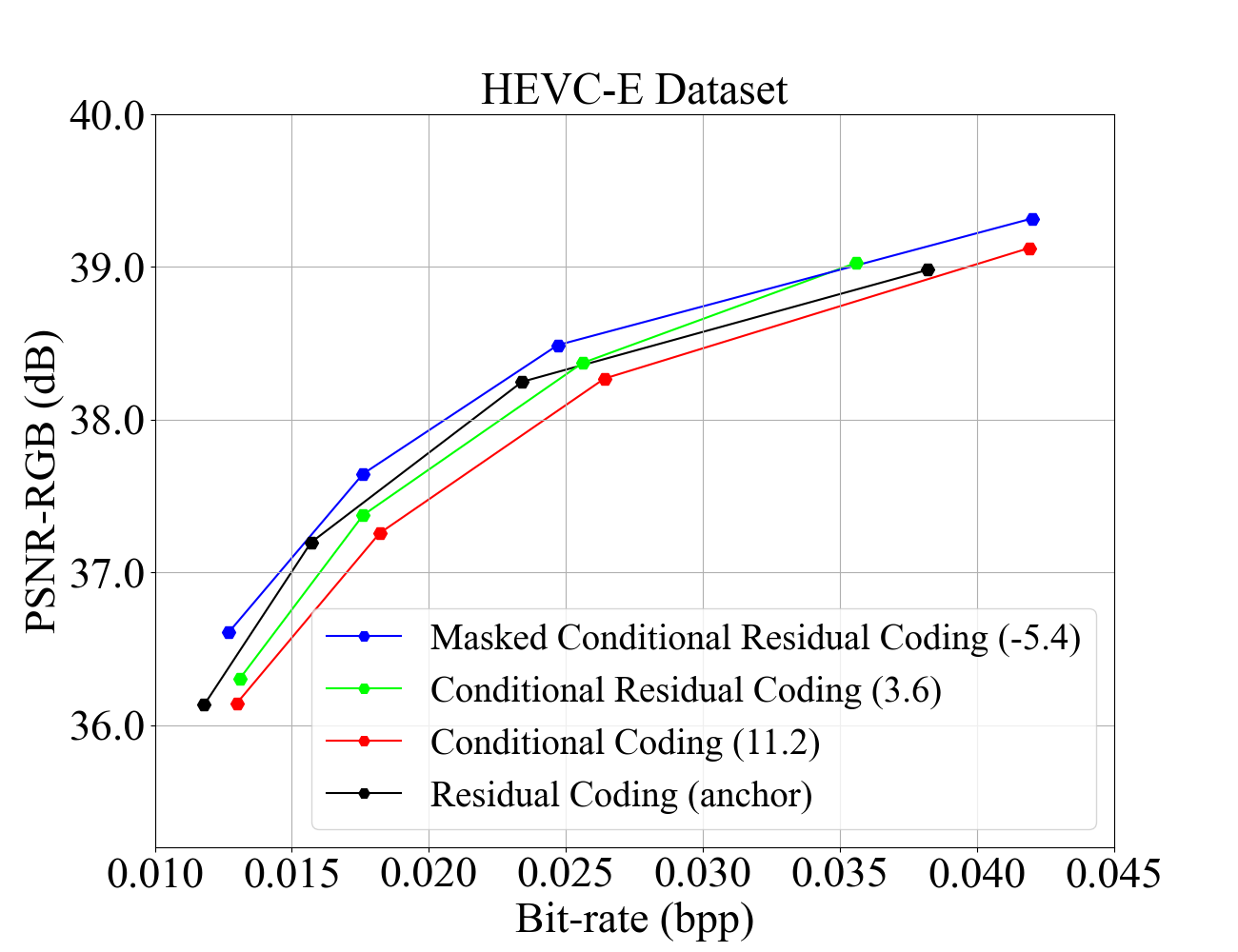}
        } \hspace{-0.4cm}
    \subfigure{
        \centering
        \includegraphics[height=0.245\linewidth, trim= 0 0 60 50, clip]{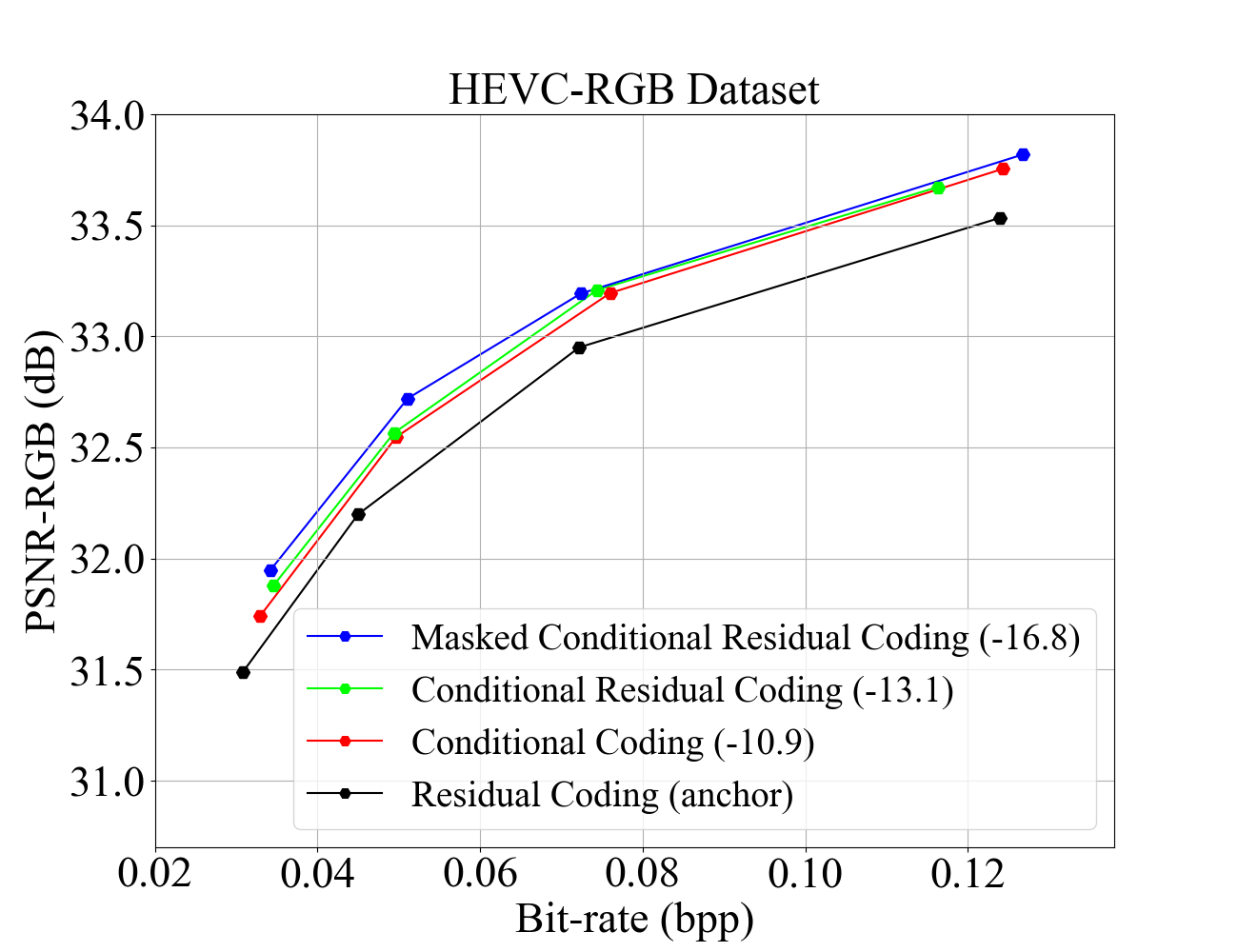}
        } \hspace{-0.4cm}
    \subfigure{
        \centering
        \includegraphics[height=0.245\linewidth, trim= 0 0 60 50, clip]{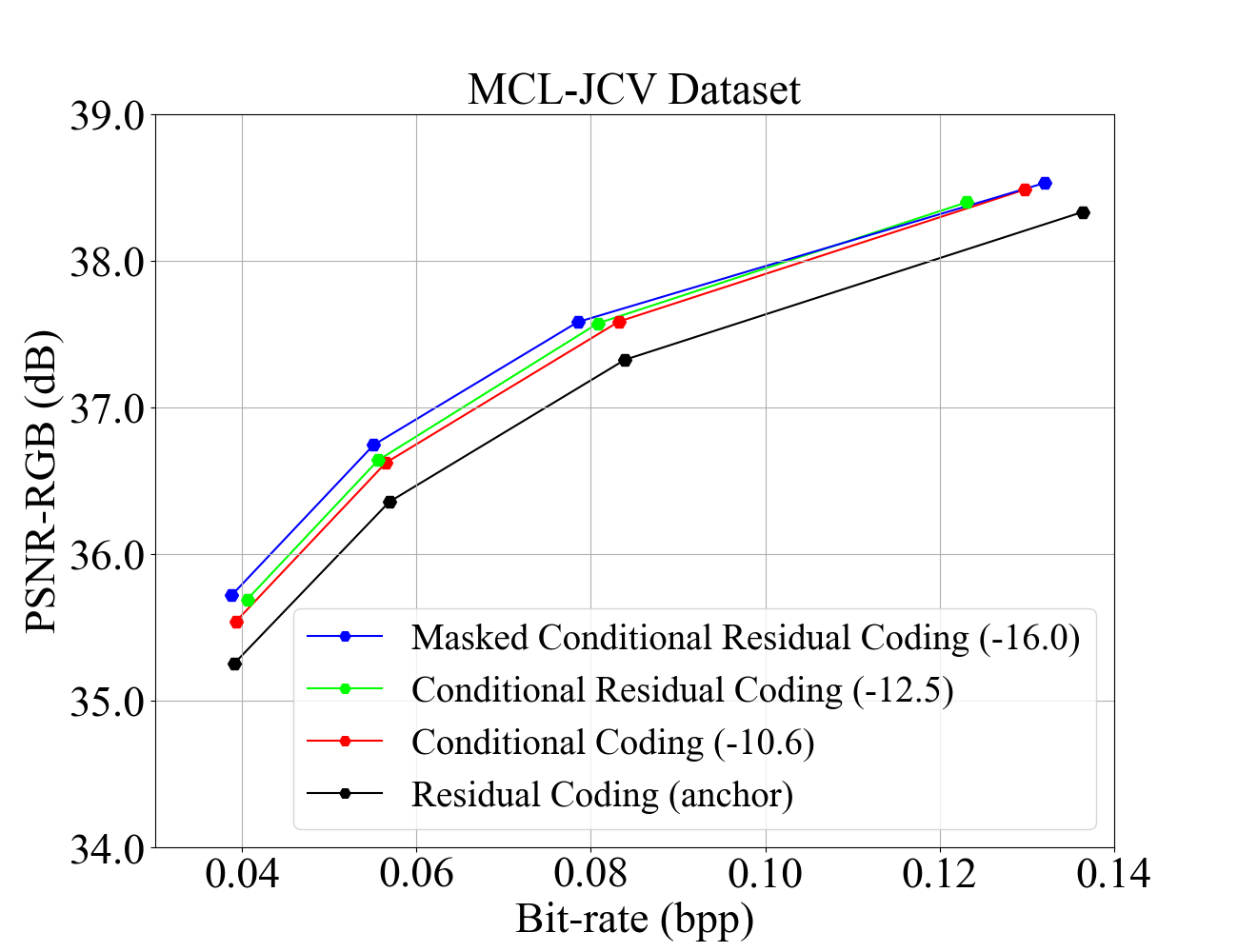}
        }
    \vspace{-3mm}
    \caption{\textcolor{black}{Rate-distortion performance comparison. The values within the parentheses represent BD-rates, with the residual coding serving as the reference. The lower the BD-rate, the better the compression performance.}}

    \label{fig:masked_cond_res}
\end{figure*}

\subsection{Settings}

\paragraph{Training Details}
\textcolor{black}{We initialize our model with the pre-trained intra codec from~\cite{dcvc_dc} and the pre-trained motion estimation network from~\cite{spynet}.} We adopt Vimeo90k dataset~\cite{vimeo}, containing 91,701 7-frame sequences of size $448 \times 256$ as training dataset, and randomly crop the training sequences into $256 \times 256$ patches. 
\textcolor{black}{Table~\ref{tab:training} summarizes our training procedure. It contains several phases, with different coding modules introduced and updated sequentially. $D$ is the distortion measured in the mean-squared error or (1 - MS-SSIM) in the RGB domain. $R^{motion}_t$ and $R_t$ are the motion and total bitrates in bits-per-pixel, respectively. $\lambda$ is set to \{1626, 845, 436, 228\} for the PSNR-RGB model and to \{5, 10, 19, 36\} for the MS-SSIM-RGB model.} 


\paragraph{Evaluation Methodologies}
We \textcolor{black}{follow the common test protocol of learned video compression to} evaluate the compression performance on UVG~\cite{uvg}, HEVC Class B~\cite{hevcctc}, \textcolor{black}{HEVC Class C~\cite{hevcctc}, HEVC Class E~\cite{hevcctc}, HEVC-RGB~\cite{hevcrgb}} and MCL-JCV~\cite{mcl} datasets. We adopt BT.601 (default in FFmpeg~\cite{ffmpeg}) to convert every test sequence from its raw format YUV420 to RGB444. We set the intra period to 32 and encode the first 96 frames for each test sequence. In particular, our Transformer-based model requires both the width and height of the input sequence to be multiples of 128. We crop all the test sequences to meet this requirement, and evaluate all the competing methods on the same cropped test sequences for a fair comparison. For HEVC-RGB~\cite{hevcrgb} and MCL-JCV~\cite{mcl} dataset, we exclude screen content sequences from evaluation; in addition, we use intra coding at scene cuts. These settings apply to all the competing methods in order to understand better the compression performance of their P-frame codecs on natural content. We report BD-rate savings~\cite{bdrate} in terms of peak single-to-noise ratio (PSNR) or MS-SSIM in the RGB domain and bits-per-pixel (bpp). 
Negative and positive BD-rate numbers suggest rate reductions and inflations, respectively. \textcolor{black}{To report the average BD-rates for different datasets, we follow the common test protocol of learned video compression. That is, each rate-distortion point for a dataset is evaluated by averaging the per-frame PSNR-RGB (or MS-SSIM-RGB) and bits-per-pixel over all the coding frames in the dataset.}



\subsection{Ablation Studies}
\label{sec:ablation}
This section presents several ablation studies to single out the contributions of different components to the final compression performance.
Unless otherwise specified, these studies are conducted without CTM and any context model to reduce training time.   

\paragraph{\textcolor{black}{Rate-distortion Comparison}}
\textcolor{black}{Fig.~\ref{fig:masked_cond_res} reports the BD-rate savings of conditional coding, conditional residual coding, and MaskCRT, with residual coding serving as the anchor.} For a fair comparison, all these competing methods share most of the components of MaskCRT, with minimal changes to its compression backbone. \textcolor{black}{As shown, our masked conditional residual coding outperforms residual coding, conditional coding, and conditional residual coding across all the test datasets. Notably, conditional coding and conditional residual coding achieve higher compression performance than residual coding on most datasets, except for HEVC-E~\cite{hevcctc}. This result is attributed to the characteristics of the HEVC-E~\cite{hevcctc} sequences, which are video conferencing-type content with the static background. In theory, such content favors residual coding. Still, there are few moving objects in the foreground. We see that our masking mechanism is able to better handle those moving objects, achieving the highest compression performance.}



\textcolor{black}{Table~\ref{tab:masked_cond_res_per_seq} further presents the per-sequence bitrate savings for UVG and HEVC-B datasets.} We observe that MaskCRT performs consistently better than conditional residual coding in all the test cases. In particular, the gains are most obvious on those fast-motion sequences, such as Jockey and ReadySteadyGo, where the motion estimates are less accurate and the assumption $H(x_t-x_c) \leq H(x_t)$ can be easily violated.

\begin{table}[t]
\centering
    \caption{\textcolor{black}{Per-sequence BD-rates of conditional residual coding and our proposed MaskCRT on UVG and HEVC-B datasets. The anchor is conditional coding based on our conditional Transformer-based codec.}}
    \label{tab:masked_cond_res_per_seq}
    \begin{tabular}{cccc}
    \hline
    \multirow{2}{*}{Datasets} & \multirow{2}{*}{Sequences} & \multicolumn{2}{c}{BD-rate (\%) PSNR-RGB}                                                                                                                 \\ \cline{3-4} 
                              &                            & \begin{tabular}[c]{@{}c@{}}Conditional \\ Residual Coding\end{tabular} & \begin{tabular}[c]{@{}c@{}}Masked Conditional\\ Residual Coding\end{tabular} \\ \hline
    \multirow{7}{*}{UVG}      & Beauty                     & -0.5                                                                   & -5.0                                                                         \\
                              & Bosphorus                  & -2.7                                                                   & -8.2                                                                         \\
                              & HoneyBee                   & -8.0                                                                   & -13.6                                                                        \\
                              & Jockey                     & -2.8                                                                   & -12.2                                                                        \\
                              & ReadySteadyGo              & -3.4                                                                   & -10.2                                                                        \\
                              & ShakeNDry                  & -0.7                                                                   & -3.3                                                                         \\
                              & YatchRide                  & -1.8                                                                   & -6.0                                                                         \\ \cline{2-4}
                              & \textcolor{black}{Average}                    & \textcolor{black}{-1.6}                                                                   & \textcolor{black}{-7.8}                                                                         \\ \hline
    \multirow{5}{*}{HEVC-B}   & BasketballDrive            & -0.9                                                                   & -5.8                                                                         \\
                              & BQTerrace                  & -1.8                                                                   & -5.4                                                                         \\
                              & Cactus                     & -2.3                                                                   & -8.1                                                                         \\
                              & Kimono1                    & 0.6                                                                    & -3.6                                                                         \\
                              & ParkScene                  & -2.0                                                                   & -3.6                                                                         \\ \cline{2-4}
                              & \textcolor{black}{Average}                    & \textcolor{black}{-1.8}                                                                   & \textcolor{black}{-5.4}                                                                        \\ \hline
    \end{tabular}
\end{table}
\begin{figure*}[h!]
\centering
\resizebox{\textwidth}{!}
{
\Large
\begin{tabular}{cccc}
    $x_t$ & $m \odot x_c$ & $ x_t - m \odot x_c$ & $m$ \\
    
    \includegraphics[width=0.3\textwidth]{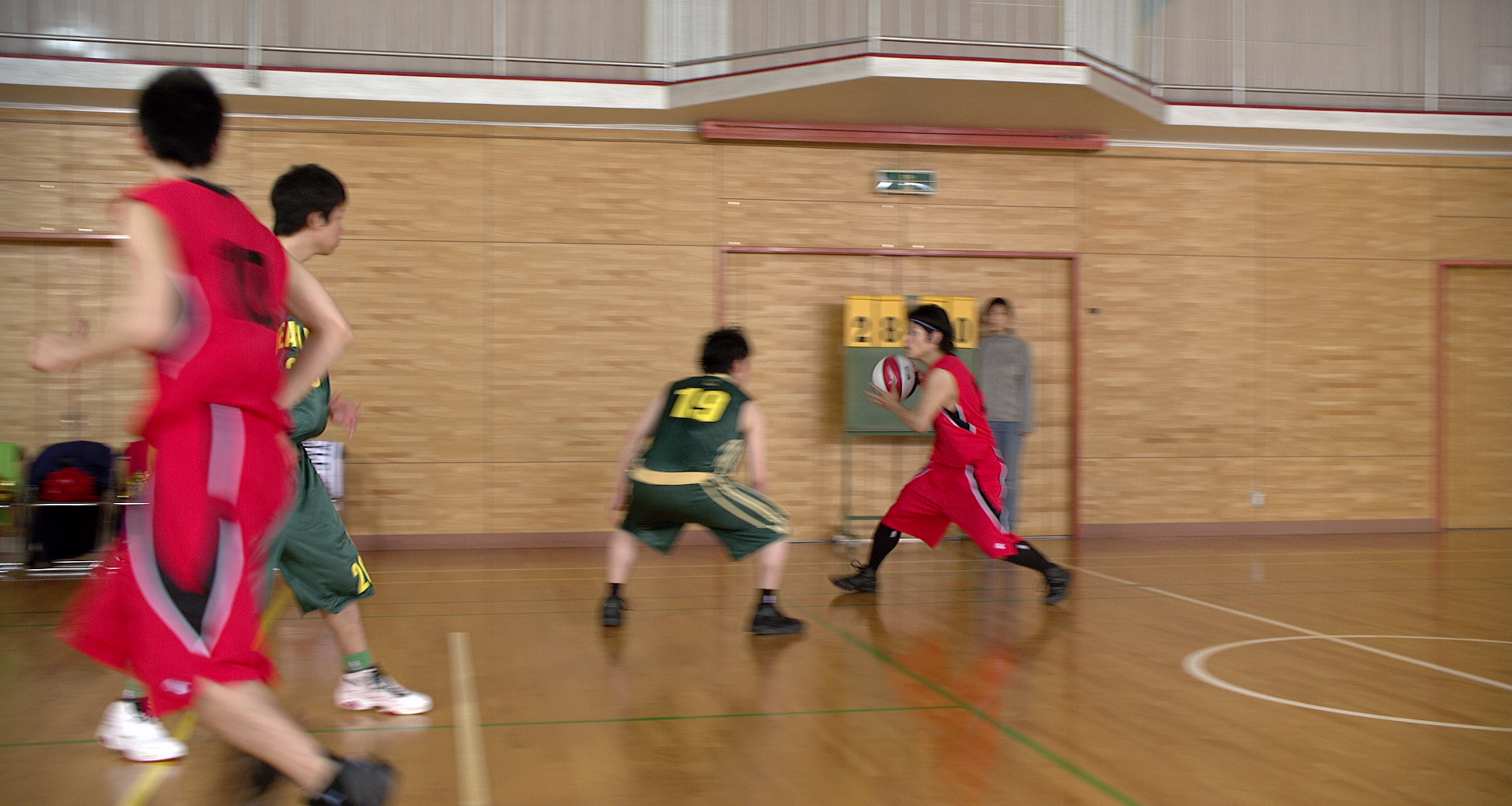}&
    \includegraphics[width=0.3\textwidth]{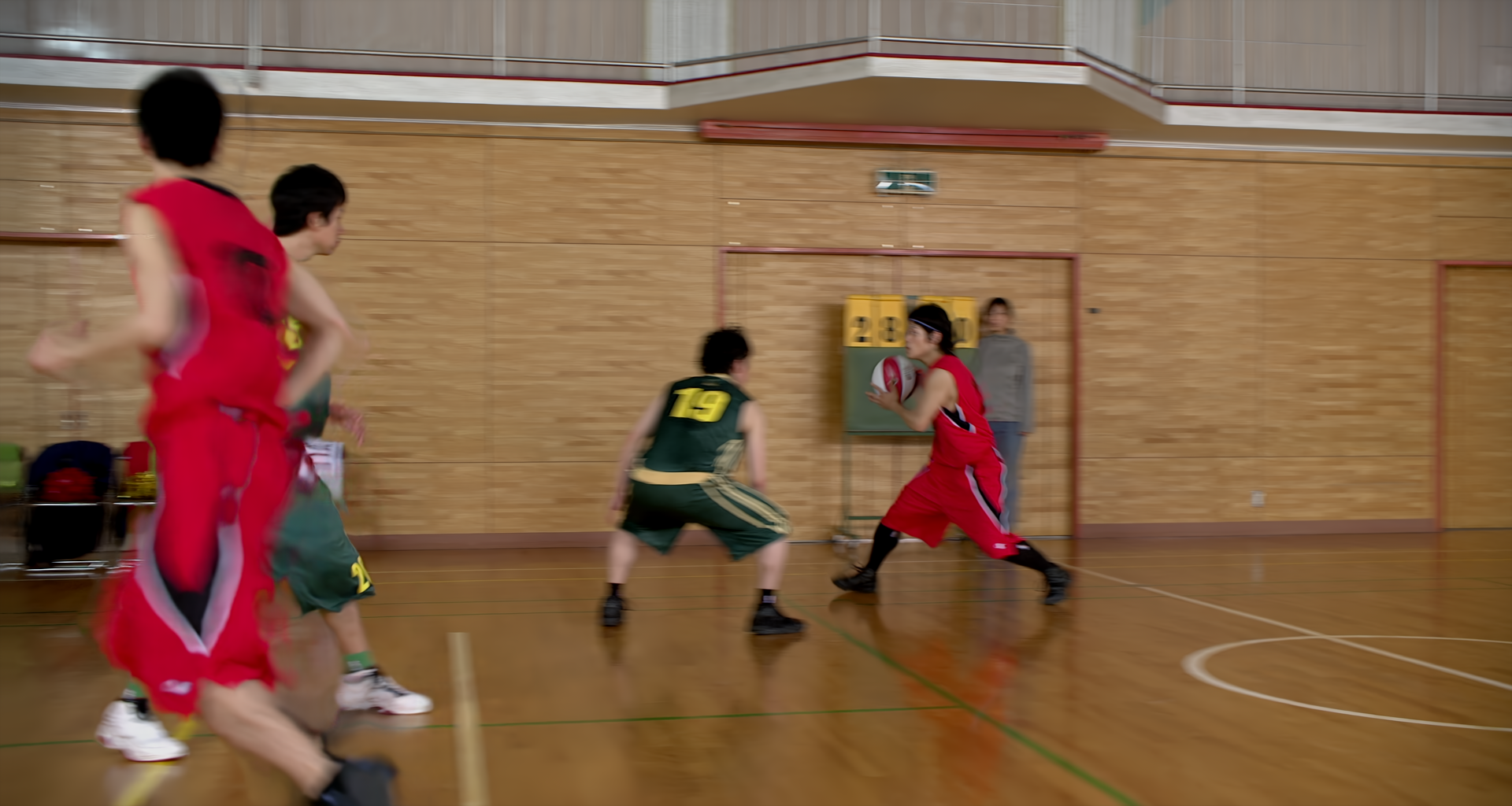}&
    \includegraphics[width=0.3\textwidth]{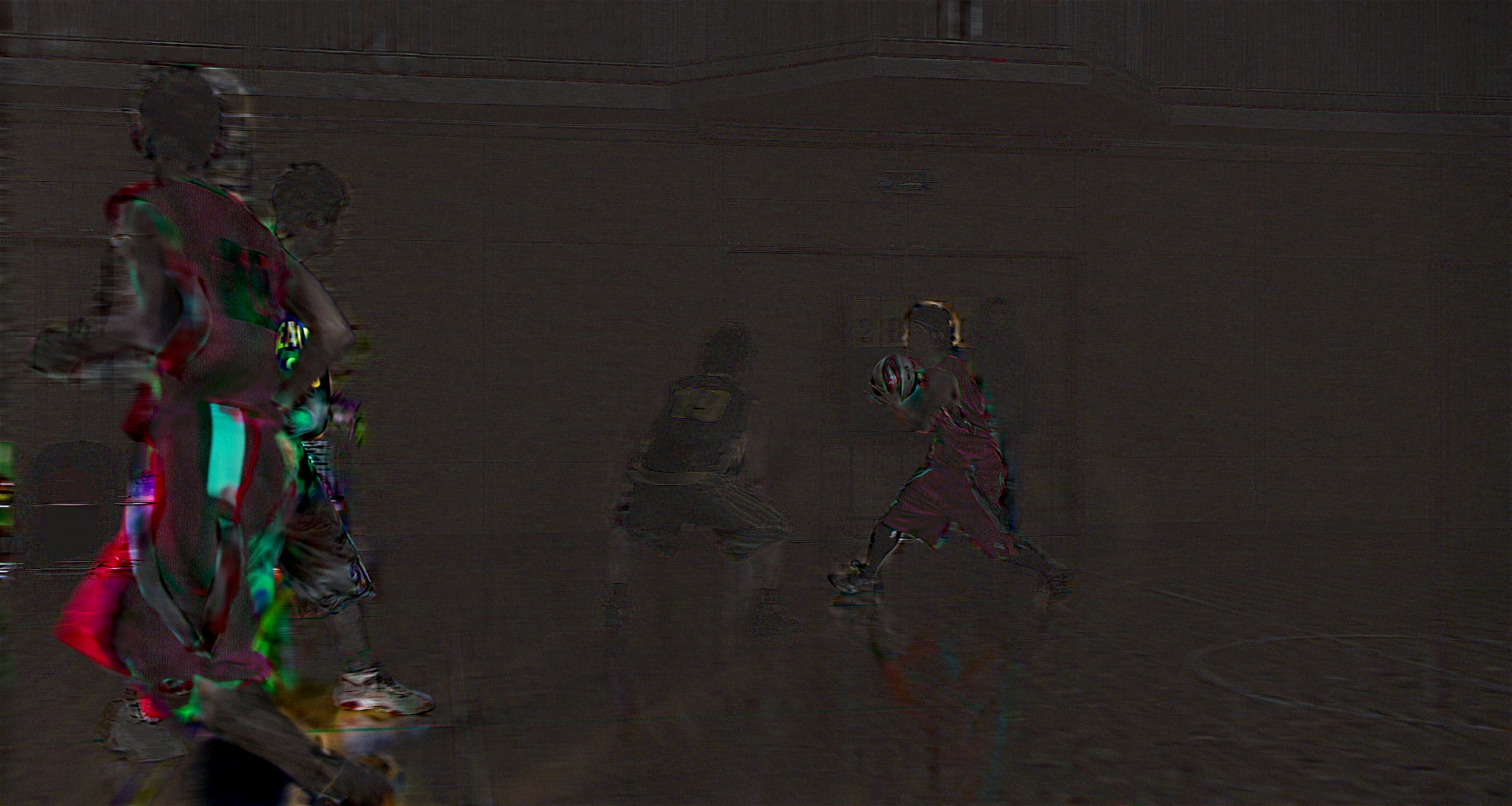}&
    \includegraphics[height=83pt]{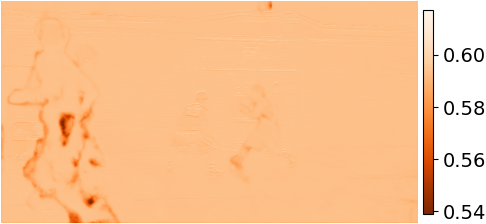} \\
    
    \includegraphics[width=0.3\textwidth]{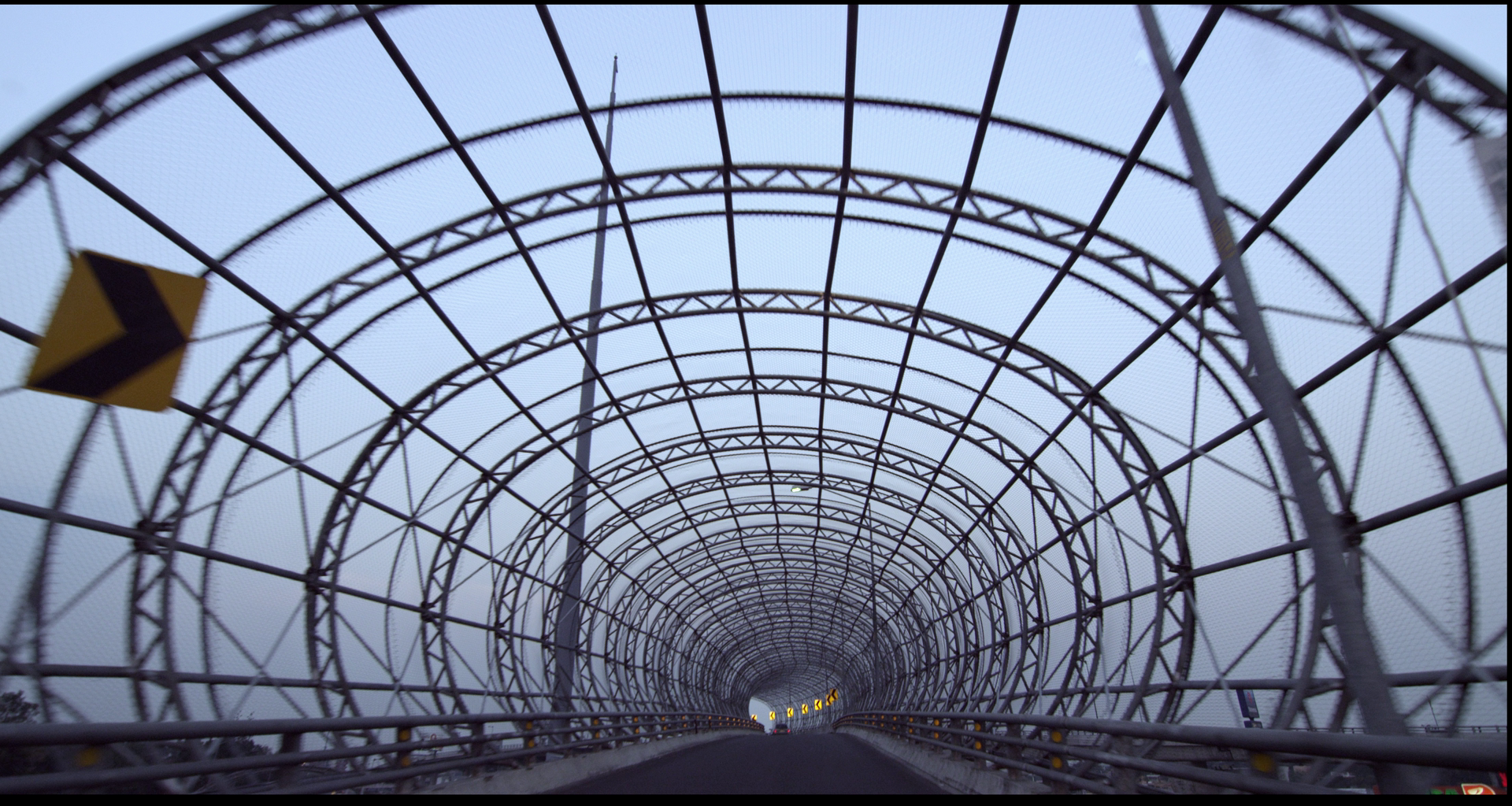}&
    \includegraphics[width=0.3\textwidth]{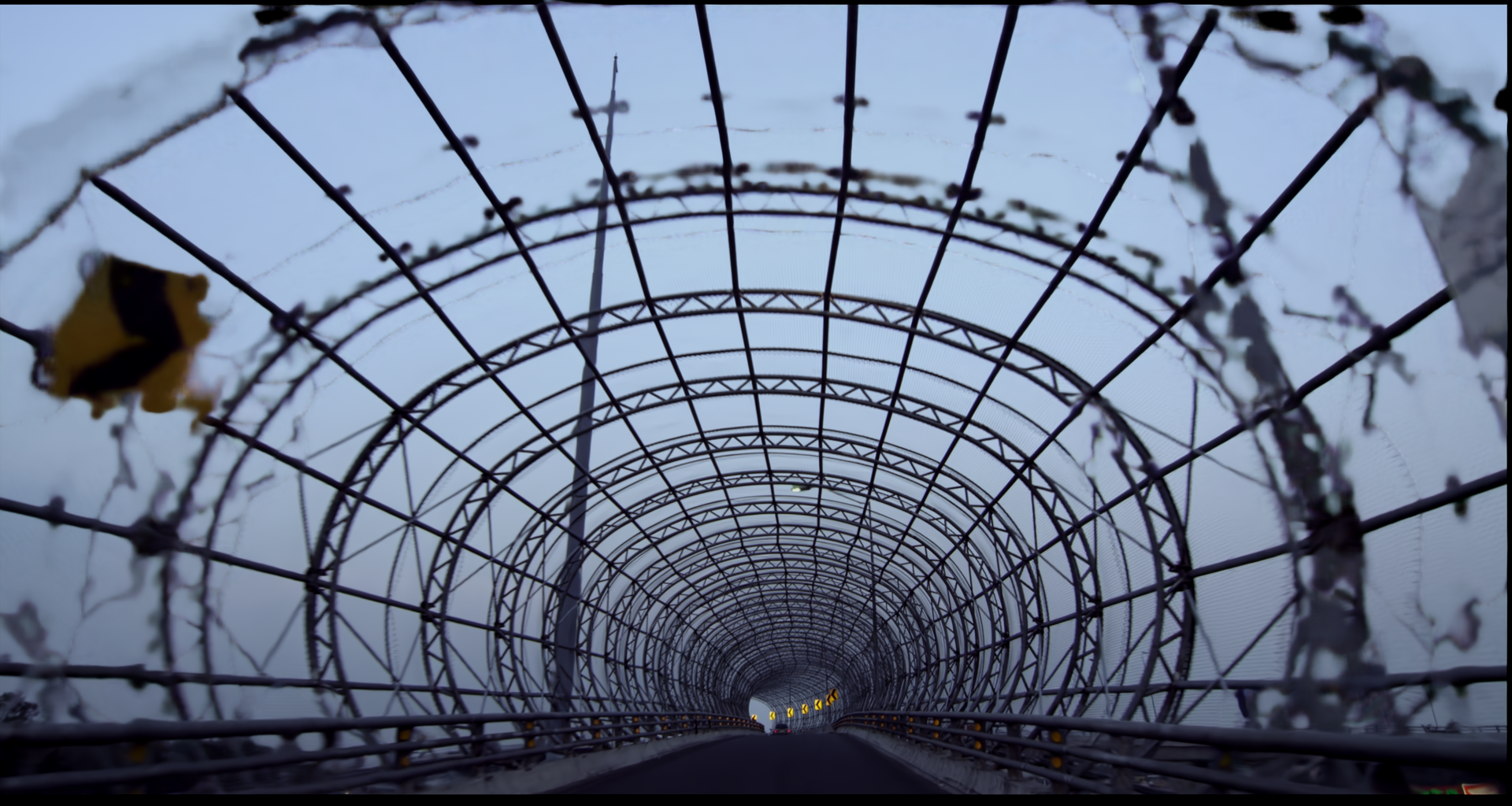}&
    \includegraphics[width=0.3\textwidth]{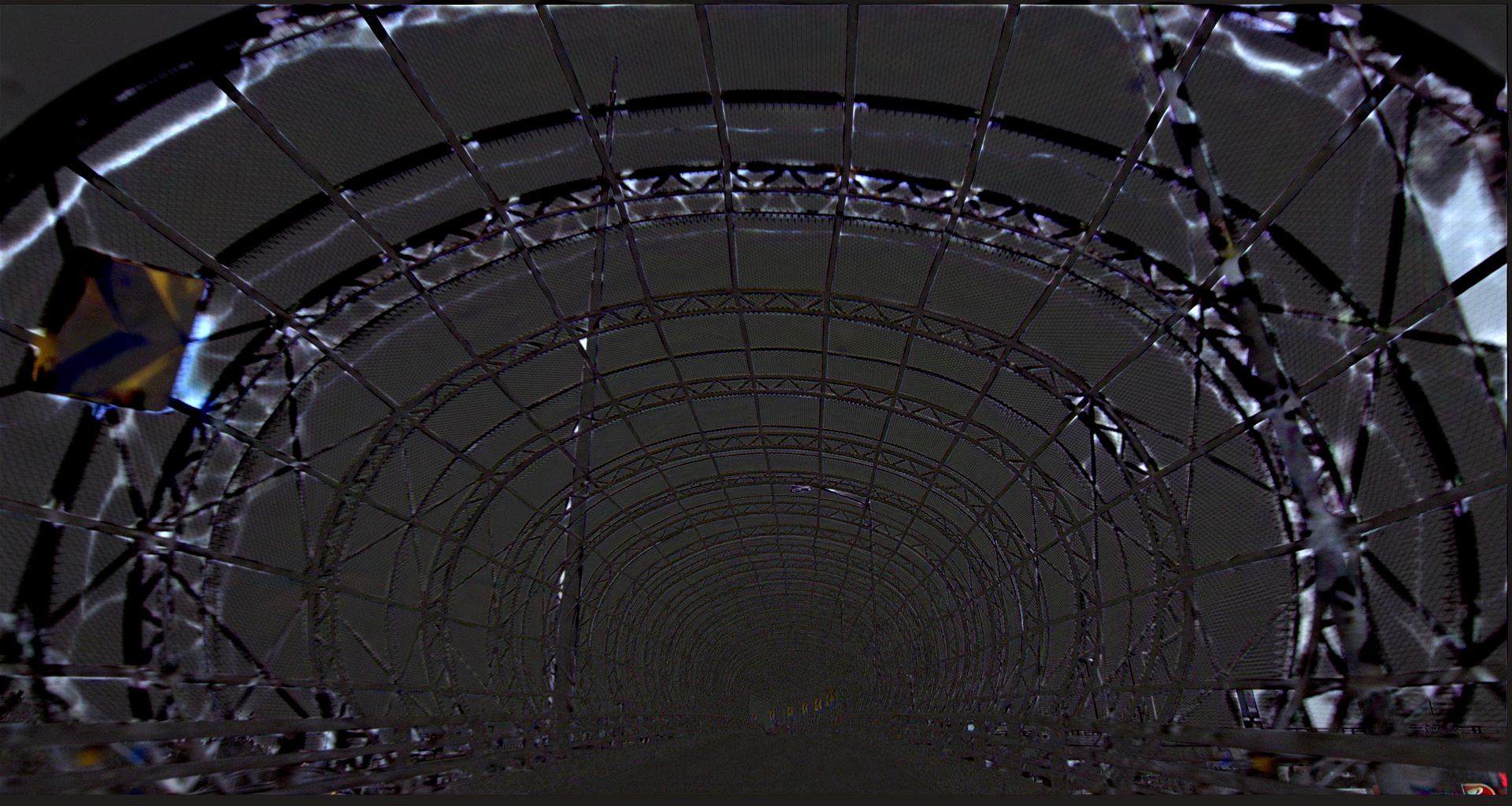}&
    \includegraphics[height=83pt]{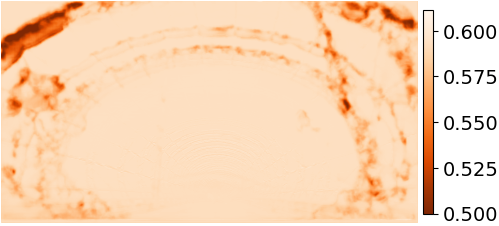} \\

    \includegraphics[width=0.3\textwidth]{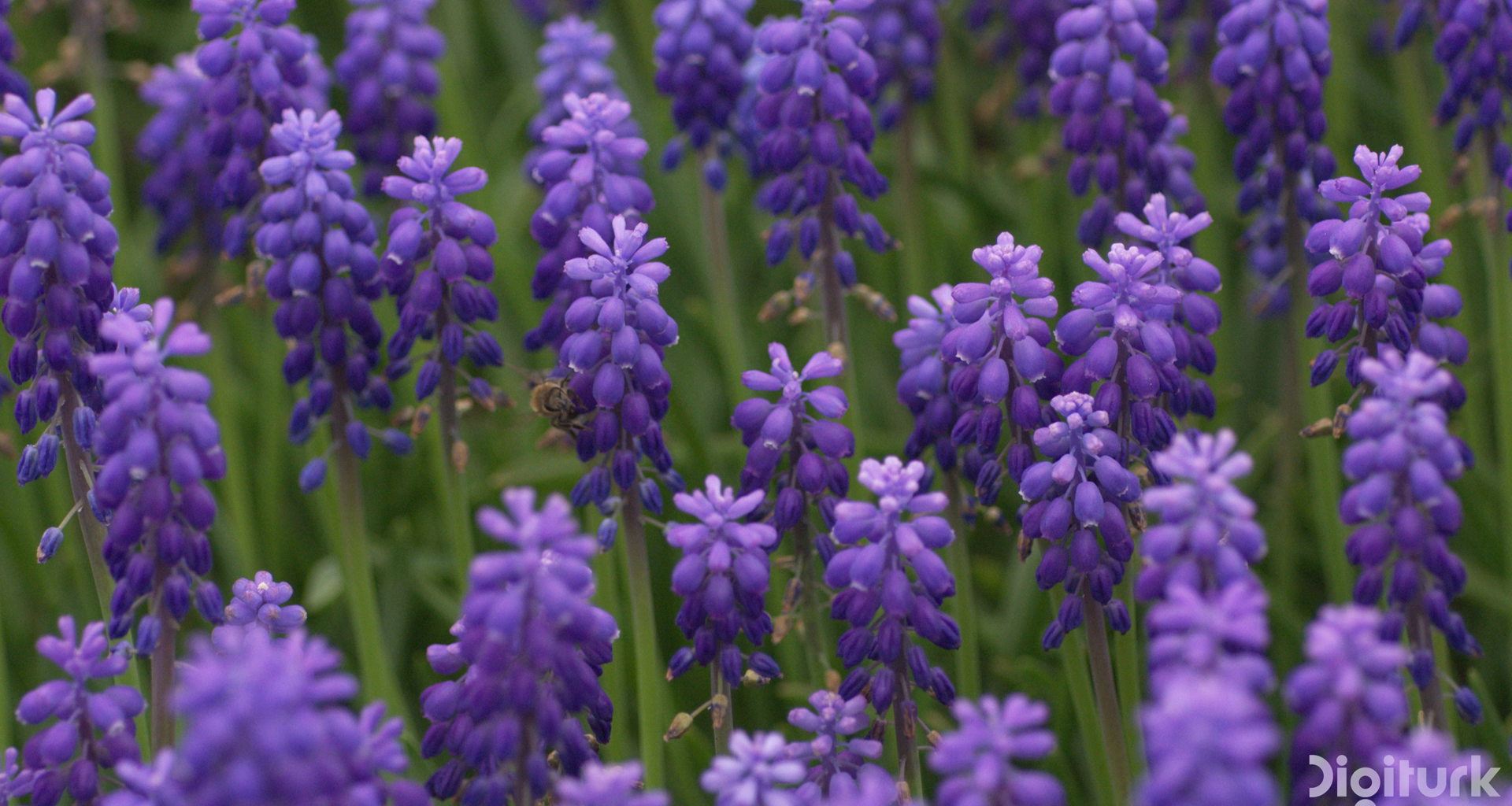}&
    \includegraphics[width=0.3\textwidth]{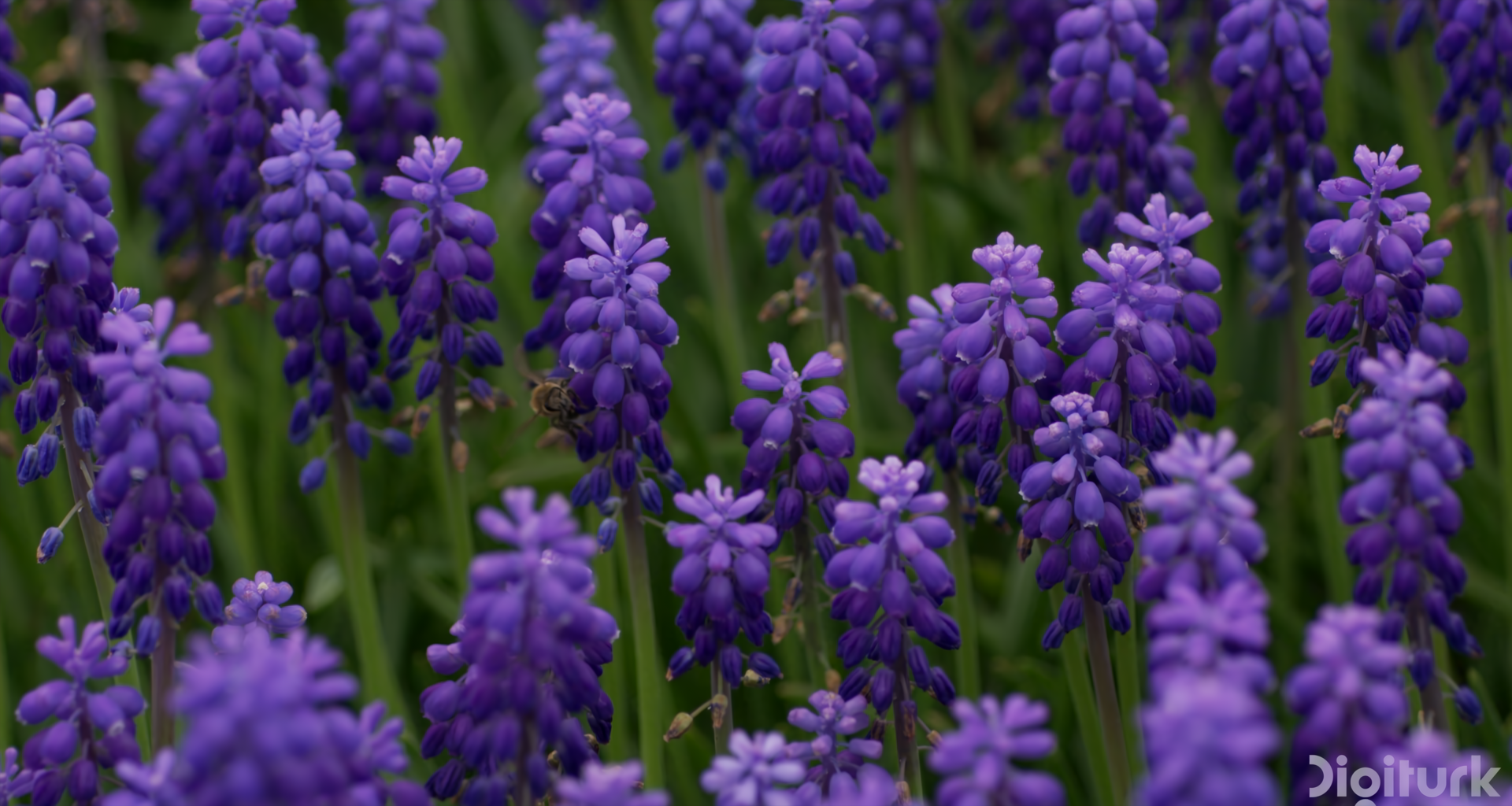}&
    \includegraphics[width=0.3\textwidth]{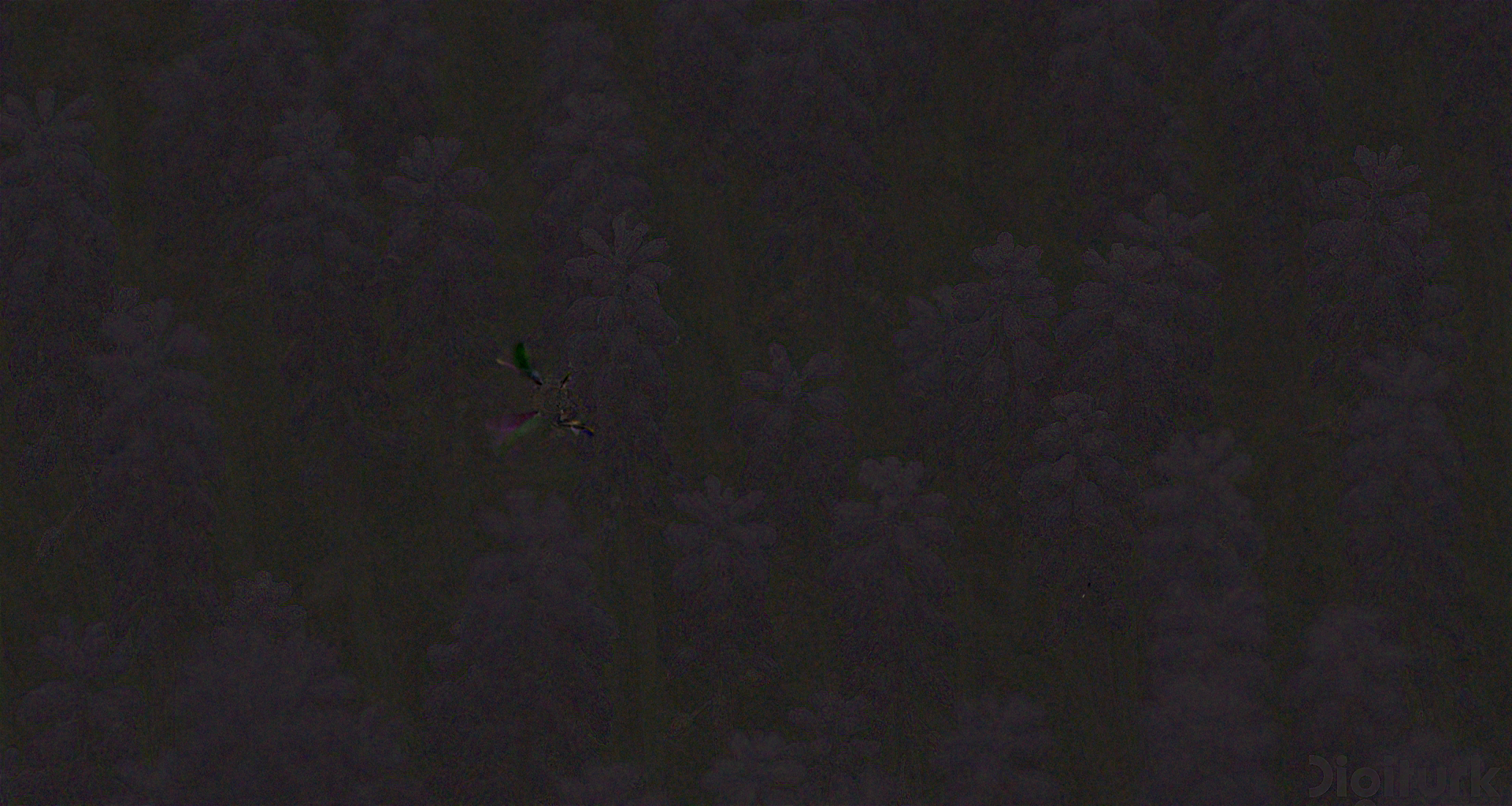}&
    \includegraphics[height=83pt]{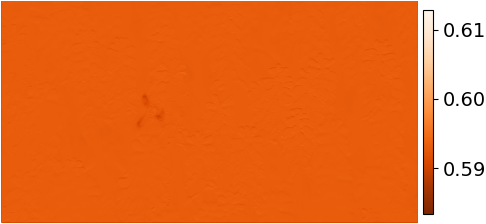} \\

    \includegraphics[width=0.3\textwidth]{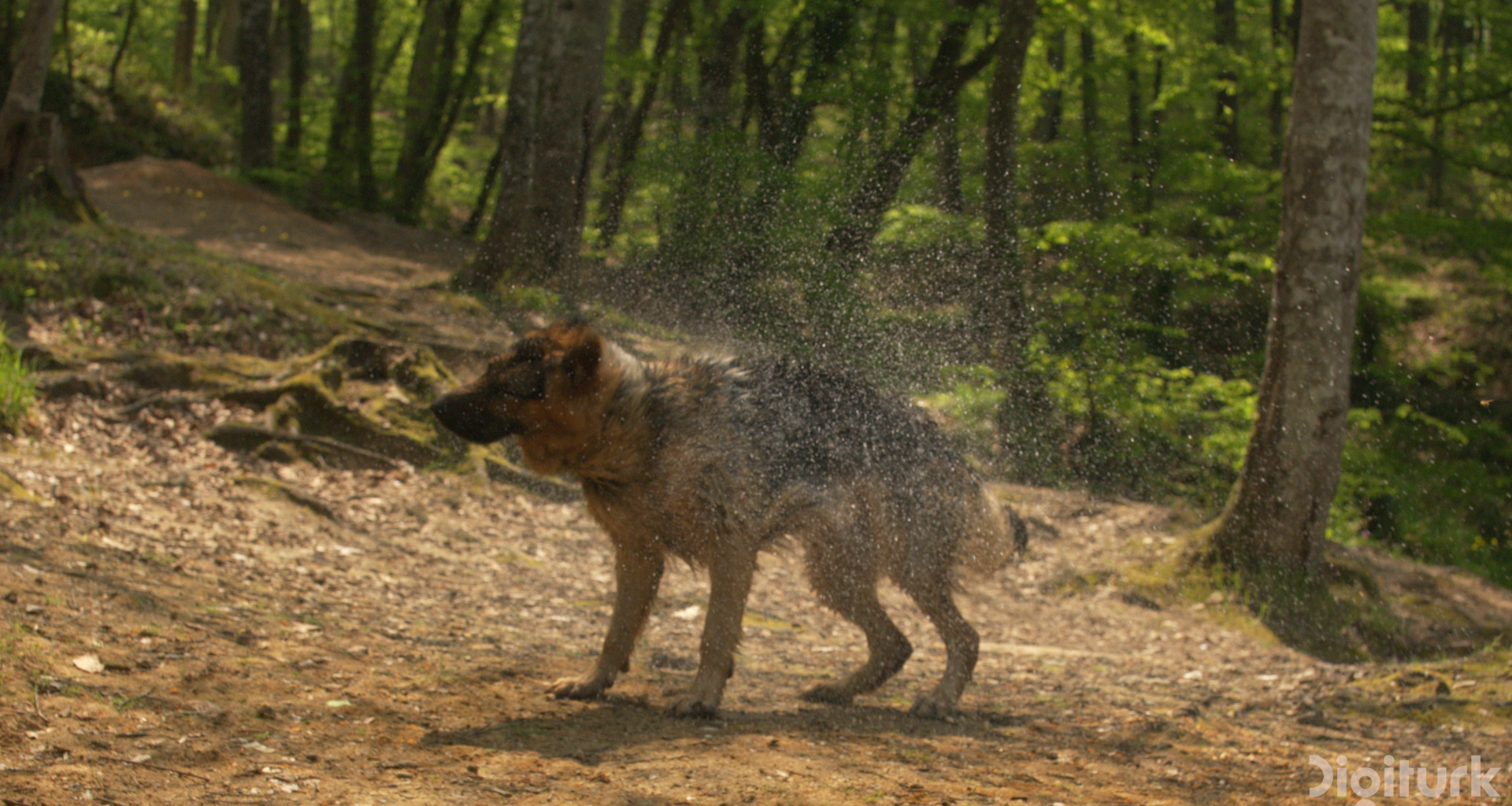}&
    \includegraphics[width=0.3\textwidth]{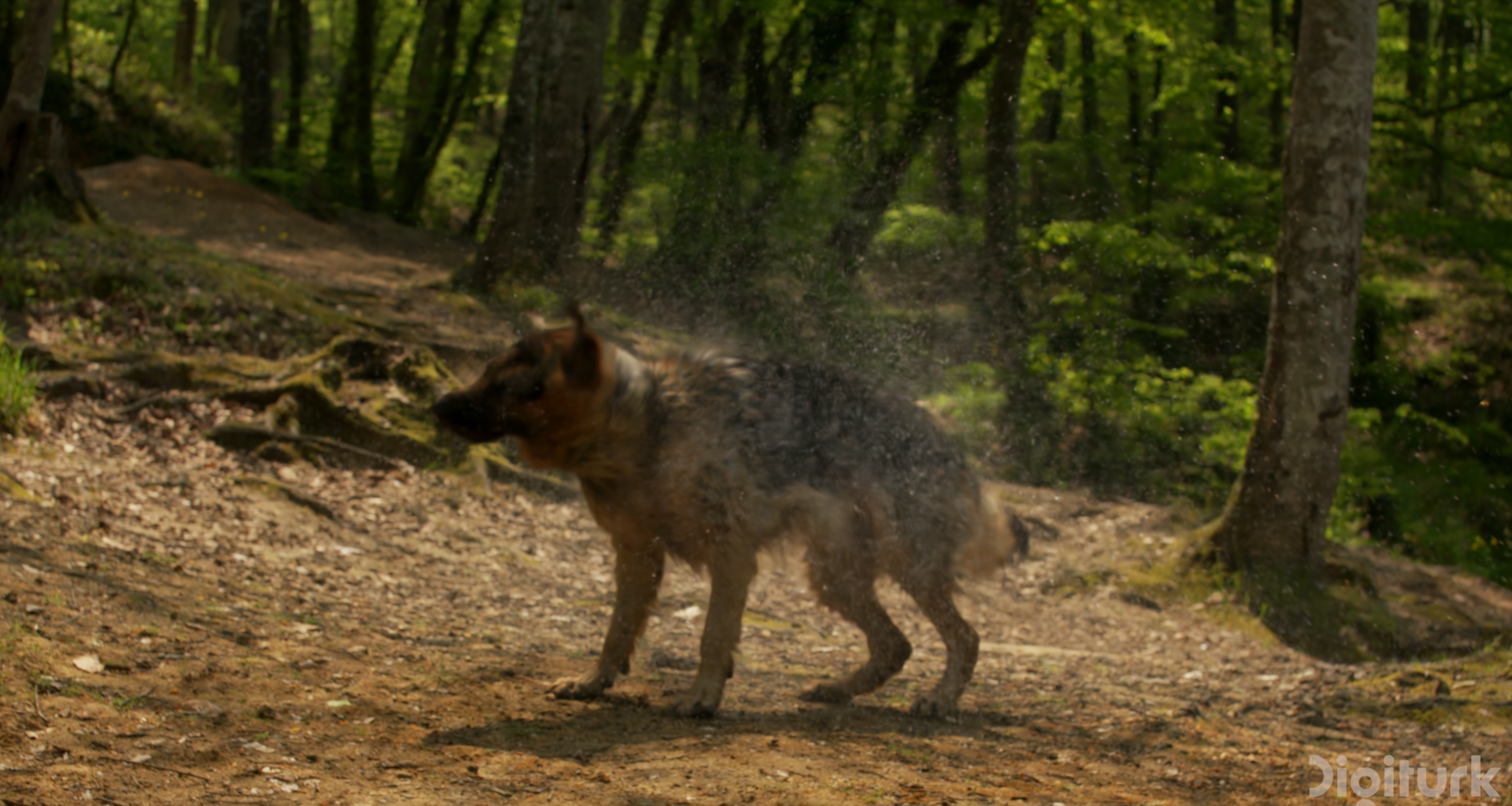}&
    \includegraphics[width=0.3\textwidth]{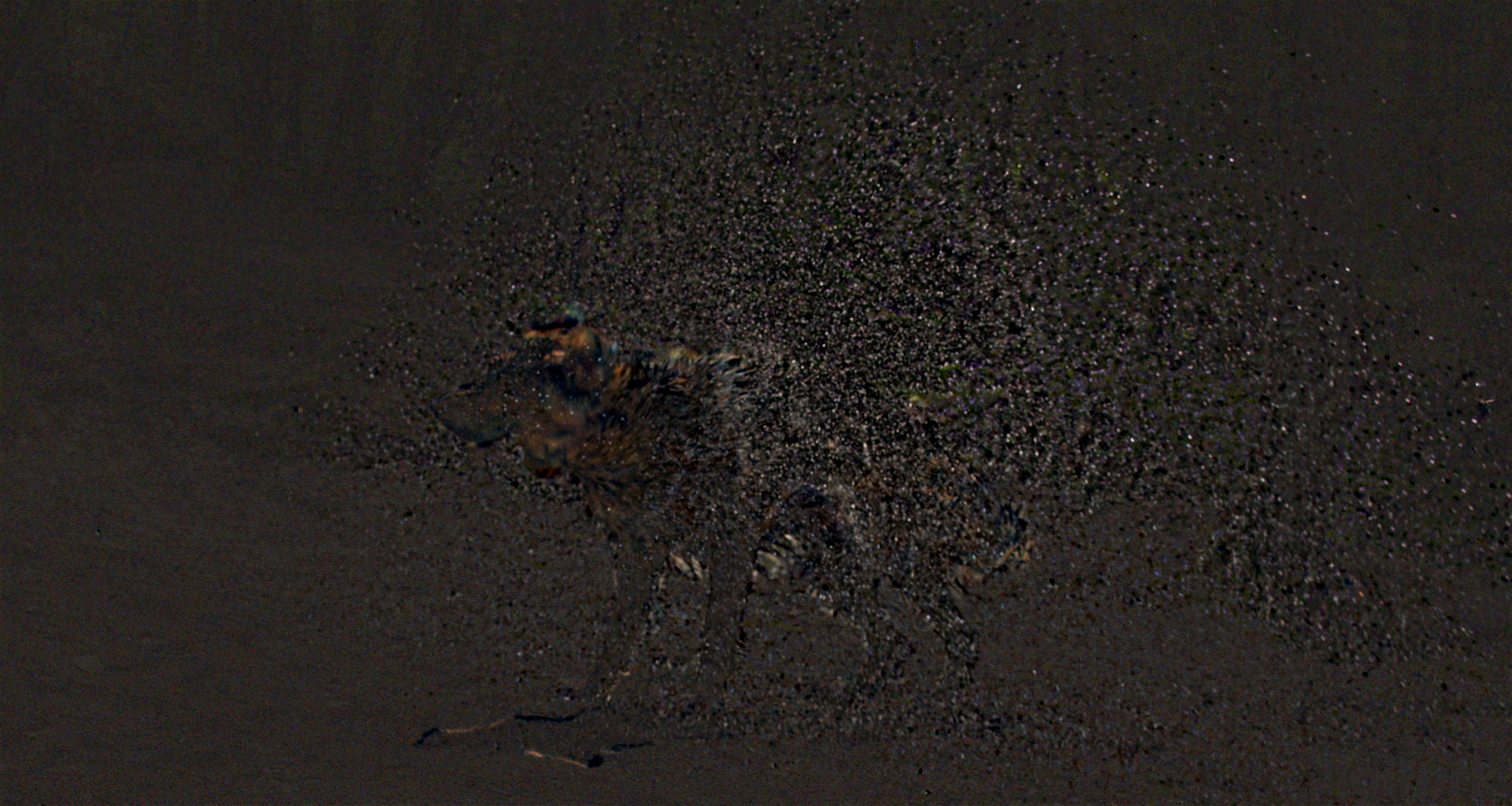}&
    \includegraphics[height=83pt]{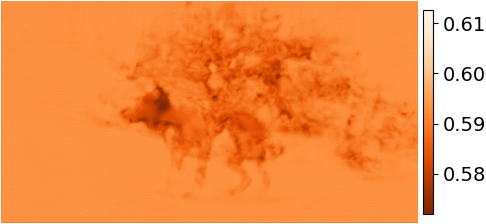} \\ 
\end{tabular}
}


\caption{Illustration of several signals in MaskCRT. Shown from left to right are the input frame $x_t$, the masked prediction signal $m \odot x_c$, the decoded prediction residue $x_t - m \odot x_c$, and the mask $m$. The darker areas in $m$ correspond to smaller mask values. That is, these areas are suppressed from being used for the prediction of $x_t$.}

\label{fig:visual}
\end{figure*}

To understand better the inner workings of MaskCRT, Fig.~\ref{fig:visual} visualizes its masks and the masked prediction residue for several test sequences having different motion characteristics. From top to bottom, the first two rows are results for two fast motion sequences, BasketballDrive and videoSRC10, whereas the last two rows correspond to HoneyBee and ShakeNDry. HoneyBee is a slow-motion sequence, and ShakeNDry has complex motion in the foreground. On all these sequences, the learned masks $m$ have relatively lower values in the regions where the motion estimates may not be reliable, e.g. the regions full of sharp edges, object boundaries, or complex motion. In these regions, the assumption $H(x_t-x_c) \leq H(x_t)$ can be easily violated. As a result,  conditional coding is more beneficial than conditional residual coding. Recall that the input to MaskCRT is $(1-m) \odot x_t + m \odot (x_t - x_c)$. Lower mask values $m$ suggest the tendency of performing conditional coding. In comparison, the mask values are seen to be relatively higher in the static background regions. In other words, conditional residual coding is preferred in these regions.

\textcolor{black}{It is worth noting that a "soft" mask value $m$ between 0 and 1 results in a multi-hypothesis reconstruction of the input image. That is, both the masked conditioning signal $m \odot x_c$ and the reconstruction of the masked input $(1-m) \odot x_t + m \odot (x_t - x_c)$ offer an approximation of the input signal $x_t$ to some extent. This is evidenced by the fact that the seemingly residual signal $(1-m) \odot x_t + m \odot (x_t - x_c)$ in Fig.~\ref{fig:visual}, when examined closely, still has much texture information. Since $x_c$ is generated by a neural network without any regularization, the mean of $x_c$ is found to be much greater than that of $x_t$. As such, $m$ are seen to have values much smaller than 1 in forming a prediction of $x_t$.}




\paragraph{\textcolor{black}{Conditional Transformer-based Codec}}
\label{ablation:cond_transformer}
\textcolor{black}{Table~\ref{tab:cond_transformer} presents the BD-rate comparison to justify our CSTB design.} We experiment with its variants mentioned in Section~\ref{sec:masked_cond_res_transformer}, with our proposed method (Fig.~\ref{fig:Variants_of_CSTB}(a)) serving as the anchor. In Table~\ref{tab:cond_transformer}, \emph{Concat.}, \emph{Cross-attn. with input as Q}, and \emph{Cross-attn. with cond. as Q} refer to the designs in Figs.~\ref{fig:Variants_of_CSTB} (b), (c), and (d), respectively. The variant \emph{Without $1 \times 1$ conv} is to justify the effectiveness of the $1 \times 1$ convolutional layer at the end of CSTB. It follows the same self-attention mechanism in our proposed scheme, but updates only the input branch without fusing both the input and conditioning branches at the end of CSTB. 

From their positive BD-rate numbers, all four variants perform worse than our proposed method. Among them, \emph{Concat.} outperforms the others. Notably, both cross-attention-based approaches are not very effective, particularly the one that formulates queries based on the input signal. We conjecture that the residual nature of the input signal inherently has little information to guide properly the attention process. In common, both the cross-attention methods are unable to allow intra-branch information exchange, leading to their inferior performance. We also see that \emph{Without $1 \times 1$ conv} yields less favorable results, implying the importance of fusing information from both the input and conditioning branches after self-attention.

\begin{table}[t]
\centering
    \caption{\textcolor{black}{Ablation experiment of CSTB for implementing the conditional Transformer-based codec. The anchor for BD-rate evaluation adopts our design in Fig.~\ref{fig:Variants_of_CSTB}(a).}}
    \label{tab:cond_transformer}
    \begin{tabular}{lccc}
    \hline
                                    & \multicolumn{3}{c}{BD-rate (\%) PSNR-RGB}                                                      \\ \cline{2-4} 
    \makecell[c]{\multirow{-2}{*}{Settings}}                                              & \hspace{-2mm}UVG   & \hspace{-2mm} HEVC-B  & \hspace{-2mm} MCL-JCV                       \\ \hline
    \textcolor{black}{Fig.~\ref{fig:Variants_of_CSTB}(b):} Concat.                         & \hspace{-2mm}5.8   & \hspace{-2mm} 1.8     & \hspace{-2mm} 2.5                           \\ \hline
    \textcolor{black}{Fig.~\ref{fig:Variants_of_CSTB}(c):} Cross-attn. with input as Q     & \hspace{-2mm}47.3  & \hspace{-2mm} 42.0    & \hspace{-2mm} 51.3                          \\ \hline
    \textcolor{black}{Fig.~\ref{fig:Variants_of_CSTB}(d):} Cross-attn. with cond. as Q     & \hspace{-2mm}15.4  & \hspace{-2mm} 14.7    & \hspace{-2mm} 19.0                          \\ \hline
    Without $1 \times 1$ conv                                                             & \hspace{-2mm}16.6  & \hspace{-2mm} 18.1    & \hspace{-2mm} 19.9 \\ \hline
    \end{tabular}
\end{table}
\begin{table*}[t]
\centering
    \caption{\textcolor{black}{Ablation experiment of CTM and several context models. The base model (also the anchor for BD-rate evaluation) is our MaskCRT without CTM and any context model.}}
    \label{tab:cdm}
    \begin{tabular}{c|ccc|ccc}
    \hline
    \multirow{2}{*}{Settings}                                                                                          & \multicolumn{3}{c|}{BD-rate (\%) PSNR-RGB} & \multicolumn{3}{c}{Complexity}                                 \\ \cline{2-7} 
                                                                                                                       & UVG       & HEVC-B      & MCL-JCV      & Model Size (M) & Encoding (kMAC/pixel) & Decoding (kMAC/pixel) \\ \hline
    Base                                                                                                         & 0.0         & 0.0          & 0.0            & 24.3           & 1376.1                & 741.8                 \\ \hline
    Base (Large)                                                                                                 & -1.3      & -1.4       & -0.4         & 26.1 (+7\%)    & 1392.2 (+1\%)         & 749.9 (+1\%)           \\ \hline
    Base + $L_{ch-corr}$~\cite{nips23}                                                                           & 0.2       & -0.8       & 0.2          & 24.3 (+0\%)    & 1376.1 (+0\%)         & 741.8 (+0\%)           \\ \hline
    Base + CTM                                                                                                   & -2.1      & -4.4       & -3.3         & 26.2 (+8\%)    & 1383.4 (+0.5\%)       & 745.4 (+0.5\%)          \\ \hline
    \begin{tabular}[c]{@{}c@{}}Base + ChARM~\cite{minnen2020channel} \end{tabular}                               & -2.2      & -4.7       & -3.3         & 30.2 (+24\%)   & 1399.1 (+2\%)         & 764.8 (+3\%)           \\ \hline
    \begin{tabular}[c]{@{}c@{}}Base + CTM \& \\ Checkerboard~\cite{multistage}\end{tabular}                      & -4.3      & -5.4       & -4.6         & 27.7 (+14\%)   & 1401.8 (+2\%)         & 767.5 (+3\%)         \\ \hline
    \begin{tabular}[c]{@{}c@{}}Base + Spatial-Channel \\ Autoregessive Model~\cite{dcvc_dc}\end{tabular}         & -5.0      & -5.6       & -6.8         & 30.9 (+28\%)   & 1445.3 (+5\%)         & 811 (+9\%)           \\ \hline
    \end{tabular}

\end{table*}
\begin{figure}[t!]
    \centering
    \subfigure{
        \centering
        \includegraphics[height=0.40\linewidth]{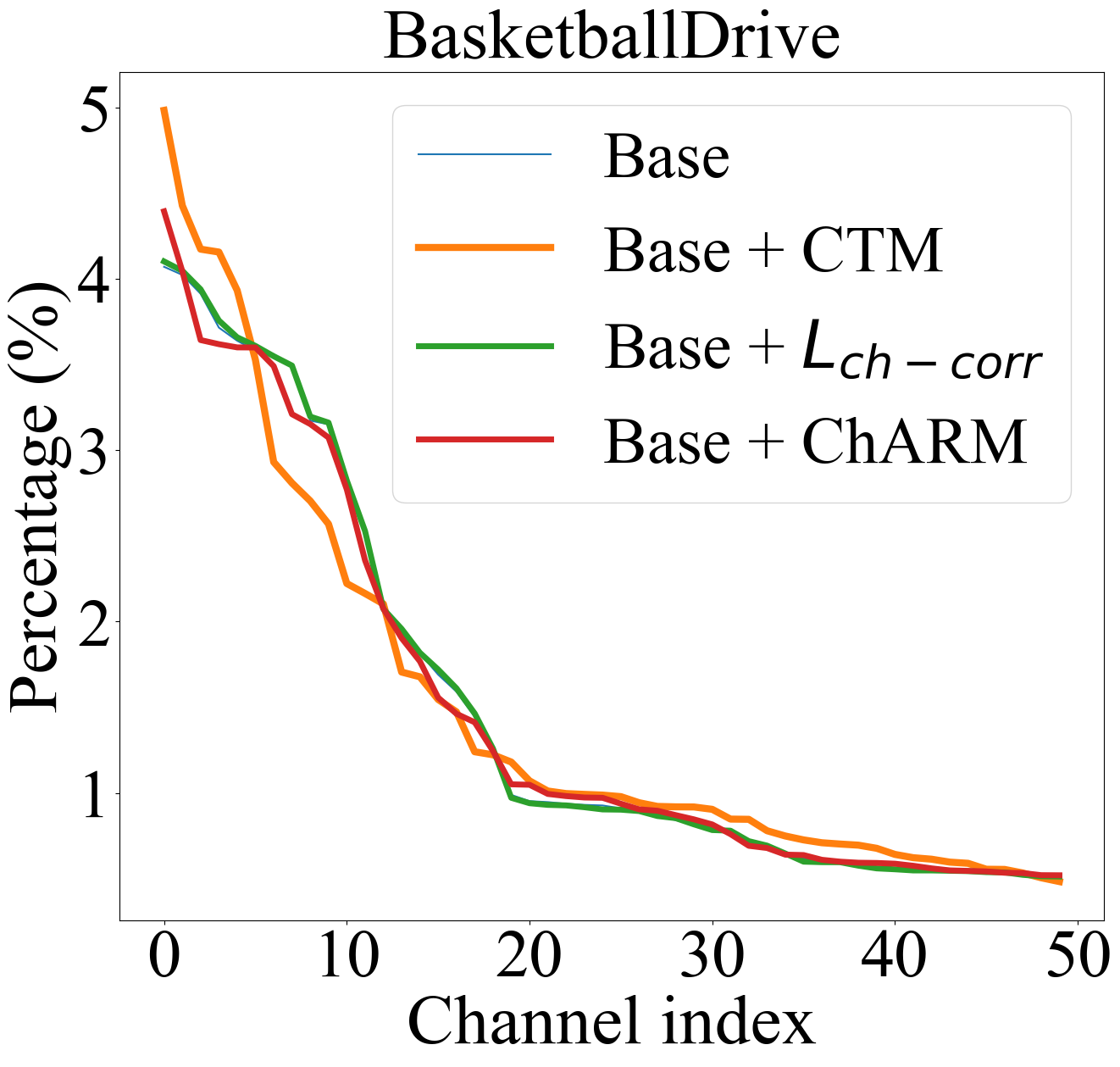}
        } 
    \subfigure{
        \centering
        \includegraphics[height=0.40\linewidth]{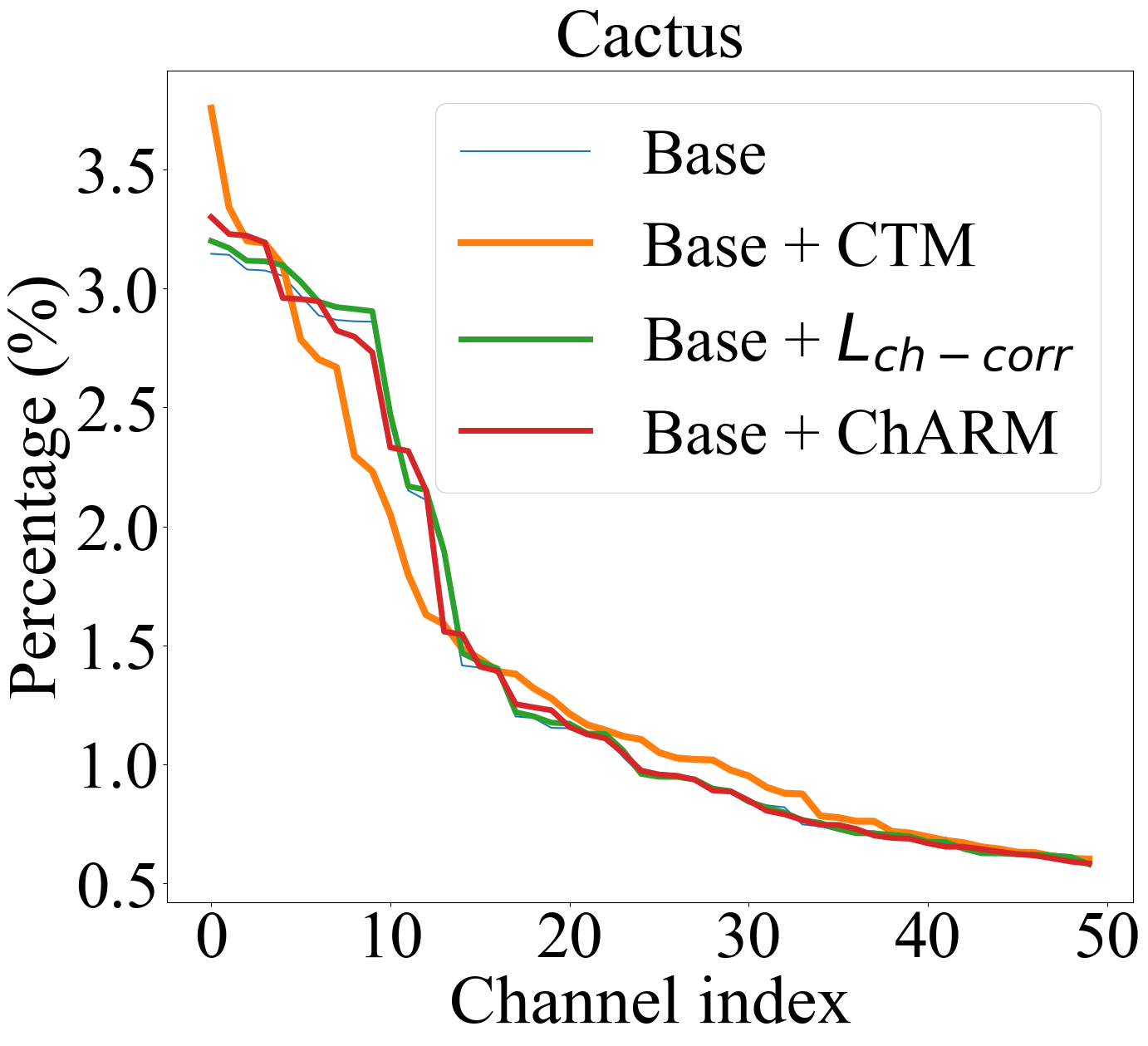}
        } 
    
    \subfigure{
        \centering
        \includegraphics[height=0.40\linewidth]{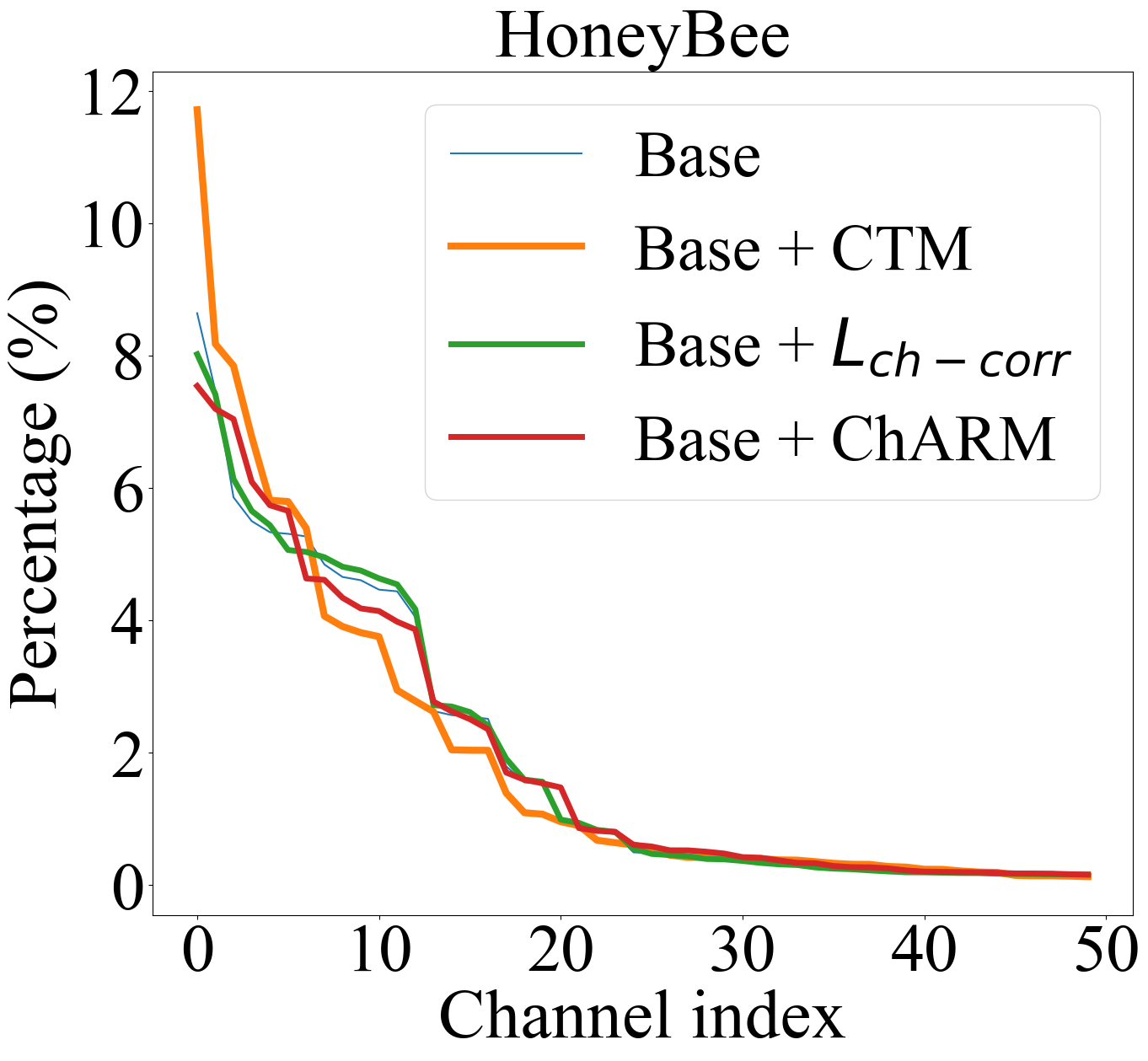}
        } 
    \subfigure{
        \centering
        \includegraphics[height=0.40\linewidth]{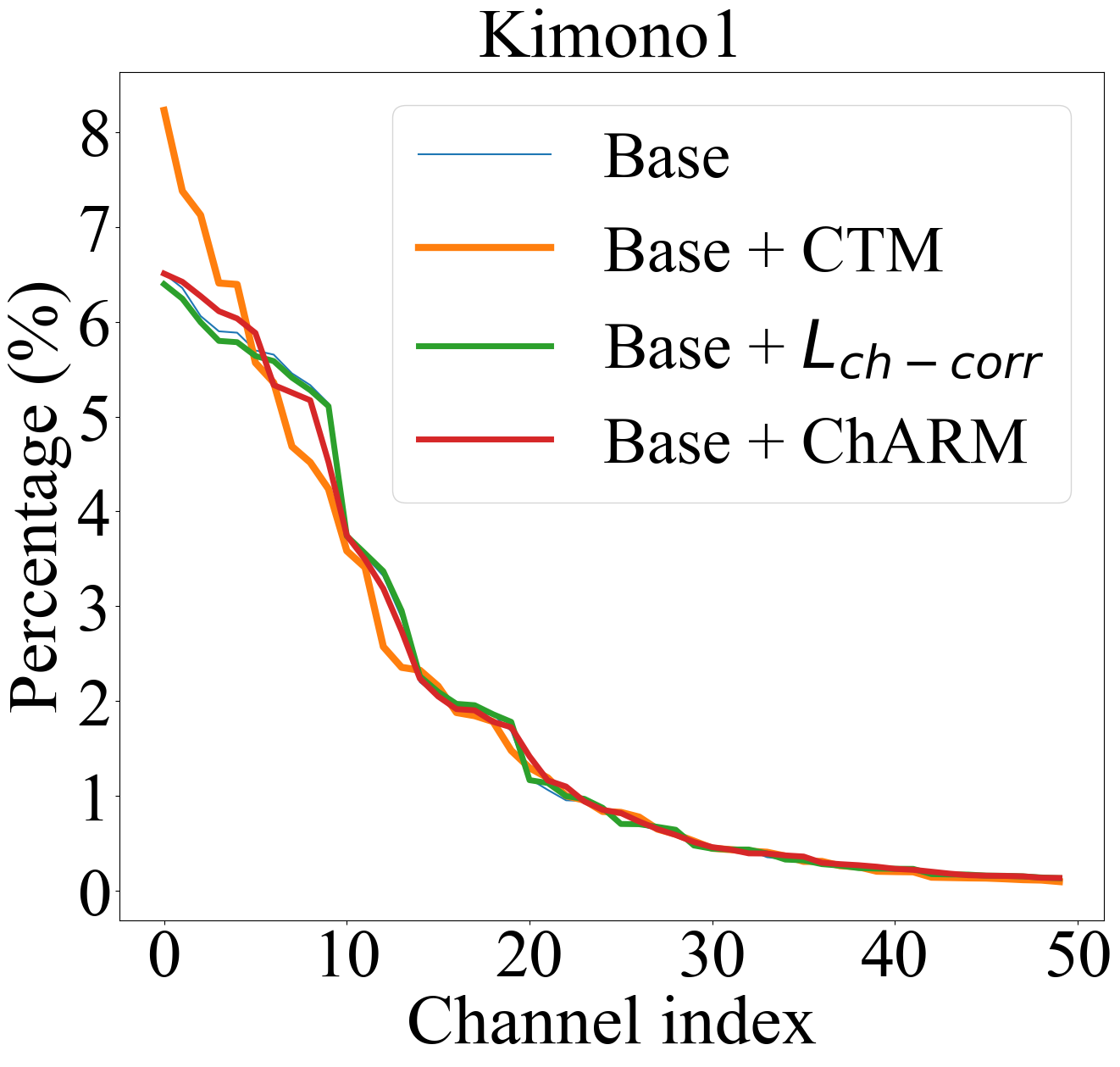}
        } 
    \caption{\textcolor{black}{The bit rate distribution over the latent channels. In each plot, the latent channels are arranged in descending order from left to right according to their average bit rates in percentage terms collected from the first 93 P-frames of each test sequence.}}

    \label{fig:energy}
\end{figure}

\paragraph{\textcolor{black}{CTM versus Context Models}}
\label{ablation:context}
\textcolor{black}{Table~\ref{tab:cdm} compares CTM with several competing context models, in terms of the BD-rate savings, model size, and kilo-Multiply-Accumulate operations per pixel (kMAC/pixel).} These context models include the channel-autoregressive model (ChARM)~\cite{minnen2020channel}, checkerboard context model~\cite{multistage}, and spatial-channel autoregressive model~\cite{dcvc_dc}. They divide the image latents into four slices along the channel, spatial, or both dimensions. We also experiment with applying the decorrelation loss~\cite{nips23} to the channel dimension, termed \emph{$L_{ch-corr}$}, and stacking more CSTBs for constructing a higher-capacity nonlinear autoencoder (termed \textit{Base (Large)}). 

From the table, we make the following observations. First, (1) although He~\emph{et al.}~\cite{elic} find that the point-wise non-linear transform, e.g. generalized divisive normalization (GDN), can be replaced with a purely convolutional neural network with sufficient capacity, our CTM (\textit{Base + CTM}) is more cost effective than this straightforward approach. This is evidenced by its higher BD-rate savings than those of \textit{Base Model (Large)} while showing comparable model size and kMAC/pixel. Second, (2) \emph{Base + CTM} performs comparably to \emph{Base + ChARM} yet with a less significant impact on the model size and encoding/decoding kMAC/pixel. Observe that \emph{Base + $L_{ch-corr}$} does not yield much gain, suggesting that linear decorrelation along the channel dimension is not as effective as it is along the spatial dimension~\cite{nips23}. In contrast, our CTM is able to decorrelate non-linearly in a way that better suits the factorial assumption of the hyperprior. Third, (3) on top of CTM, incorporating additionally the checkerboard context model has a marginal BD-rate saving of 1\%-2\%. Notably, the performance gap between \emph{Base + CTM + Checkerboard} and \emph{Base + Spatial-channel Autoregressive Model} on UVG and HEVC-B appears modest. However, the complexity increase of \emph{Base + Spatial-channel Autoregressive Model} is nearly two times that of \emph{CTM + Checkerboard}. 

\textcolor{black}{To shed light on the inner working of CTM, we further investigate the bit rate distribution over the latent channels in Fig.~\ref{fig:energy}. We see that with CTM, the first few channels of the image latents $y_t$ represent a larger portion of the entire compressed bit rate than the other cases without CTM. That is, CTM results in an energy compaction effect, with more information captured by the first few channels. Such an energy compaction effect justifies the good coding performance of CTM.}




\begin{figure*}[h!]
    \centering
    \subfigure{
        \centering
        \includegraphics[height=0.245\linewidth, trim= 0 0 60 50, clip]{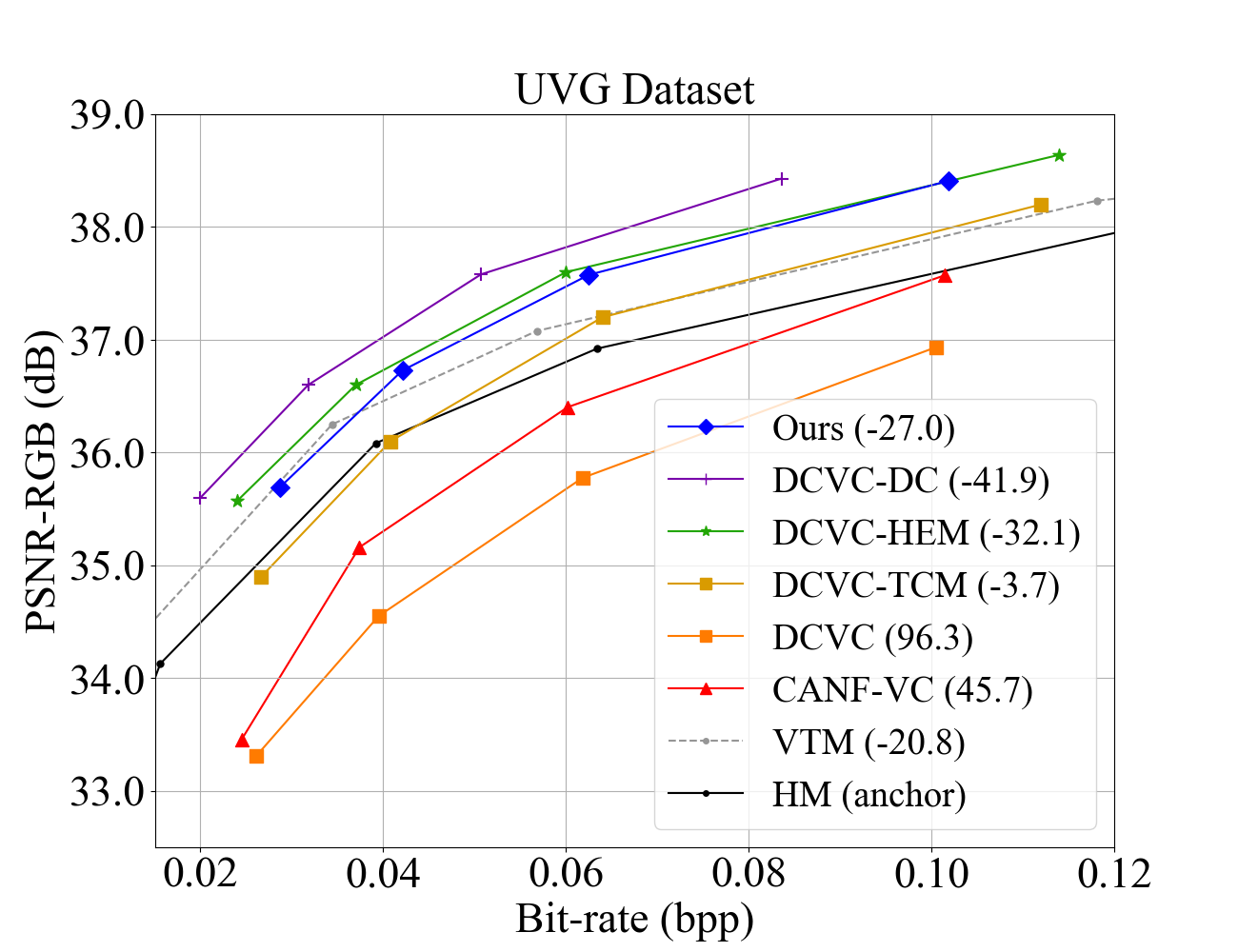}
        } \hspace{-0.4cm}
    \subfigure{
        \centering
        \includegraphics[height=0.245\linewidth, trim= 0 0 60 50, clip]{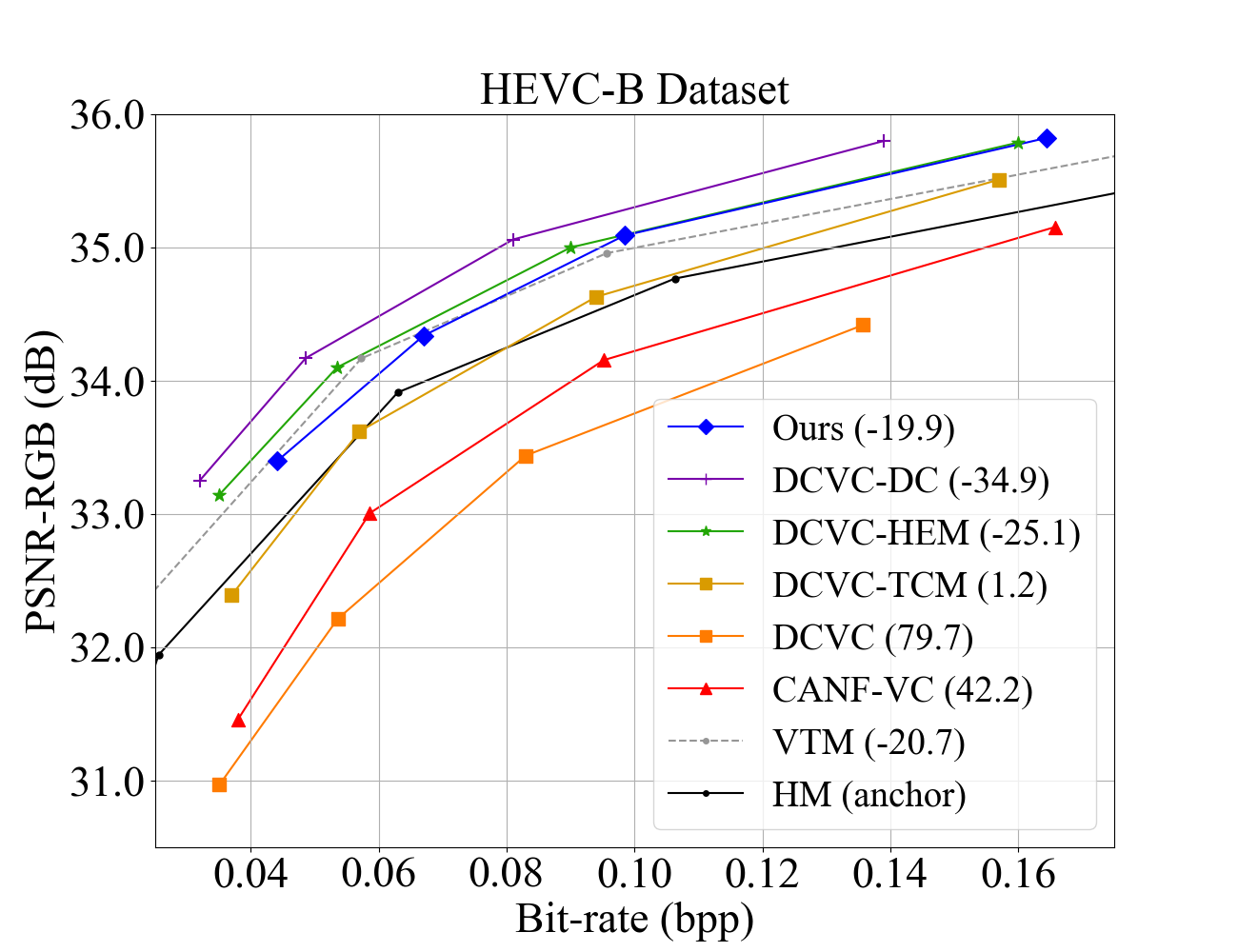}
        } \hspace{-0.4cm}
    \subfigure{
        \centering
        \includegraphics[height=0.245\linewidth, trim= 0 0 60 50, clip]{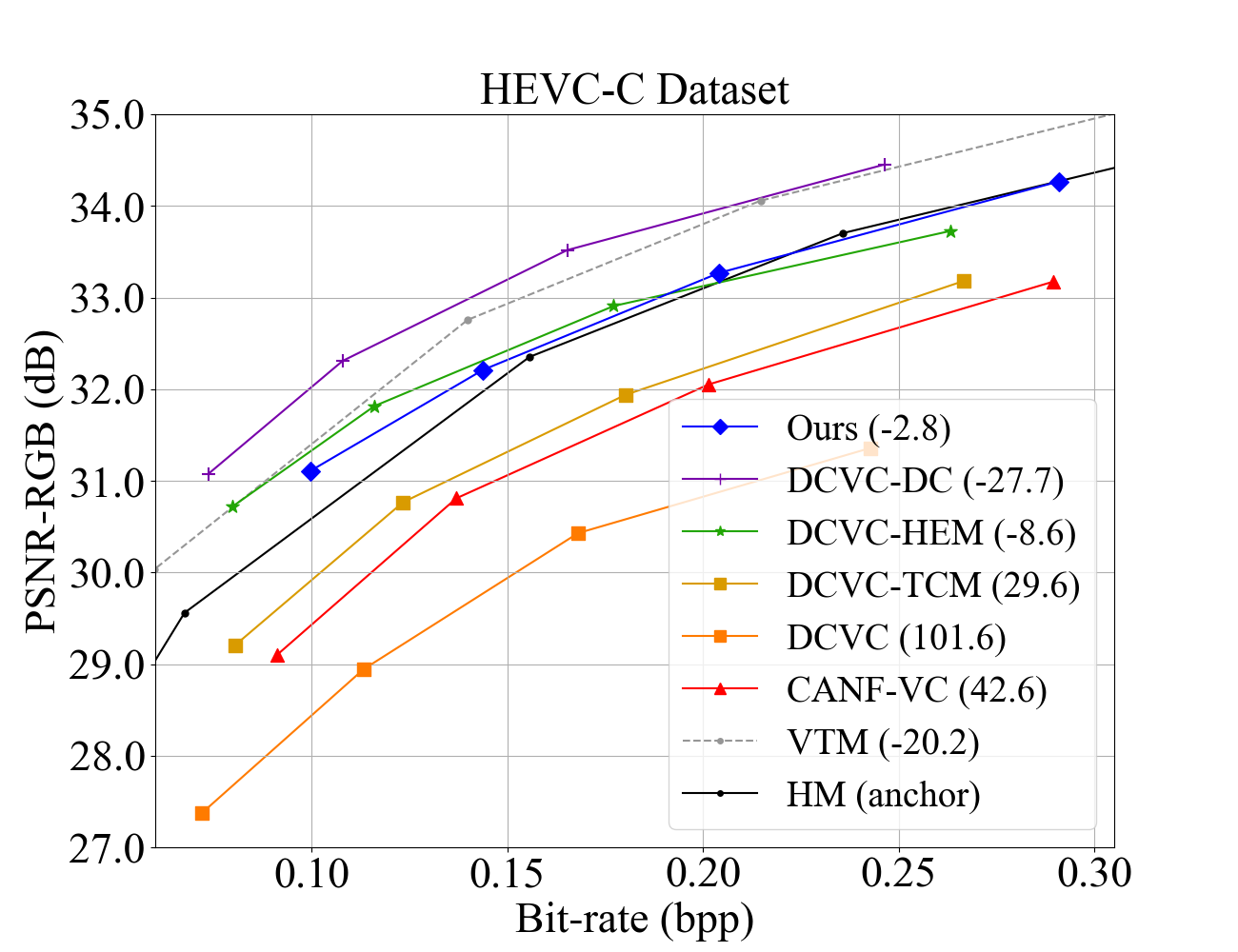}
        } \hspace{-0.4cm}
    \subfigure{
        \centering
        \includegraphics[height=0.245\linewidth, trim= 0 0 60 50, clip]{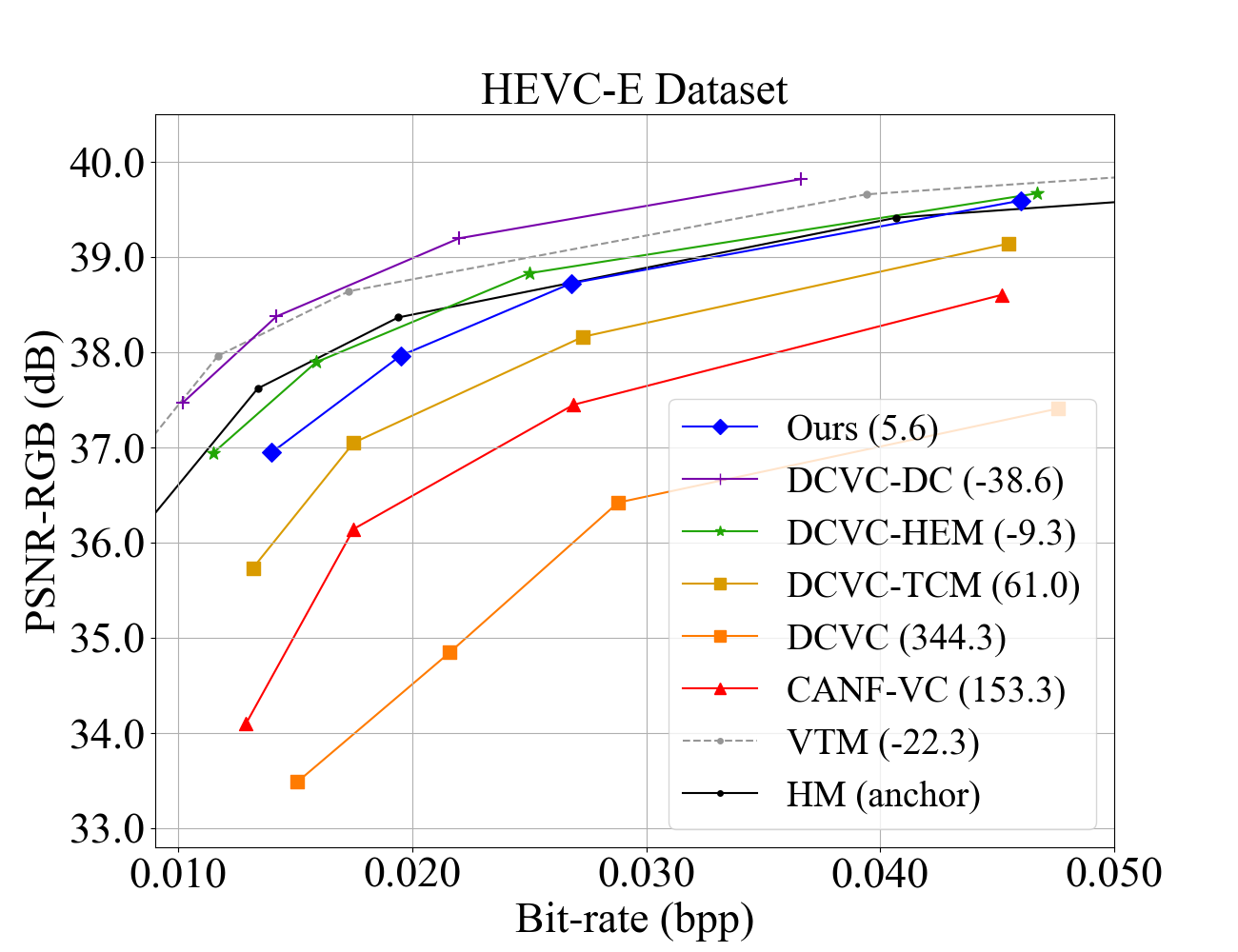}
        } \hspace{-0.4cm}
    \subfigure{
        \centering
        \includegraphics[height=0.245\linewidth, trim= 0 0 60 50, clip]{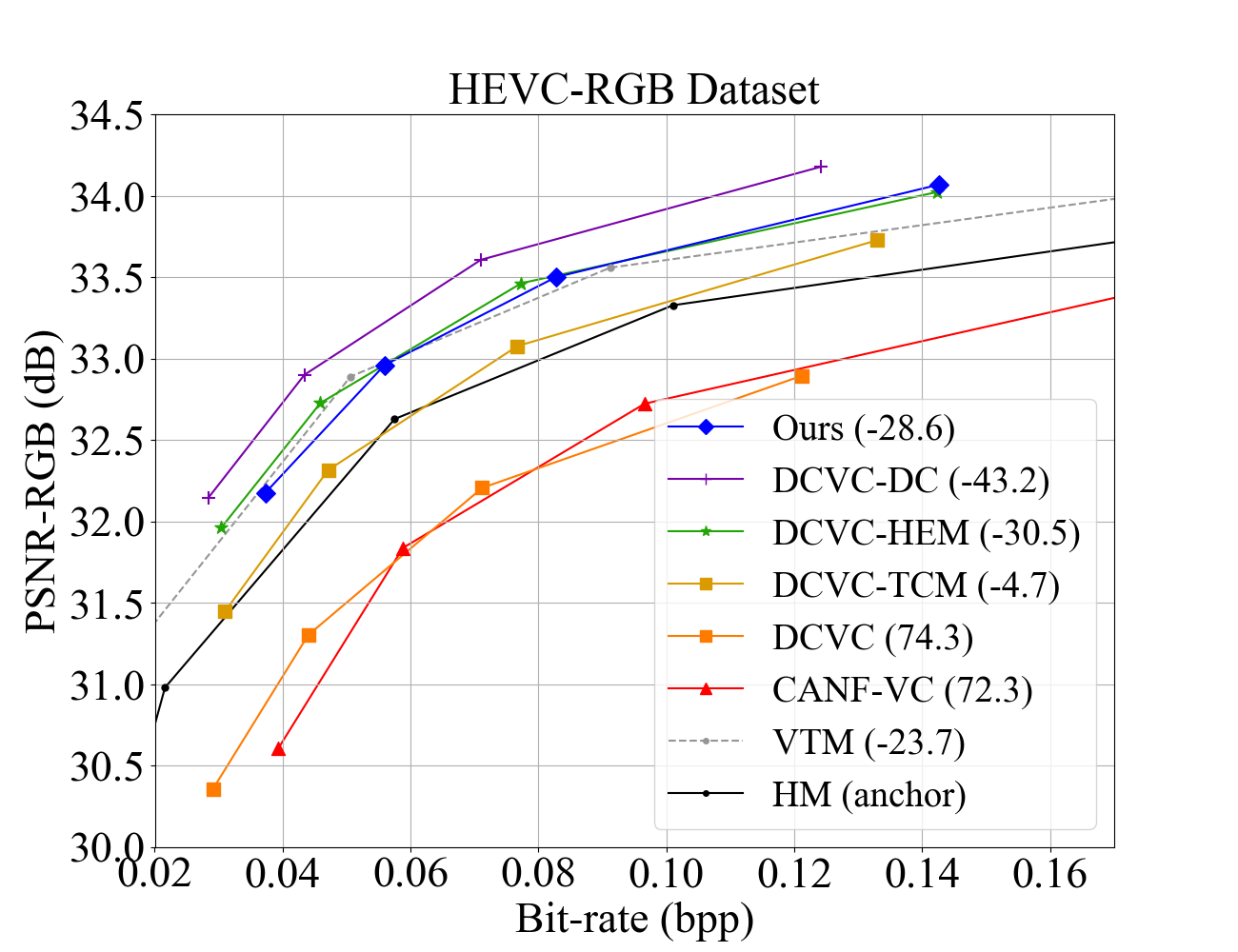}
        } \hspace{-0.4cm}
    \subfigure{
        \centering
        \includegraphics[height=0.245\linewidth, trim= 0 0 60 50, clip]{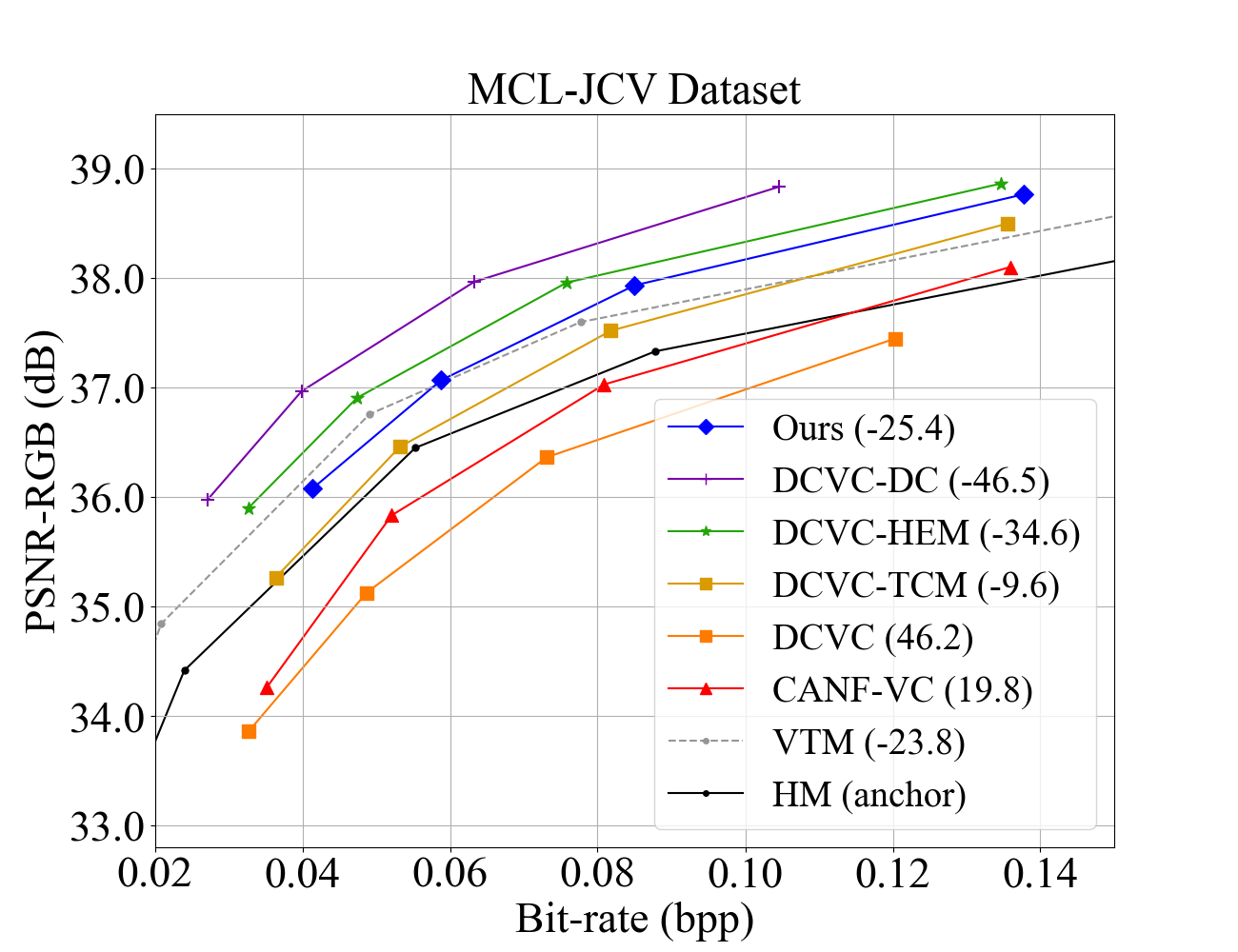}
        }
        
    \caption{\textcolor{black}{Rate-distortion performance comparison in terms of PSNR-RGB. The values within the parentheses represent BD-rates with HM-16.25 (low delay)~\cite{hm} serving as the anchor. Negative BD-rates suggest bitrate savings.}}
    \label{fig:p_frame_rd}
\end{figure*}
\begin{figure*}[h!]
    \centering
    \subfigure{
        \centering
        \includegraphics[height=0.245\linewidth, trim= 0 0 60 50, clip]{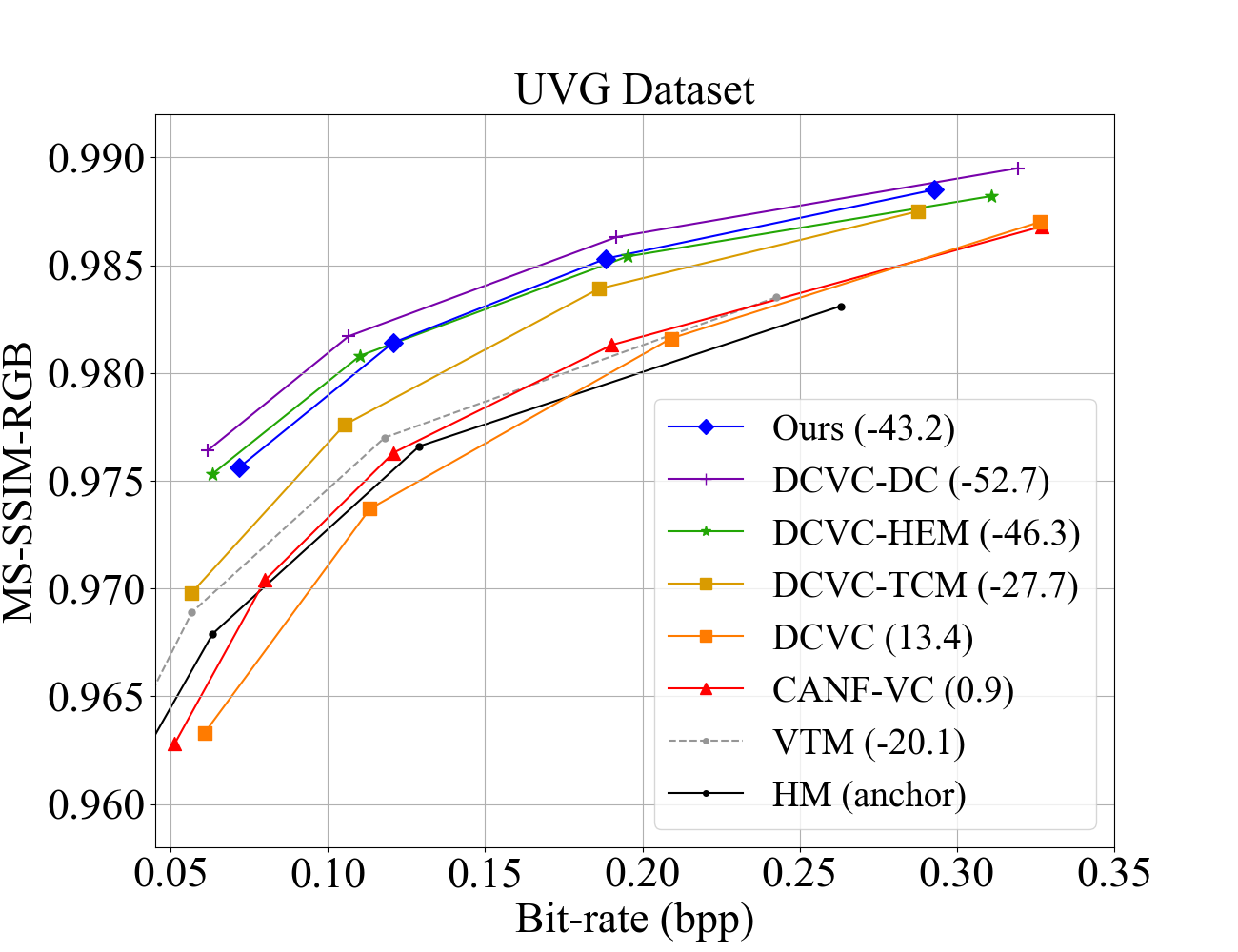}
        }  \hspace{-0.4cm}
    \subfigure{
        \centering
        \includegraphics[height=0.245\linewidth, trim= 0 0 60 50, clip]{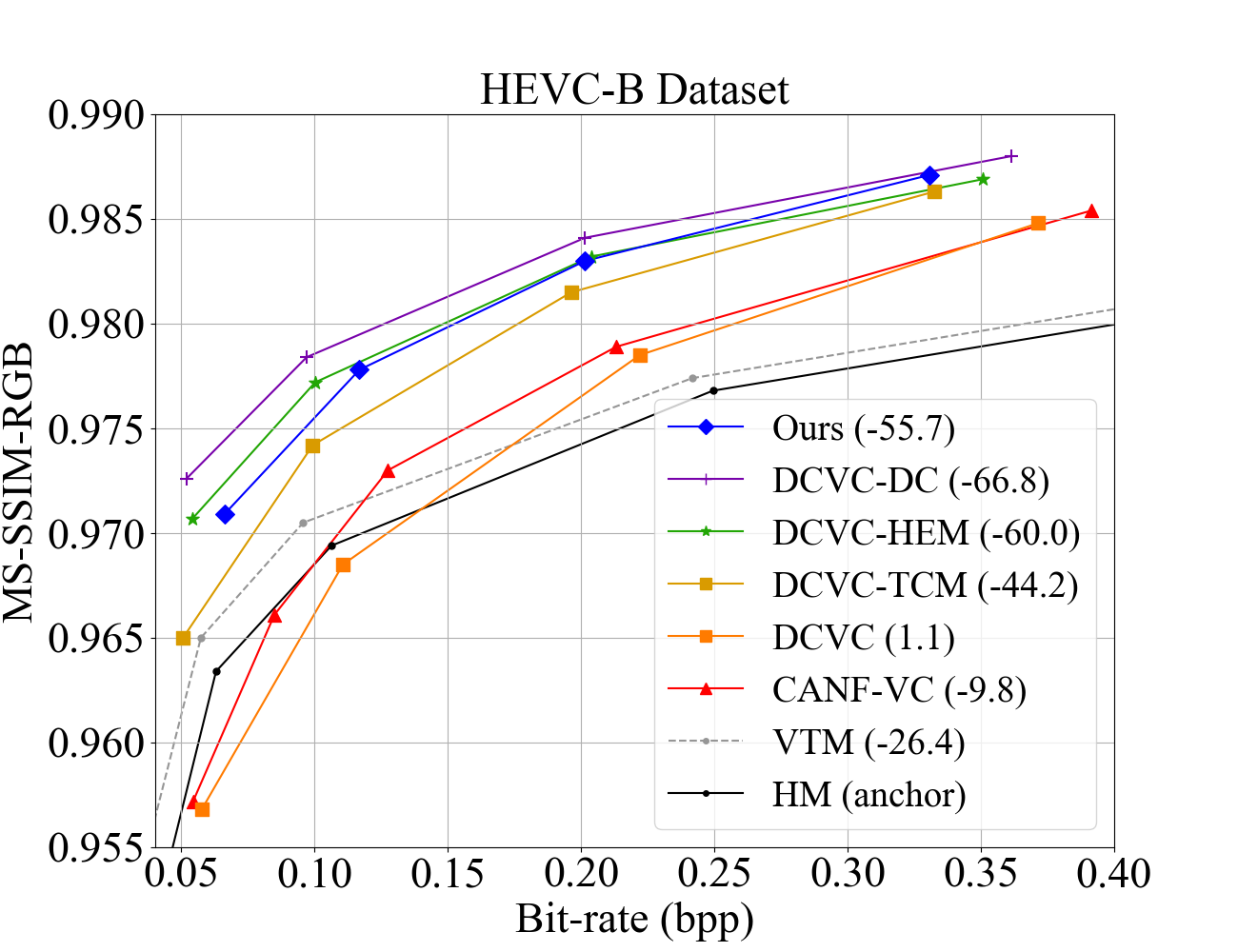}
        }  \hspace{-0.4cm}
    \subfigure{
        \centering
        \includegraphics[height=0.245\linewidth, trim= 0 0 60 50, clip]{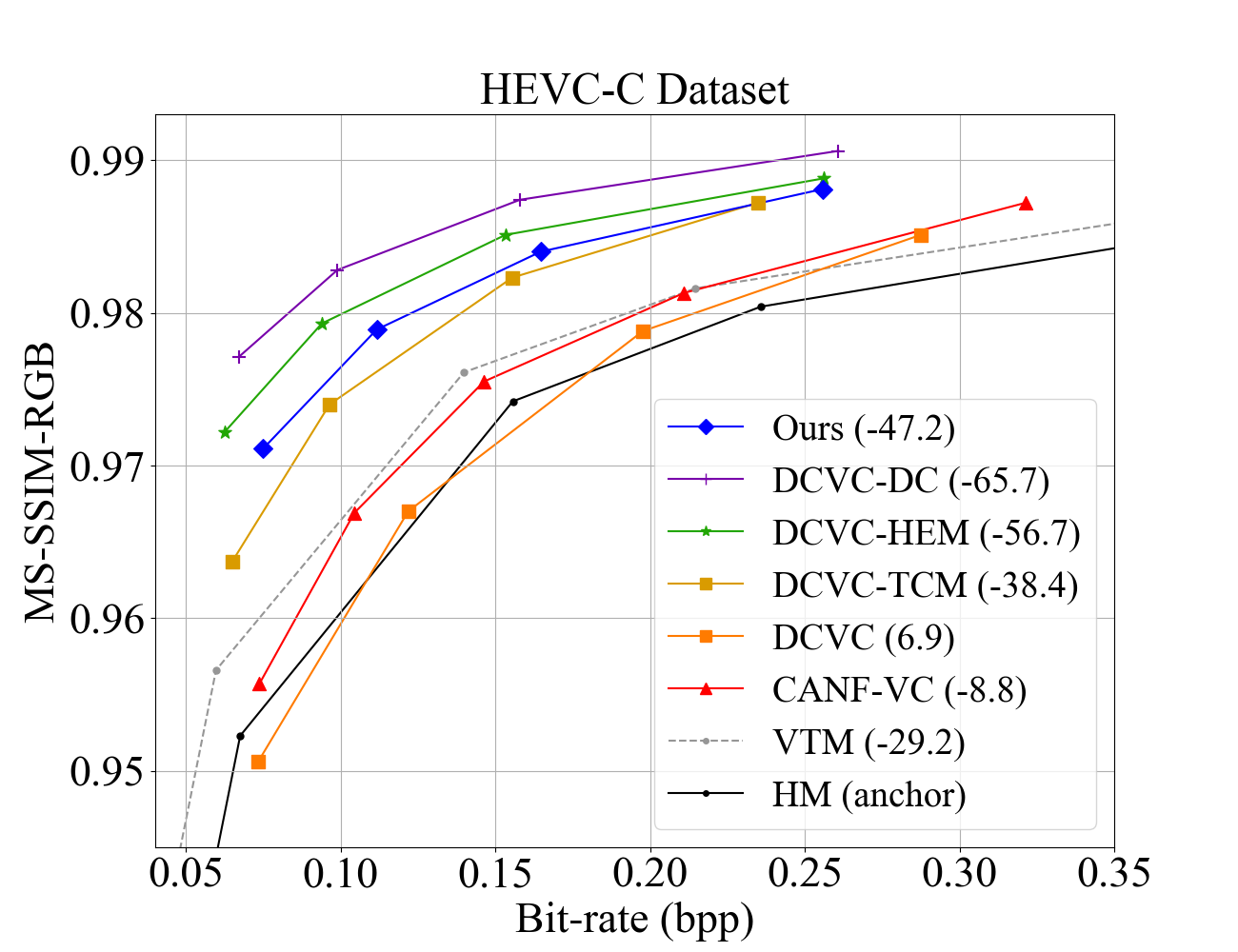}
        }  \hspace{-0.4cm}
    \subfigure{
        \centering
        \includegraphics[height=0.245\linewidth, trim= 0 0 60 50, clip]{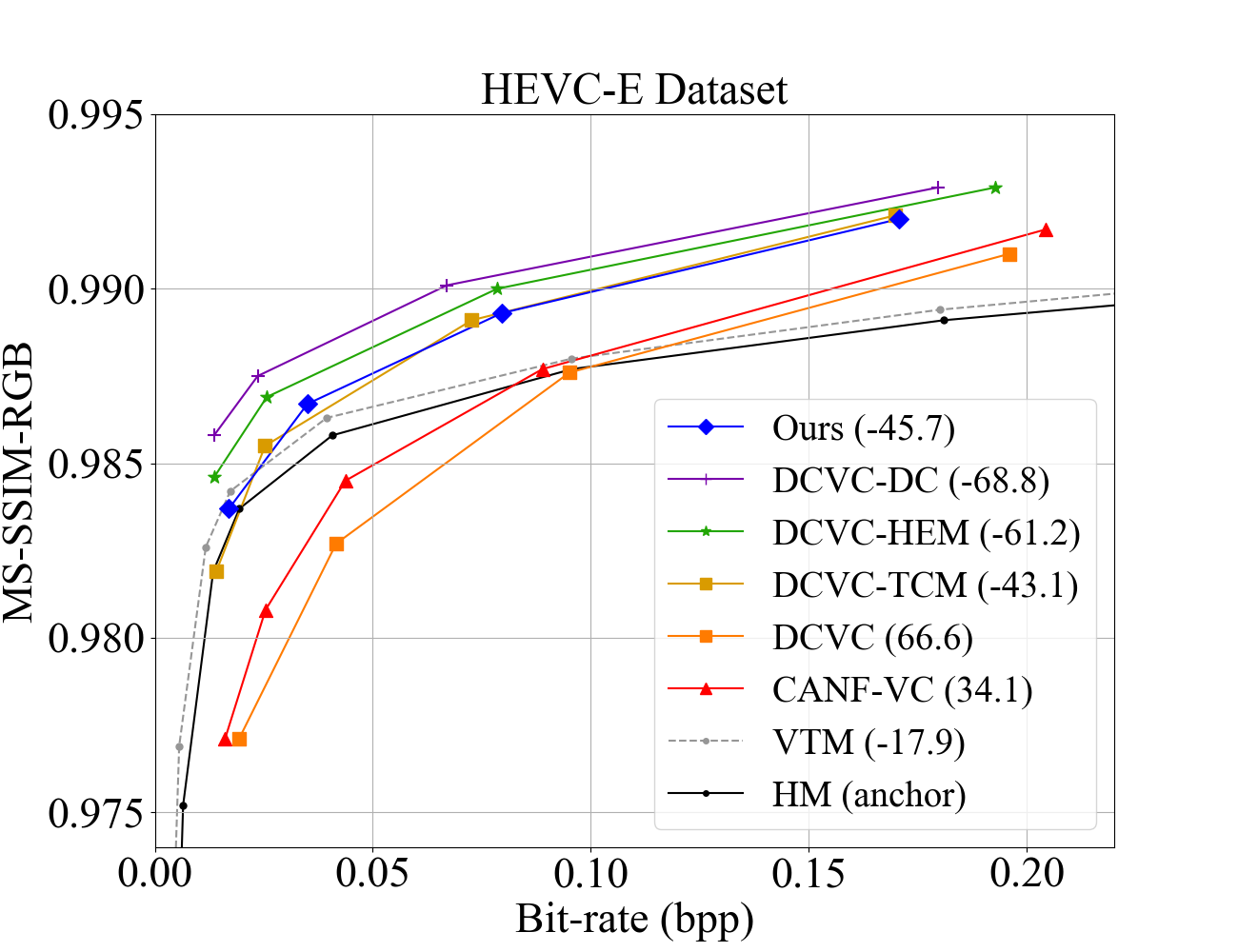}
        }  \hspace{-0.4cm}
    \subfigure{
        \centering
        \includegraphics[height=0.245\linewidth, trim= 0 0 60 50, clip]{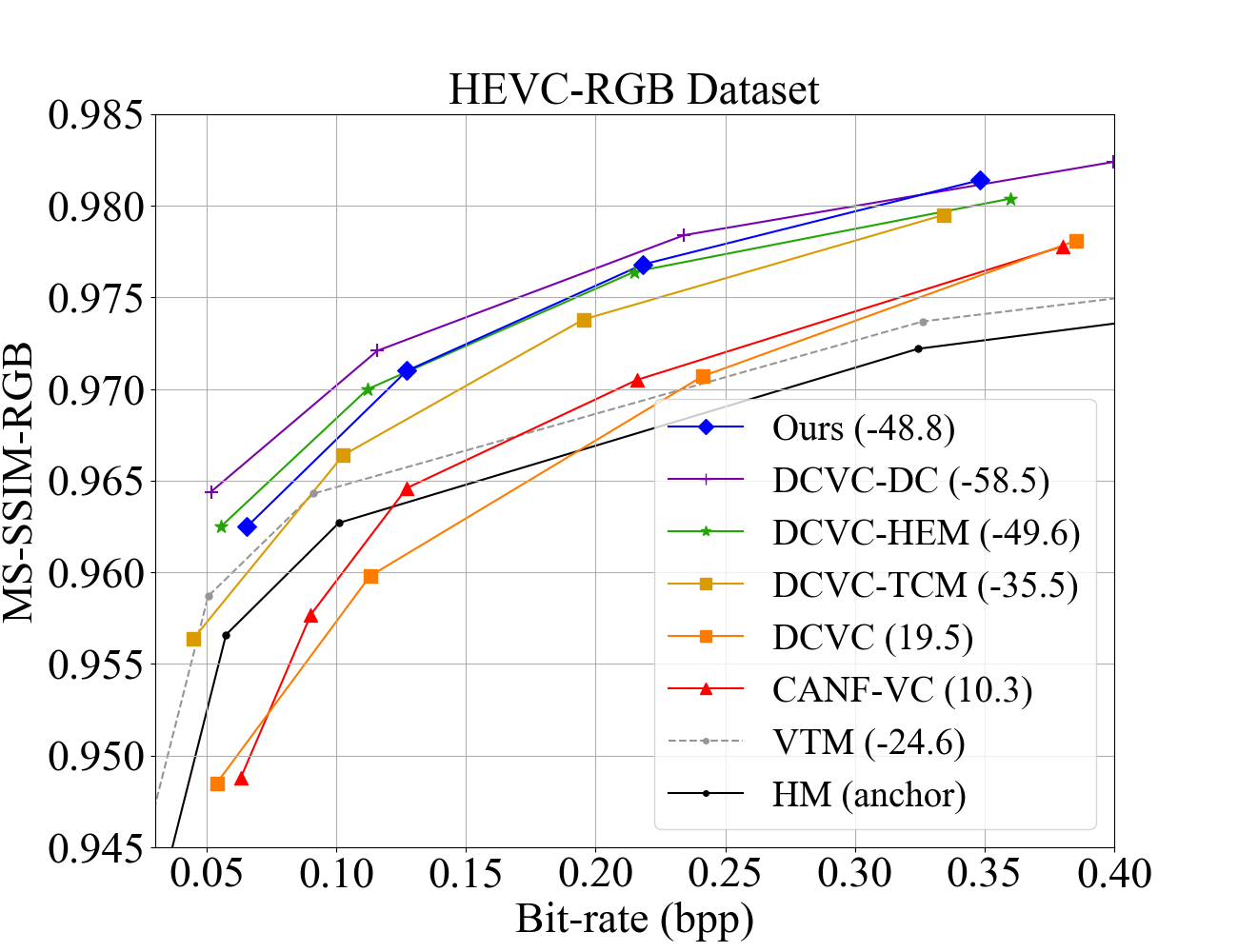}
        }  \hspace{-0.4cm}
    \subfigure{
        \centering
        \includegraphics[height=0.245\linewidth, trim= 0 0 60 50, clip]{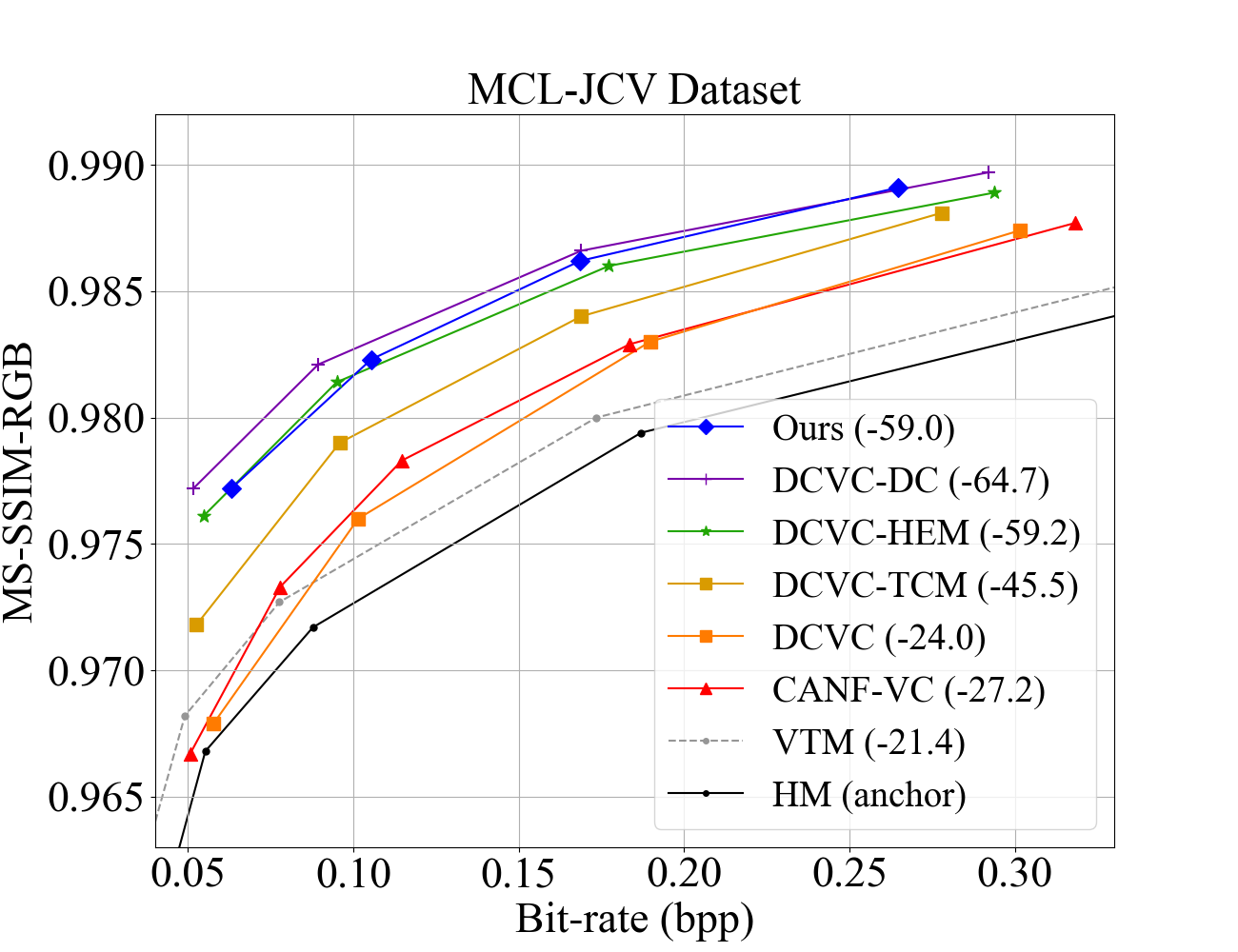}
        }
        
    \caption{\textcolor{black}{Rate-distortion performance comparison in terms of MS-SSIM-RGB. The values within the parentheses represent BD-rates with HM-16.25 (low delay)~\cite{hm} serving as the anchor. Negative BD-rates suggest bitrate savings.}}
    \label{fig:p_frame_ssim_rd}
\end{figure*}

\subsection{Comparison with the State-of-the-Art Methods}

We compare our MaskCRT with the state-of-the-art traditional codecs and learned P-frame codecs. The traditional codecs include HM-16.25~\cite{hm} and VTM-17.0~\cite{vtm} with \emph{encoder\_lowdelay\_main\_rext.cfg} and \emph{encoder\_lowdelay\_P\_vtm.cfg}, respectively. Following the recommendation from~\cite{dcvc_dc}, both codecs encode the input video in YUV444 format, because they perform \textcolor{black}{better than directly encoding in YUV420} under the common test conditions created for learned codecs. The learned codecs for comparison include \textcolor{black}{DCVC~\cite{dcvc}}, CANF-VC~\cite{canf}, DCVC-TCM~\cite{tcm}, DCVC-HEM~\cite{hem}, and DCVC-DC~\cite{dcvc_dc}. 

\begin{figure}[t]
    \centering
    \includegraphics[width=\linewidth]{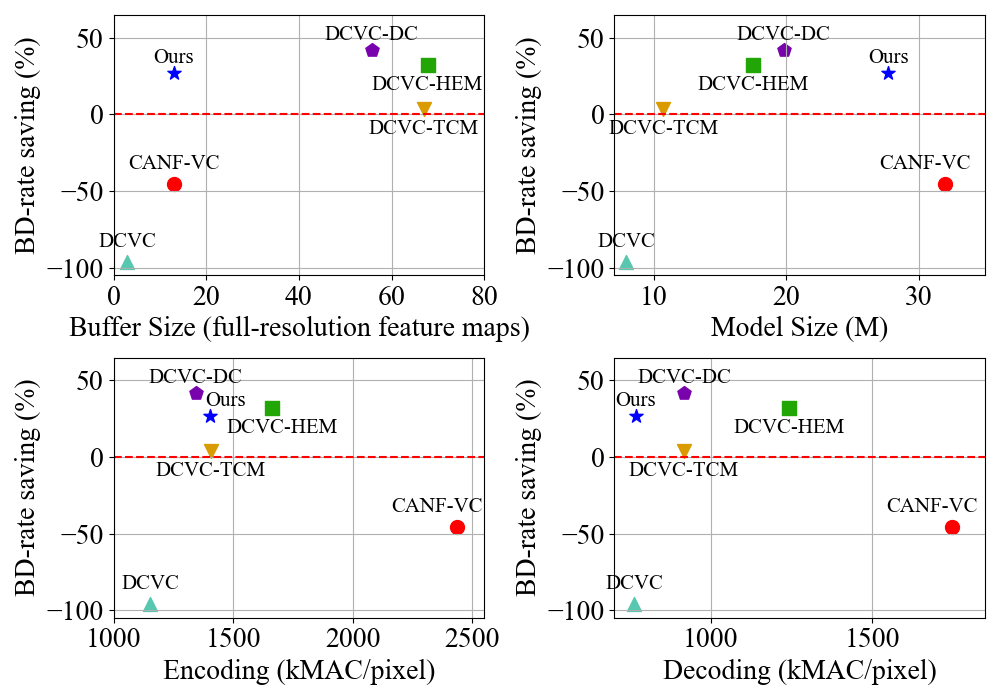}
    \caption{Characterization of complexity-performance trade-offs. The complexity metrics include the decoded picture buffer size, model size, and kMAC/pixel for encoding and decoding. The vertical axis is the BD-rate savings \textcolor{black}{in terms of PSNR-RGB} evaluated on UVG dataset with HM-16.25 (low delay) serving as the anchor. The higher the saving, the better the compression performance. The horizontal axis is the complexity metrics.}
    \label{fig:complexity_visual}
\end{figure}

Fig.~\ref{fig:p_frame_rd} and \textcolor{black}{Fig.~\ref{fig:p_frame_ssim_rd}} represent the rate-distortion comparison \textcolor{black}{in terms of PSNR-RGB and MS-SSIM-RGB, respectively}. Table~\ref{tab:complexity} and Fig.~\ref{fig:complexity_visual} report how the tested methods trade off between computational complexity and BD-rate savings. The model complexity is analyzed in four platform-independent metrics: the decoded picture buffer size, model size, and kMAC/pixel for encoding and decoding. 

MaskCRT is seen to outperform \textcolor{black}{DCVC~\cite{dcvc}}, CANF-VC~\cite{canf} and DCVC-TCM~\cite{tcm} in terms of PSNR-RGB \textcolor{black}{and MS-SSIM-RGB} across all the test datasets. Although it is inferior to DCVC-HEM~\cite{hem} and DCVC-DC~\cite{dcvc_dc}, MaskCRT uses only about one-fifth of the buffer size needed by \textcolor{black}{DCVC-TCM~\cite{tcm}, DCVC-HEM~\cite{hem} and DCVC-DC~\cite{dcvc_dc}} to achieve comparable compression performance. To be precise, \textcolor{black}{DCVC~\cite{dcvc} buffers 1 reconstructed frame while} both MaskCRT and CANF-VC~\cite{canf} buffer 3 reconstructed frames and 2 reconstructed optical flow maps, requiring the equivalent of 13 full-resolution feature maps. In contrast, DCVC-TCM~\cite{tcm} stores 64 full-resolution feature maps and 1 reconstructed frame, requiring the equivalent of 67 full-resolution feature maps. DCVC-HEM~\cite{hem} additionally stores the latents of the previously coded frame and coded flow as the latent prior for entropy coding, having the equivalent of 67.75 full-resolution feature maps. Similarly, DCVC-DC~\cite{dcvc_dc} buffers 48 full-resolution feature maps, 1 reconstructed frame, the latent prior, and 4 full-resolution motion feature maps, requiring the equivalent of 55.75 full-resolution feature maps. \textcolor{black}{For benchmarking, both VVC and HEVC use 4 reference frames in RGB format, which amount to 12 full-resolution feature maps.} We stress that the decoded picture buffer size has implications for practicality. On resource-limited devices, the decoded picture buffer usually sits in the off-chip memory. Accessing a large number of feature maps in the off-chip memory at high frame rates incurs high memory bandwidth, which becomes an obstacle to real-time applications. As compared with CANF-VC, MaskCRT shows much higher BD-rate savings while reducing the encoding and decoding kMAC/pixel considerably. In addition, it has comparable encoding kMAC/pixel to the DCVC family, but the lowest decoding kMAC/pixel among all the competing methods. It is worth noting that DCVC-DC~\cite{dcvc_dc} has made great effort to reduce the model complexity. To our best knowledge, the other methods have not received the same level of optimization. Last but not least, \textcolor{black}{MaskCRT outperforms HM~\cite{hm} and VTM~\cite{vtm} in terms of MS-SSIM, as these traditional codecs are not optimized for MS-SSIM. In terms of PSNR-RGB, MaskCRT} is one of the few learned video codecs that shows comparable BD-rate results to VTM-17.0 across all the test datasets.

\begin{table}[t]
\centering
    \caption{Complexity comparison between our MaskCRT and the state-of-the-art learned P-frame codecs.}
    \label{tab:complexity}
    \begin{tabular}{cccccc}
    \hline
    Methods  & \begin{tabular}[c]{@{}c@{}}Model\\ Size (M)\end{tabular} & \begin{tabular}[c]{@{}c@{}}Encoding\\ kMAC/pixel\end{tabular} & \begin{tabular}[c]{@{}c@{}}Decoding\\ kMAC/pixel\end{tabular} & \begin{tabular}[c]{@{}c@{}}Buffer\\ Size\tablefootnote{\textcolor{black}{The buffer size is measured in terms of the number of full-resolution feature maps needed to be stored for coding one input frame. That is, these feature maps have the same spatial resolution as the input frame. As an example, one full-resolution RGB frame is equivalent to 3 full-resolution feature maps.}}\end{tabular} \\ \hline
    \textcolor{black}{DCVC~\cite{dcvc}}        & \textcolor{black}{7.9}       & \textcolor{black}{1153.43}    & \textcolor{black}{762.41}  & \textcolor{black}{3}        \\ \hline
    CANF-VC~\cite{canf}     & 32        & 2435.8     & 1748.3  & 13       \\ \hline
    DCVC-TCM~\cite{tcm}     & 10.7      & 1406.6     & 915.5   & 67       \\ \hline
    DCVC-HEM~\cite{hem}     & 17.5      & 1662.8     & 1241.1  & 67.75    \\ \hline
    DCVC-DC~\cite{dcvc_dc}  & 19.8      & 1343.7     & 917.1   & 55.75    \\ \hline
    Ours (MaskCRT)          & 27.7      & 1401.8     & 767.5   & 13       \\ \hline
    \end{tabular}
\end{table}

\section{Conclusion}
\label{sec:conclusion}
In this work, we propose a Transformer-based conditional video codec, termed MaskCRT. It features a Transformer-based conditional autoencoder, a spatially adaptive hybrid of conditional coding and conditional residual coding, and a channel transform module (CTM) to decorrelate the image latents along the channel dimension. Experimental results confirm the superiority of MaskCRT to the single use of conditional coding or conditional residual coding. Our CTM is also seen to be a cost effective solution that strikes a good balance between the coding efficiency and model complexity. Compared to the state-of-the-art learned video codecs, MaskCRT exhibits competitive coding performance while requiring significantly smaller buffer size.

\textcolor{black}{\textbf{Limitations:} The key to the improved coding efficiency of MaskCRT is the efficiency of temporal prediction, which depends on the pixel-level soft mask $m$. Currently, the soft mask $m$ is predicted based solely on the causal information, such as the decoded flow map and the reconstructed temporal predictor, without having access to the current input frame in order to save the overhead needed to signal the mask. How to better predict or even signal the mask for video sequences with scene change remains an open issue.} 

\textcolor{black}{\textbf{Future Work:} MaskCRT presents a new research direction for learned video compression. While most conditional coding schemes are focused on how to formulate a better conditioning signal at the cost of increased buffer size or complexity in order to remedy the bottleneck issue, MaskCRT exemplifies that formulating a good temporal predictor in a conditional residual coding framework is able to achieve better coding efficiency in a cost effective manner. It also allows the traditional temporal prediction techniques, such as bi-directional prediction, long-term prediction, and multiple reference frames, to be easily incorporated. We shall extend MaskCRT by considering these temporal prediction techniques in the future work.} 

\bibliography{egbib}
\bibliographystyle{IEEEtran}

\begin{IEEEbiography}[{\includegraphics[width=1in,height=1.25in,clip,keepaspectratio]{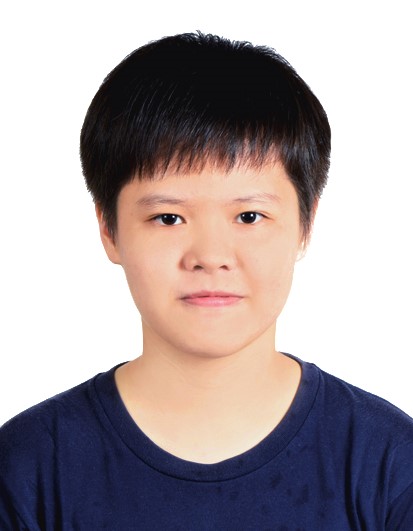}}]{Yi-Hsin Chen} 
received her B.S. degree in applied mathematics from National Chung Hsing University (NCHU), Taiwan, and her M.S. degree in data science and engineering from National Chiao Tung University (NCTU), Taiwan, in 2018 and 2020, respectively. She is currently pursuing her Ph.D. degree in computer science and engineering, National Yang Ming Chiao Tung University (NYCU), Taiwan. Her research interests include learning-based image/video coding, image/video restoration, computer vision, and deep learning.
\end{IEEEbiography}
\vspace{-3em}

\begin{IEEEbiography}[{\includegraphics[width=1in,height=1.25in,clip,keepaspectratio]{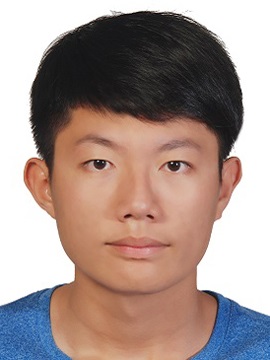}}]{Hong-Sheng Xie} 
received his B.S. degree in applied mathematics from National Chung Hsing University (NCHU), Taiwan, and his M.S. degree in computer science and engineering, National Yang Ming Chiao Tung University (NYCU), Taiwan, in 2021 and 2024, respectively. His research interests include learning-based image/video coding, computer vision, and deep learning.
\end{IEEEbiography}
\vspace{-3em}

\begin{IEEEbiography}[{\includegraphics[width=1in,height=1.25in,clip,keepaspectratio]{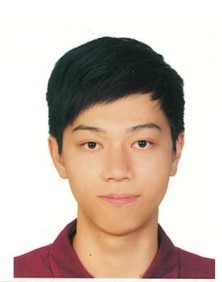}}]{Cheng-Wei Chen} 
received his B.S. degree in applied mathematics from National University of Kaohsiung (NUK), Taiwan, in 2022. He is currently pursuing his M.S. degree in computer science and engineering, National Yang Ming Chiao Tung University (NYCU), Taiwan. His research interests include learning-based image/video coding, computer vision, and deep learning.
\end{IEEEbiography}
\vspace{-3em}

\begin{IEEEbiography}[{\includegraphics[width=1in,height=1.25in,clip,keepaspectratio]{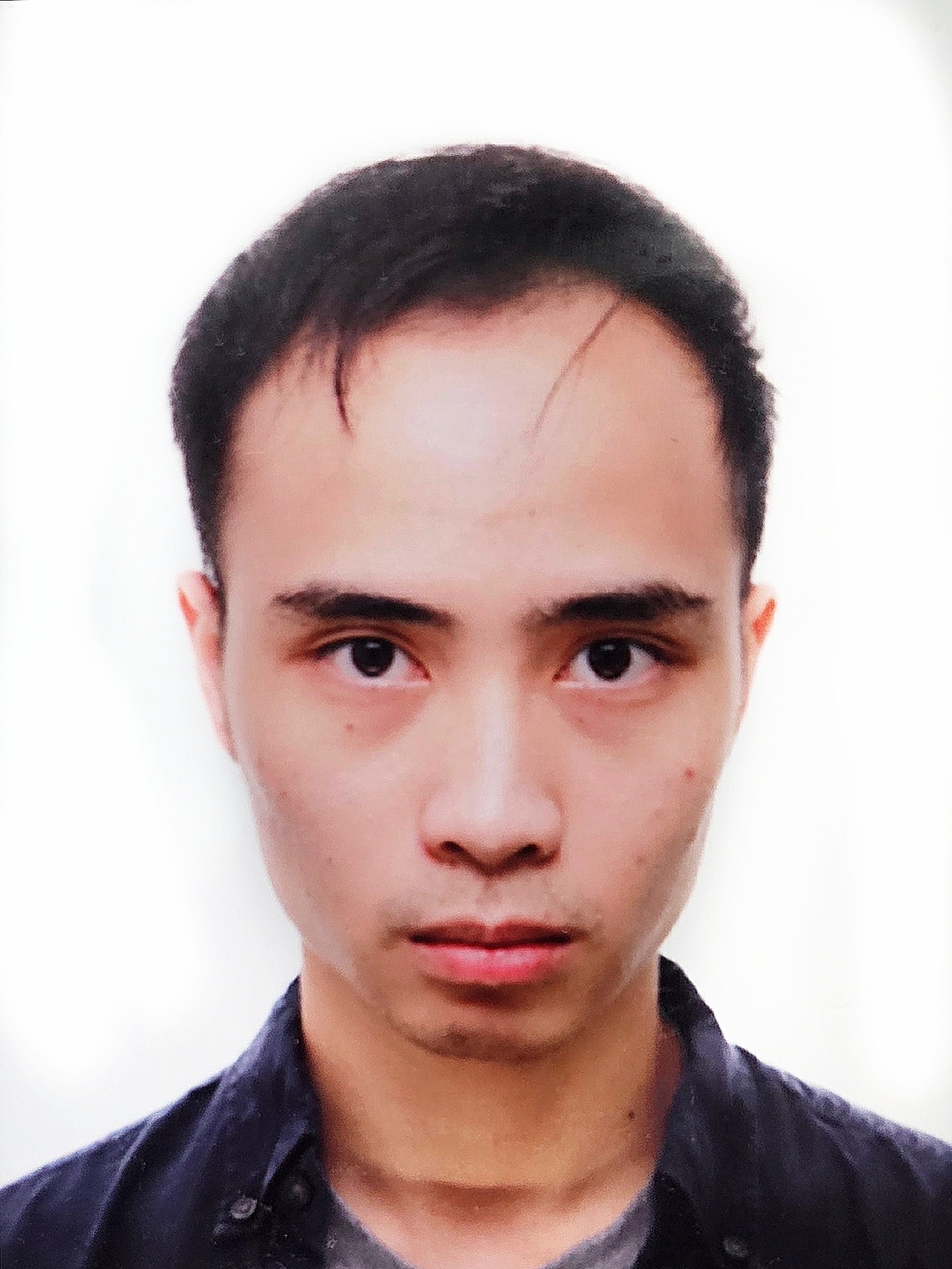}}]{Zong-Lin Gao} 
received his B.S. degree in applied Industrial Engineering from National Yang Ming Chiao Tung University (NYCU), Taiwan, in 2022. He is currently pursuing his M.S. degree in computer science and engineering, National Yang Ming Chiao Tung University (NYCU), Taiwan. His research interests include learning-based image/video coding, computer vision, and deep learning.
\end{IEEEbiography}
\vspace{-3em}

\begin{IEEEbiography}[{\includegraphics[width=1in,height=1.25in,clip,keepaspectratio]{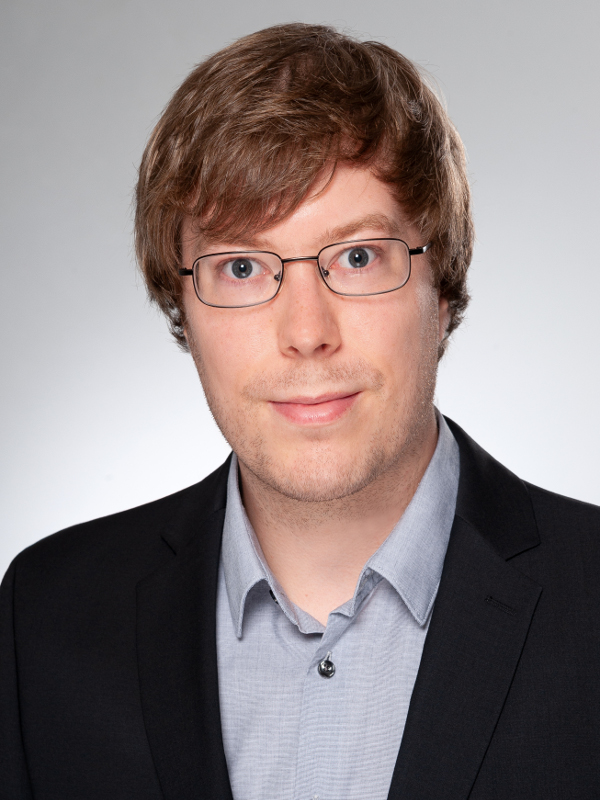}}]{Martin Benjak} 
studied electrical engineering at the University of Applied Sciences of Osnabrück and at the Leibniz University of Hannover. He received his B.Sc. and M.Sc. in 2017 and 2019, respectively. After graduating, he joined the Institut für Informationsverarbeitung at Leibniz University Hannover where he currently works as a research assistant towards his PhD. His current research interests are learning-based video coding and video coding for machines.
\end{IEEEbiography}
\vspace{-3em}

\begin{IEEEbiography}[{\includegraphics[width=1in,height=1.25in,clip,keepaspectratio]{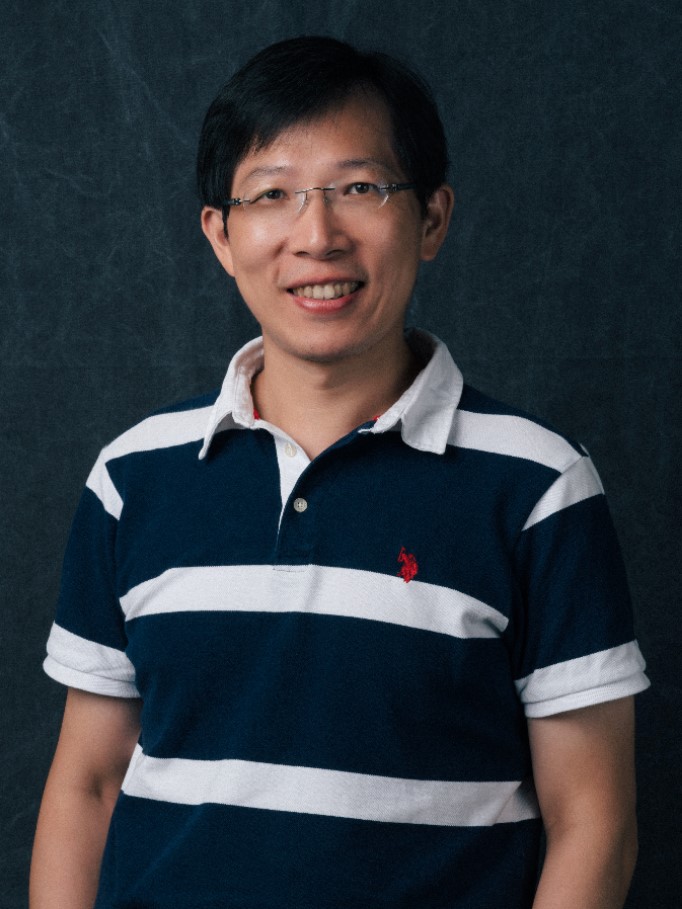}}]{Wen-Hsiao Peng} 
received his Ph.D. degree from National Chiao Tung University (NCTU), Taiwan, in 2005. He was with the Intel Microprocessor Research Laboratory, USA, from 2000 to 2001, where he was involved in the development of ISO/IEC MPEG-4 fine granularity scalability. Since 2003, he has actively participated in the ISO/IEC and ITU-T video coding standardization process and contributed to the development of SVC, HEVC, and SCC standards. He was a Visiting Scholar with the IBM Thomas J. Watson Research Center, Yorktown Heights, NY, USA, from 2015 to 2016. He has authored over 100+ technical papers in the field of video/image processing and communications and over 60 standards contributions. His research interests include neural network-based image/video coding, ISO/IEC \& ITU-T video coding standards, visual signal processing , computer vision, etc. He is Past Chair of the IEEE Circuits and Systems Society (CASS) Visual Signal Processing and Communications (VSPC) Technical Committee. During his term, VSPC Technical Committee received the IEEE CASS Outstanding Technical Committee Recognition in 2021. He served as AEiC for Digital Communications/Lead Guest Editor/Guest Editor/SEB Member for IEEE Journal on Emerging and Selected Topics in Circuits and Systems (JETCAS), Associate Editor/Special Session Organizer for IEEE TCSVT, and Guest Editor for IEEE TCAS-II. He was Distinguished Lecturer of the IEEE CASS (2022-2023) and APSIPA (2017-2018). He was appointed Editor-in-Chief of the IEEE JETCAS for 2024-2025.
\end{IEEEbiography}

\begin{IEEEbiography}[{\includegraphics[width=1in,height=1.25in,clip,keepaspectratio]{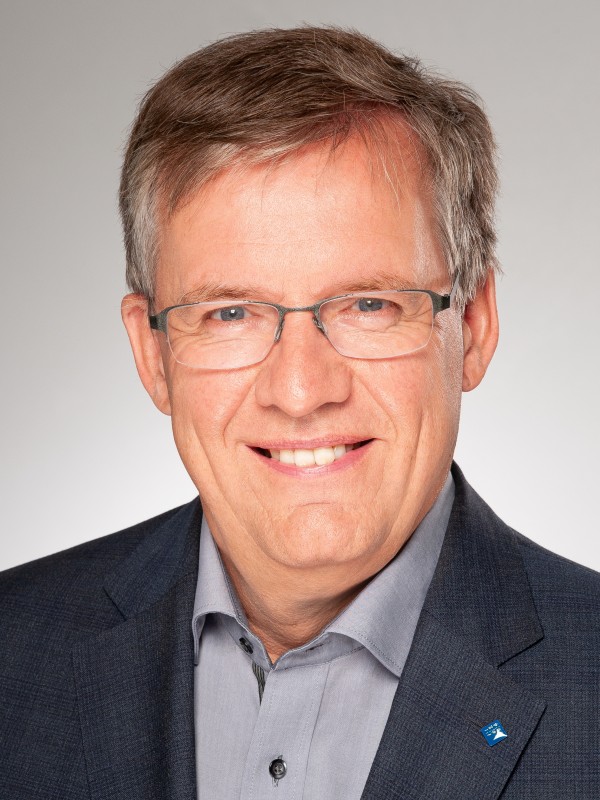}}]{Prof. Dr.-Ing. Jörn Ostermann} 
studied Electrical Engineering and Communications Engineering at the University of Hannover and Imperial College London, respectively. In 1994 he received a Dr.-Ing. from the University of Hannover for his work on low bit-rate and object-based analysis-synthesis video coding. Since 2003 he is Full Professor and Head of the Institut für Informationsverarbeitung at Leibniz Universität Hannover, Germany. In July 2020 he was appointed Convenor of MPEG Technical Coordination. He is a Fellow of the IEEE (class of 2005) and member of the IEEE Technical Committee on Multimedia Signal Processing and past chair of the IEEE CAS Visual Signal Processing and Communications (VSPC) Technical Committee. He is named as inventor on more than 30 patents. His current research interests are video coding and streaming, computer vision, machine learning, 3D modeling, face animation, and computer-human interfaces.
\end{IEEEbiography}

\vfill
\end{document}